\documentclass[times, twocolumn, trackchanges]{aastex63} % turn on change highlighting

\usepackage{xcolor}
\usepackage{amsmath}
\usepackage{graphicx}
\usepackage{natbib}
\usepackage{multirow}
\usepackage{booktabs}
\usepackage{tabularx}
\usepackage{savesym}
\savesymbol{tablenum}
\usepackage{siunitx}
\restoresymbol{SIX}{tablenum}

\usepackage{xspace} % handle spacing after newcommand macros

\newcommand{\I}{$\mathcal{I}$\xspace}
\newcommand{\Q}{$\mathcal{Q}$\xspace}
\newcommand{\U}{$\mathcal{U}$\xspace}
\newcommand{\Qr}{$\mathcal{Q}_\phi$\xspace}
\newcommand{\Ur}{$\mathcal{U}_\phi$\xspace}
\newcommand{\Lir}{$L_{\mathrm{IR}}/L_\star$\xspace}
\newcommand{\Ntarg}{185\xspace} % total number of disk targets on final list
\newcommand{\Ndebris}{26\xspace} % total number of debris disk detections.
\newcommand{\Nnondet}{75\xspace} % total number of non-detections.
\newcommand{\Nobs}{104\xspace} % total number of targets observed.
\newcommand{\Npolonly}{8\xspace} % number of pol-only detections.

\definecolor{lightgray}{gray}{0.9}

\accepted{27 April 2020}
\begin{document}

%% If you wish, you may supply running head information, although
%% this information may be modified by the editorial offices.
%% The left head contains a list of authors,
%% usually a maximum of three (otherwise use et al.).  The right
%% head is a modified title of up to roughly 44 characters.
%% Running heads will not print in the manuscript style.

\shorttitle{GPIES Debris Disks}
\shortauthors{Esposito et al.}


% TITLE

\title{Debris Disk Results from the Gemini Planet Imager Exoplanet Survey's Polarimetric Imaging Campaign}

%% Use \author, \affiliation, and the \and command to format

% AUTHORS

\correspondingauthor{Thomas M. Esposito}
\email{tesposito@berkeley.edu}

% Only format to set author list as of aastex6.1.

\author[0000-0002-0792-3719]{Thomas M. Esposito}
\affiliation{Astronomy Department, University of California, Berkeley, CA 94720, USA}

\author[0000-0002-6221-5360]{Paul Kalas}
\affiliation{Astronomy Department, University of California, Berkeley, CA 94720, USA}
\affiliation{SETI Institute, Carl Sagan Center, 189 Bernardo Ave.,  Mountain View, CA 94043, USA}
\affiliation{Institute of Astrophysics, FORTH, GR-71110 Heraklion, Greece}

\author[0000-0002-0176-8973]{Michael P. Fitzgerald}
\affiliation{Department of Physics \& Astronomy, 430 Portola Plaza, University of California, Los Angeles, CA 90095, USA}

\author[0000-0001-6205-9233]{Maxwell A. Millar-Blanchaer}
\altaffiliation{NASA Hubble Fellow}
\affiliation{NASA Jet Propulsion Laboratory, California Institute of Technology, Pasadena, CA 91109, USA}

\author[0000-0002-5092-6464]{Gaspard Duch\^ene}
\affiliation{Astronomy Department, University of California, Berkeley, CA 94720, USA}
\affiliation{Universit\'e Grenoble Alpes / CNRS, Institut de Plan\'etologie et d'Astrophysique de Grenoble, 38000 Grenoble, France}

\author{Jennifer Patience}
\affiliation{School of Earth and Space Exploration, Arizona State University, PO Box 871404, Tempe, AZ 85287, USA}

\author[0000-0001-9994-2142]{Justin Hom}
\affiliation{School of Earth and Space Exploration, Arizona State University, PO Box 871404, Tempe, AZ 85287, USA}

\author[0000-0002-3191-8151]{Marshall D. Perrin}
\affiliation{Space Telescope Science Institute, 3700 San Martin Drive, Baltimore, MD 21218, USA}

\author[0000-0002-4918-0247]{Robert J. De Rosa}
\affiliation{Kavli Institute for Particle Astrophysics and Cosmology, Stanford University, Stanford, CA 94305, USA}

\author{Eugene Chiang}
\affiliation{Astronomy Department, University of California, Berkeley, CA 94720, USA}
\affiliation{Earth and Planetary Science Department, University of California, Berkeley, CA 94720, USA}

\author[0000-0002-1483-8811]{Ian Czekala}
\altaffiliation{NASA Hubble Fellowship Program Sagan Fellow}
\affiliation{Astronomy Department, University of California, Berkeley, CA 94720, USA}

\author[0000-0003-1212-7538]{Bruce Macintosh}
\affiliation{Kavli Institute for Particle Astrophysics and Cosmology, Stanford University, Stanford, CA 94305, USA}

\author{James R. Graham}
\affiliation{Astronomy Department, University of California, Berkeley, CA 94720, USA}

\author[0000-0003-4142-9842]{Megan Ansdell}
\affiliation{Astronomy Department, University of California, Berkeley, CA 94720, USA}

\author[0000-0001-6364-2834]{Pauline Arriaga}
\affiliation{Department of Physics \& Astronomy, 430 Portola Plaza, University of California, Los Angeles, CA 90095, USA}

\author{Sebastian Bruzzone}
\affiliation{Department of Physics and Astronomy, Centre for Planetary Science and Exploration, The University of Western Ontario, London, ON N6A 3K7, Canada}

\author{Joanna Bulger}
\affiliation{Pan-STARRS Observatory, Institute for Astronomy, University of Hawai’i, 2680 Woodlawn Drive, Honolulu, HI 96822, USA}

\author[0000-0002-8382-0447]{Christine H. Chen}
\affiliation{Space Telescope Science Institute, 3700 San Martin Drive, Baltimore, MD 21218, USA}

\author{Tara Cotten}
\affiliation{Physics and Astronomy, University of Georgia, 240 Physics, Athens, GA 30602, USA}

\author[0000-0001-9290-7846]{Ruobing Dong}
\affiliation{University of Victoria, Department of Physics and Astronomy, 3800 Finnerty Rd, Victoria, BC V8P 5C2, Canada}

\author{Zachary H. Draper}
\affiliation{University of Victoria, Department of Physics and Astronomy, 3800 Finnerty Rd, Victoria, BC V8P 5C2, Canada}
\affiliation{National Research Council of Canada Herzberg, 5071 West Saanich Road, Victoria, BC V9E 2E7, Canada}

\author[0000-0002-7821-0695]{Katherine B. Follette}
\affiliation{Physics and Astronomy Department, Amherst College, 21 Merrill Science Drive, Amherst, MA 01002, USA}

\author{Li-Wei Hung}
\affiliation{Department of Physics \& Astronomy, 430 Portola Plaza, University of California, Los Angeles, CA 90095, USA}

\author{Ronald Lopez}
\affiliation{Department of Physics \& Astronomy, 430 Portola Plaza, University of California, Los Angeles, CA 90095, USA}

\author{Brenda C. Matthews}
\affiliation{National Research Council of Canada Herzberg, 5071 West Saanich Road, Victoria, BC V9E 2E7, Canada}
\affiliation{University of Victoria, Department of Physics and Astronomy, 3800 Finnerty Rd, Victoria, BC V8P 5C2, Canada}

\author[0000-0002-9133-3091]{Johan Mazoyer}
\altaffiliation{NASA Hubble Fellowship Program Sagan Fellow}
\affiliation{NASA Jet Propulsion Laboratory, California Institute of Technology, Pasadena, CA 91109, USA}
\affiliation{LESIA, Observatoire de Paris, Université PSL, CNRS, Sorbonne Université, Université de Paris, 5 place Jules Janssen, 92195 Meudon, France}

\author[0000-0003-3050-8203]{Stan Metchev}
\affiliation{Department of Physics and Astronomy, Centre for Planetary Science and Exploration, The University of Western Ontario, London, ON N6A 3K7, Canada}
\affiliation{Department of Physics and Astronomy, Stony Brook University, Stony Brook, NY 11794-3800, USA}

\author[0000-0003-0029-0258]{Julien Rameau}
\affiliation{Institut de Recherche sur les Exoplan{\`e}tes, D{\'e}partement de Physique, Universit{\'e} de Montr{\'e}al, Montr{\'e}al, QC, H3C 3J7, Canada}

\author[0000-0003-1698-9696]{Bin Ren}
\affiliation{Department of Physics and Astronomy, Johns Hopkins University, Baltimore, MD 21218, USA}
\affiliation{Space Telescope Science Institute, 3700 San Martin Drive, Baltimore, MD 21218, USA}

\author[0000-0002-7670-670X]{Malena Rice}
\affiliation{Department of Astronomy, Yale University, New Haven, CT 06511, USA}

\author[0000-0002-5815-7372]{Inseok Song}
\affiliation{Physics and Astronomy, University of Georgia, 240 Physics, Athens, GA 30602, USA}

\author{Kevin Stahl}
\affiliation{Department of Physics \& Astronomy, 430 Portola Plaza, University of California, Los Angeles, CA 90095, USA}

\author[0000-0003-0774-6502]{Jason Wang}
\affiliation{Department of Astronomy, California Institute of Technology, Pasadena, CA 91125, USA}
\affiliation{Astronomy Department, University of California, Berkeley, CA 94720, USA}

\author[0000-0002-9977-8255]{Schuyler Wolff}
\affiliation{Leiden Observatory, Leiden University, P.O. Box 9513, 2300 RA Leiden, The Netherlands}

\author{Ben Zuckerman}
\affiliation{Department of Physics \& Astronomy, 430 Portola Plaza, University of California, Los Angeles, CA 90095, USA}

\author[0000-0001-5172-7902]{S. Mark Ammons}
\affiliation{Lawrence Livermore National Laboratory, 7000 East Ave, Livermore, CA 94550, USA}

\author[0000-0002-5407-2806]{Vanessa P. Bailey}
\affiliation{NASA Jet Propulsion Laboratory, California Institute of Technology, Pasadena, CA 91109, USA}

\author[0000-0002-7129-3002]{Travis Barman}
\affiliation{Lunar and Planetary Laboratory, University of Arizona, Tucson, AZ 85721, USA}

\author[0000-0001-6305-7272]{Jeffrey Chilcote}
\affiliation{Kavli Institute for Particle Astrophysics and Cosmology, Stanford University, Stanford, CA 94305, USA}
\affiliation{Department of Physics, University of Notre Dame, 225 Nieuwland Science Hall, Notre Dame, IN 46556, USA}

\author{Rene Doyon}
\affiliation{Institut de Recherche sur les Exoplan{\`e}tes, D{\'e}partement de Physique, Universit{\'e} de Montr{\'e}al, Montr{\'e}al, QC, H3C 3J7, Canada}

\author[0000-0003-3978-9195]{Benjamin L. Gerard}
\affiliation{University of Victoria, Department of Physics and Astronomy, 3800 Finnerty Rd, Victoria, BC V8P 5C2, Canada}
\affiliation{National Research Council of Canada Herzberg, 5071 West Saanich Road, Victoria, BC V9E 2E7, Canada}

\author[0000-0002-4144-5116]{Stephen J. Goodsell}
\affiliation{Gemini Observatory, 670 N. A'ohoku Place, Hilo, HI 96720, USA}

\author[0000-0002-7162-8036]{Alexandra Z. Greenbaum}
\affiliation{Department of Astronomy, University of Michigan, Ann Arbor, MI 48109, USA}

\author[0000-0003-3726-5494]{Pascale Hibon}
\affiliation{Gemini Observatory, Casilla 603, La Serena, Chile}

\author[0000-0001-8074-2562]{Sasha Hinkley}
\affiliation{University of Exeter, Astrophysics Group, Physics Building, Stocker Road, Exeter, EX4 4QL, UK}

\author{Patrick Ingraham}
\affiliation{Large Synoptic Survey Telescope, 950N Cherry Ave., Tucson, AZ 85719, USA}

\author[0000-0002-9936-6285]{Quinn Konopacky}
\affiliation{Center for Astrophysics and Space Science, University of California San Diego, La Jolla, CA 92093, USA}

\author{J\'er\^ome Maire}
\affiliation{Center for Astrophysics and Space Science, University of California San Diego, La Jolla, CA 92093, USA}

\author[0000-0001-7016-7277]{Franck Marchis}
\affiliation{SETI Institute, Carl Sagan Center, 189 Bernardo Ave.,  Mountain View, CA 94043, USA}

\author[0000-0002-5251-2943]{Mark S. Marley}
\affiliation{Space Science Division, NASA Ames Research Center, Mail Stop 245-3, Moffett Field, CA 94035, USA}

\author[0000-0002-4164-4182]{Christian Marois}
\affiliation{National Research Council of Canada Herzberg, 5071 West Saanich Road, Victoria, BC V9E 2E7, Canada}
\affiliation{University of Victoria, Department of Physics and Astronomy, 3800 Finnerty Rd, Victoria, BC V8P 5C2, Canada}

\author[0000-0001-6975-9056]{Eric L. Nielsen}
\affiliation{SETI Institute, Carl Sagan Center, 189 Bernardo Ave.,  Mountain View, CA 94043, USA}
\affiliation{Kavli Institute for Particle Astrophysics and Cosmology, Stanford University, Stanford, CA 94305, USA}

\author[0000-0001-7130-7681]{Rebecca Oppenheimer}
\affiliation{Department of Astrophysics, American Museum of Natural History, New York, NY 10024, USA}

\author{David Palmer}
\affiliation{Lawrence Livermore National Laboratory, 7000 East Ave, Livermore, CA 94550, USA}

\author{Lisa Poyneer}
\affiliation{Lawrence Livermore National Laboratory, 7000 East Ave, Livermore, CA 94550, USA}

\author{Laurent Pueyo}
\affiliation{Space Telescope Science Institute, 3700 San Martin Drive, Baltimore, MD 21218, USA}

\author[0000-0002-9246-5467]{Abhijith Rajan}
\affiliation{Space Telescope Science Institute, 3700 San Martin Drive, Baltimore, MD 21218, USA}

\author[0000-0002-9667-2244]{Fredrik T. Rantakyr\"o}
\affiliation{Gemini Observatory, Casilla 603, La Serena, Chile}

\author[0000-0003-2233-4821]{Jean-Baptiste Ruffio}
\affiliation{Kavli Institute for Particle Astrophysics and Cosmology, Stanford University, Stanford, CA 94305, USA}

\author[0000-0002-8711-7206]{Dmitry Savransky}
\affiliation{Sibley School of Mechanical and Aerospace Engineering, Cornell University, Ithaca, NY 14853, USA}

\author{Adam C. Schneider}
\affiliation{School of Earth and Space Exploration, Arizona State University, PO Box 871404, Tempe, AZ 85287, USA}

\author[0000-0003-1251-4124]{Anand Sivaramakrishnan}
\affiliation{Space Telescope Science Institute, 3700 San Martin Drive, Baltimore, MD 21218, USA}

\author[0000-0003-2753-2819]{R\'{e}mi Soummer}
\affiliation{Space Telescope Science Institute, 3700 San Martin Drive, Baltimore, MD 21218, USA}

\author[0000-0002-9121-3436]{Sandrine Thomas}
\affiliation{Large Synoptic Survey Telescope, 950N Cherry Ave., Tucson, AZ 85719, USA}

\author[0000-0002-4479-8291]{Kimberly Ward-Duong}
\affiliation{Five College Astronomy Department, Amherst College, Amherst, MA 01002, USA}


% ABSTRACT
\begin{abstract}

We report the results of a ${\sim}4$-year direct imaging survey of \Nobs stars to resolve and characterize circumstellar debris disks in scattered light as part of the Gemini Planet Imager Exoplanet Survey. We targeted nearby (${\lesssim}150$~pc), young (${\lesssim}500$~Myr) stars with high infrared (IR) excesses (\Lir$>10^{-5}$), including 38 with previously resolved disks. Observations were made using the Gemini Planet Imager high-contrast integral field spectrograph in $H$-band (1.6~$\mu$m) coronagraphic polarimetry mode to measure both polarized and total intensities. We resolved \Ndebris debris disks and three protoplanetary/transitional disks. Seven debris disks were resolved in scattered light for the first time, including newly presented HD~117214 and HD~156623, and we quantified basic morphologies of five of them using radiative transfer models. All of our detected debris disks except HD~156623 have dust-poor inner holes, and their scattered-light radii are generally larger than corresponding radii measured from resolved thermal emission and those inferred from spectral energy distributions. To assess sensitivity, we report contrasts and consider causes of non-detections. Detections were strongly correlated with high IR excess and high inclination, although polarimetry outperformed total intensity angular differential imaging for detecting low inclination disks (${\lesssim}70\degr$). Based on post-survey statistics, we improved upon our pre-survey target prioritization metric predicting polarimetric disk detectability. We also examined scattered-light disks in the contexts of gas, far-IR, and millimeter detections. Comparing $H$-band and ALMA fluxes for two disks revealed tentative evidence for differing grain properties. Finally, we found no preference for debris disks to be detected in scattered light if wide-separation substellar companions were present.

\end{abstract}

% KEYWORDS
%% Keywords should appear after the \end{abstract} command. The uncommented example has been keyed in ApJ style. See the instructions to authors for the journal to which you are submitting your paper to determine what keyword punctuation is appropriate.

%\keywords{circumstellar matter - infrared: planetary systems - techniques: high angular resolution}

% INTRODUCTION
\section{Introduction} \label{sect:intro}

Debris disks are the extrasolar analogs of our interplanetary and Kuiper Belt dust complex \citep{mann2006}, and often represent the brightest non-stellar component of a young planetary system due to the relatively large cumulative surface area of emitting and scattering grains. The dust optical depth declines with age as grains are lost to collisional erosion, Poynting-Robertson drag, sublimation, and ejection due to radiation pressure \citep{backman1993} or stellar winds \citep{augereau2006, strubbe2006}. The ratio of dust detected in the thermal infrared (IR) to the total stellar bolometric luminosity, \Lir, may be roughly characterized as $10^{-3}$ at 10 Myr, $10^{-4}$ at 100 Myr and $10^{-5}$ at 1 Gyr, with roughly 1 dex of scatter at any age \citep{spangler2001, wyatt2008}. However, the detectability of colder dust in optical/near-IR scattered light strongly depends on other parameters such as the spatial distribution of dust and the viewing geometry. For example, an edge-on disk extending to ${>}1000$ au such as $\beta$ Pic \citep{smith1984} is a best-case scenario due to the disk's line-of-sight dust density and angular size, but this information is not known a priori from infrared observations. Therefore, when designing a scattered-light survey for debris disks, it is relatively difficult to predict which observations will be successful.

Nevertheless, debris disk surveys are motivated by the fact that each detection reveals the physical properties of the constituent dust grains and the overall system architecture. Basic aspects of debris disk morphology (e.g., position angle, inclination to the line of sight, and the radial extents of one or more belts containing dust and planetesimals) constrain the likely locations of planetary bodies orbiting the host star \citep{roques1994}. Debris disks that are well resolved (i.e., with $\lesssim1''$ angular resolution) often show radial, azimuthal, and vertical structure that can be attributed to dynamical interactions with planets \citep{liou1999, ozernoy2000, kuchner2003, quillen2006, thebault2012, rodigas2014_shepherd, lee2016}.

For example, in the case of $\beta$ Pic, a 24 Myr old \citep{mamajek2014} A6V star at 19.4 pc (Gaia DR2; \citealt{gaia_mission, gaia_dr2}), a subtle vertical warp first detected with the Hubble Space Telescope (\textit{HST}) along the midplane of its edge-on debris disk \citep{burrows1995} is now associated with a directly imaged planet in the system \citep{mouillet1997, lagrange2009a}. In the case of Fomalhaut, a 440 Myr old \citep{mamajek2012} A4V star at 7.7 pc (Gaia DR2), a narrow debris belt detected with \textit{HST} is offset by 15 au from the central star \citep{kalas2005_fomb}. This offset could arise from the secular perturbation of a planet on an eccentric orbit \citep{wyatt1999}, though the orbit of Fomalhaut~b is likely too eccentric ($e\sim0.9$) to account for the observed offset \citep{kalas2013, beust2014}. Thus, the overall picture is that accurate measurements of debris disk morphology indicate the possible location and properties of planets before they are directly detected.

The Gemini Planet Imager (GPI) is one of the latest generation ground-based adaptive-optics (AO) instruments that are dedicated to the direct detection of extrasolar planetary systems \citep{macintosh2014}. Located at the Cassegrain focus of the Gemini South telescope (7.8-m effective diameter primary), GPI uses a high-order AO system with two deformable mirrors in a woofer-tweeter configuration to minimize residual atmospheric turbulence \citep{poyneer2014_gpi, bailey2016}. An apodized-pupil Lyot coronagraph \citep{soummer2009, savransky2014} then suppresses diffracted starlight before the incident light is fed to an integral field spectrograph (IFS; \citealt{chilcote2012, larkin2014_gpi}) with high spatial resolution over a ${\sim}2\farcs8\times2\farcs8$ field of view. Broadband filters enable observations in $Y$ (central wavelength 1.05 $\micron$), $J$ (1.24 $\micron$), $H$ (1.65 $\micron$), $K1$ (2.05 $\micron$), and $K2$ (2.26 $\micron$) bands \citep{maire2014}, where $K$ is split into two bands to avoid spectral crosstalk on the detector. The IFS can either disperse the light into its spectral components (``spec-mode'') or into two orthogonal linear polarization states (``pol-mode'') before it reaches the 2048$\times$2048 HAWAII-2RG infrared array.

The Gemini Planet Imager Exoplanet Survey (GPIES) is an 890 hour campaign awarded by Gemini Observatory to search $\sim$600 young, nearby stars for giant planets ($>1$ $\mathrm{M_J}$ orbiting at $\gtrsim5$ au) and light-scattering debris disks \citep{macintosh2018}.  The scientific results from the first 300 stars surveyed for giant planets are reported in \citet{nielsen2019}.

In this work we report observational results from the GPIES debris disk survey from 2014 November through 2018 December, including some pre-survey data from the GPI instrument commissioning period in 2013 and 2014. We describe the target sample (Section 2), observing strategy (Section 3), and data reduction methods (Section 4). Then we present our disk detections, non-detections, and findings from the combined sample (Section 5). Finally, we discuss the implications of these results and the impact of this survey on our understanding of debris disks (Section 6).

% AVOID ORPHANED SECTION HEADING
%\vspace{40pt}
%%%

\section{Target Sample}

\subsection{Selection Criteria} \label{sect:targ_selection}

We selected our initial list of circumstellar disk targets for the GPIES survey in 2014 February. This sample consisted of two groups, both of which satisfied the practical observing constraints of a stellar magnitude limit ($m_I\leq9$ mag) required for closed-loop operation of GPI and a suitable declination range for Gemini South ($-87\degr\leq\delta\leq25\degr$).  We also rejected binaries with angular separations $0\farcs02$--$3\farcs0$ and $\Delta m_I\leq5$ mag because the companion would degrade GPI's contrast performance. Properties for the targets that we observed are given in Table \ref{tab:obs_targs}.

The first group comprised 38 circumstellar disks previously resolved in either scattered light or thermal emission. The second group consisted of 141 A--M main-sequence stars that met criteria of having unresolved IR excesses with \Lir$\geq10^{-5}$, heliocentric distances $d \lesssim 150$ pc, and age $\leq$700 Myr. Systems older than 700 Myr are not expected to have GPI-detectable levels of circumstellar dust. That said, four of the targets in our sample with disks resolved by scattered-light and/or \textit{Herschel} imaging also have age estimates firmly over this limit (albeit with large uncertainties). The upper limit on distance was intended to retain disk flux (which scales like $d^{-2}$) and ensure that even relatively radially compact disks would still have angular sizes larger than GPI's occulted region (which has a radius of ${\sim}19$ au at 150 pc). IR excesses were calculated for this survey from their spectral energy distributions (SEDs) using catalog data from \textit{IRAS}, \textit{Spitzer}, \textit{Herschel}, and \textit{WISE} using methods similar to those described in \citet{cotten2016}: the SEDs stellar photosphere component was fit with a PHOENIX NextGen model stellar photosphere \citep{hauschildt1999} and then a single blackbody emission curve was fit to the photosphere-subtracted SEDs residual IR fluxes. Of the 179 targets in these two groups, 19 were subsequently pruned from the list because they were found to be evolved giant stars, false IR excesses (e.g., from background-contaminated photometry), visual binaries of the type described above, or older than previously thought (e.g., Hyades cluster members with an upwardly revised cluster age ${>}700$ Myr; \citealt{brandt2015}).

We began the survey with the resulting list of 160 targets and augmented it twice during the campaign. The first set of additions, in 2015 September, included 16 new disk targets to replace those that were pruned, selected by the same initial criteria. A second set of nine targets was added in 2016 December after they were shown to have resolved disks with \textit{HST} or ALMA\footnote{The \textit{HST} NICMOS detections were HD 377 and TWA 25 \citep{choquet2016}. The ALMA detections were AK Sco, HD 111161, HD 112810, HD 129590, HD 138813, HD 156623 \citep{lieman-sifry2016}, and $\tau$ Cet \citep{macgregor2016b}.}. After these additions, our final list contained a total of \Ntarg targets.

All \Ntarg disk targets in the final sample were reserved for $H$-band pol-mode observations with GPI and nearly all were selected for $H$ spec-mode observations. These two observing modes will be discussed in the next section. Here we note that pol-mode was generally considered the GPIES disk detection method whereas spec-mode was used for planet detection. Most disk targets were observed in spec-mode simply because the stellar properties also satisfied the criteria for the planet-detection component of GPIES \citep{nielsen2019}. The exceptions were older stars ($>$0.5 Gyr) with IR excesses that were unlikely to host detectable planets\footnote{The 12 pol-mode-only targets were HR~506, HR~1254, HD~50554, $o^1$~Cnc, $o^2$~Cnc, $\eta$~Crv, 61~Vir, HD~157587, $\phi^1$~Pav, HR~8323, 39~Peg, and $\tau$~Cet.}.

% Target histograms
\begin{figure*}[ht]
\centering
%\epsscale{1.0}
\includegraphics[width=5.5in]{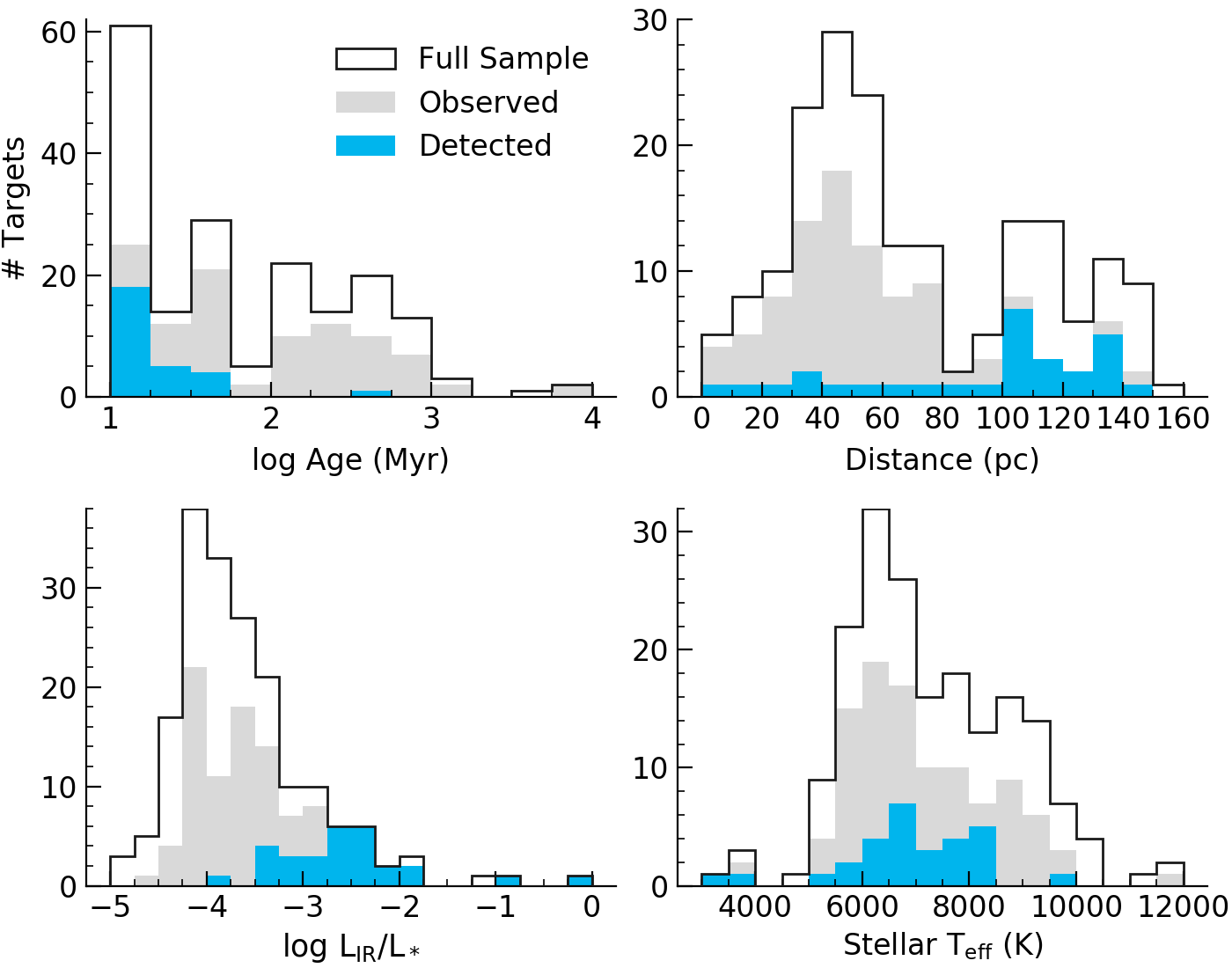}
\caption{Host star properties for GPIES disk targets. Overlapping distributions are shown for all stars in our target list (185), observed stars (104), and stars with detected disks of any class (29).}
\label{fig:targ_histograms}
\end{figure*}

% Target Age v Distance
\begin{figure}[h]
\centering
%\epsscale{1.0}
\includegraphics[width=\columnwidth]{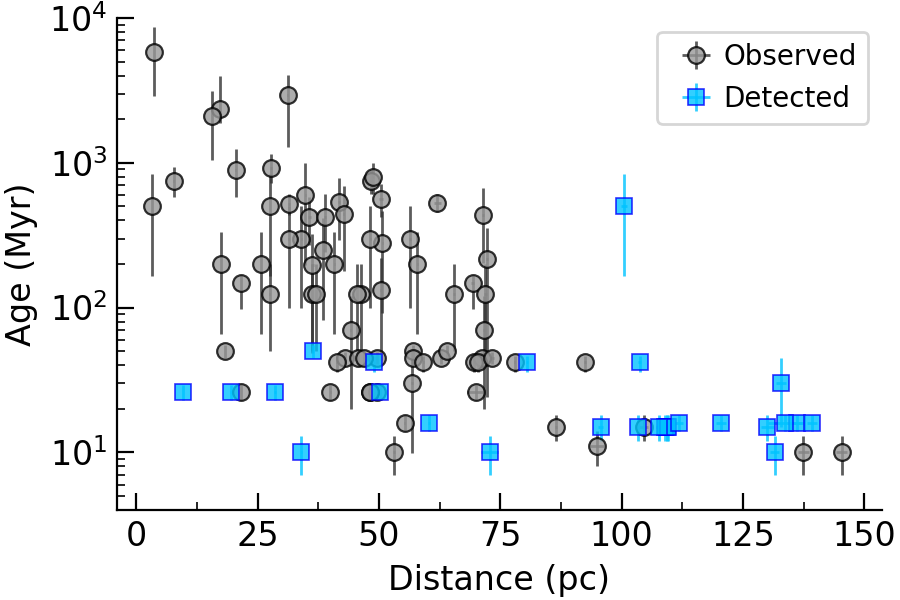}
\caption{Observed GPIES disk target age versus distance. Nearly all GPIES disk detections (blue squares) are around stars younger than 50 Myr. The apparent preference for finding disks around more distant stars is heavily skewed by the Scorpius--Centaurus OB association, which contains many young and dusty stars at 100--140 pc. Distance error bars are all smaller than the symbols, as are some age error bars.}
\label{fig:targ_age_dist}
\end{figure}

\subsection{Target Prioritization} \label{sect:metric}

The \Ntarg stars that passed our disk selection criteria exceeded the number that we expected to observe within the ${\sim}$40 hr of telescope time initially allocated for pol-mode observations. We therefore developed a metric of disk detectability in polarized light in order to prioritize observations of these targets.

The metric was based on the expected polarized scattered-light brightness of each disk. To ensure a uniform approach for all targets, we adopted a simple disk model, specifically a narrow circular ring. The disk radius was set to the stellocentric ``blackbody radius'' $R_\mathrm{bb}$ (in astronomical units) where grains emitting as blackbodies would account for the observed excess IR flux of the SEDs (over that of the star's photosphere). We set the inner and outer disk radii to 2.5\% below and above $R_\mathrm{bb}$ for a total fractional width of 5\%.

We adopted a single grain radius equal to the system's blowout radius ($s_\mathrm{blow}$) calculated from the mass ($M_\star$) and bolometric luminosity ($L_\star$) of each star according to \citet{burns1979} (Eq. 19). The radiation pressure efficiency factor for each grain (denoted $Q_{\mathrm{pr}}$ as in \citealt{burns1979}) was $Q_\mathrm{pr}=Q_{\mathrm{abs}}+Q_{\mathrm{sca}}(1-\langle\cos{\theta}\rangle)$, where $Q_{\mathrm{abs}}$ and $Q_{\mathrm{sca}}$ are the grain's dimensionless absorption and scattering efficiency factors, respectively, and $\langle\cos{\theta}\rangle$ is the scattering anisotropy factor. For this calculation, we assumed isotropic scattering ($\langle\cos{\theta}\rangle{=}0$), which is not strictly consistent with the anisotropic scattering phase function that we use later, but is tolerable considering the magnitudes of other uncertainties involved here. We also approximated grains with radii ${\sim}s_\mathrm{blow}$ to be within the geometrical optics limit --- i.e., the grains were larger than the incident wavelength of light and $Q_{\mathrm{abs}}{\approx}1$ and $Q_{\mathrm{sca}}{\approx}1$ --- so that $Q_\mathrm{pr}{=}2$. We also adopted an average grain mass density of $\rho{=}2.7$ g cm$^{-3}$; i.e., a general rocky composition with density between that of graphite (${\sim}2.2$ g cm$^{-3}$) and silicate/SiO$_2$ (${\sim}3.5$ g cm$^{-3}$; \citealt{draine2003}).

Finally, assuming optically thin emission, we determined the total disk mass from the amount of infrared emission. Specifically, we converted the disk luminosity of $L_\mathrm{IR}$ into a surface area of blackbody emission as $A=L_\mathrm{IR} / (\sigma_B T_\mathrm{dust}^4)$, where $\sigma_B$ is the Stefan-Boltzmann constant. This area divided by the surface area of a spherical grain with $s=s_\mathrm{blow}$ then yielded the number of grains in the disk. In turn, this led to an estimate of the disk mass via the same average grain mass density adopted earlier.

From this setup, we computed polarized scattered-light model disk images at 10 inclinations evenly spaced in $\cos{i}$ between $0\degr$ and $90\degr$ (inclusive) with the radiative transfer code MCFOST that assumes Mie theory is valid for homogeneous, spherical particles \citep{pinte2006}. For a rough approximation of the model grains' scattering and polarization phase functions, we adopted refractive indices of nonporous astrosilicate particles \citep{draine1984_si}. At each inclination, we computed the ratio of the model's polarized intensity ($\sqrt{\mathcal{Q}^2 + \mathcal{U}^2}$) to the expected total intensity stellar point-spread function (PSF; scaled by the stellar flux) as a function of stellocentric separation across the GPI field of view and identified the maximum ratio. The total intensity radial profile of the PSF was measured from stacked $H$-band, occulted, pol-mode images of HD 12759  (SpT = G5V, $m_H$=5.6 mag) taken during commissioning and with a binary companion masked out. Our disk metric value for each target is the average of the maximum ratios from all model inclinations.

In some cases, $R_\mathrm{bb}$ was smaller than GPI's $H$-band focal plane mask (FPM; radius of $0\farcs123$), so we forced $R_\mathrm{bb}$ equal to the physical equivalent of $0\farcs13$, i.e., just outside the FPM. In the singular case of Fomalhaut, the $R_\mathrm{bb}$ of $10\arcsec$ was much larger than GPI's field of view, so we forced it to be equivalent to $1\farcs4$, i.e., the effective maximum projected separation visible.

The assumptions and methods underlying the metric calculation were simple yet adequate for our operational goal of prioritizing the observations.  Moreover, the metric became one of the testable aspects of our experiment. Now, with the conclusion of our survey, we can quantify whether our detection rates correlated with these metric scores or if some other method has more predictive power. We present our conclusions in Section \ref{sect:discuss_metric}.  In brief, we found the metric to be a less effective quantity than simple \Lir values for predicting scattered-light disk detectability.

% AVOID ORPHANED SECTION HEADING
\vspace{40pt}

\subsection{Final Sample Demographics} \label{sect:final_sample}

Breakdowns of the final full sample by system age, distance, \Lir, and stellar effective temperature are shown in Figure \ref{fig:targ_histograms}, where they are further broken down by targets observed and disks detected. For the full sample, the stars have a median age of 45 Myr, and only seven have estimates ranging to 1 Gyr or older. Age ranges quoted in Table \ref{tab:obs_targs} are the ${\sim}68$\% confidence ranges. The ages are well established for 106 stars through their membership in stellar moving groups. The remaining targets generally have age range estimates based on lithium abundance and chromospheric activity for lower-mass stars, and evolution across the color-magnitude diagram for higher-mass stars; see \citet{nielsen2019} for details. We note that 13 stars\footnote{The stars assigned to moving groups after kinematic re-evaluation are g~Lup and NZ~Lup ($\beta$~Pic); HD~3888 and HD~8813 (TucHor); HR~506, $\nu$~Hor, $\pi^1$~Ori, V435~Car, and HD~205674 (AB Dor); HD~10472 (Columba); and HD~31392 and HD~84075 (Argus).} from this latter group were recently associated with moving groups based on a re-evaluation of their kinematic information via the methods presented in \citet{lee2019_arxiv} and may have different ages (mostly younger) from those we report here.

Stellar masses ($M_\star$) and effective temperatures ($T_{\mathrm{eff}}$) quoted in Table \ref{tab:obs_targs} were newly computed for GPIES and described in detail in \citet{nielsen2019}. Briefly, optical photometry measurements were fit to synthetic photometry derived from model stellar atmospheres, and a mass was then estimated based on the star's position on the color-magnitude diagram. That position was also used to estimate the age for higher-mass stars, as noted in the previous paragraph. Optical photometry measurements ($G_B$, $G$, $G_R$ bands) were selected from the \textit{Gaia} catalog \citep{gaia_dr2}, combining catalog uncertainties with the systematic uncertainties reported in \citet{evans2018}. For the handful of stars too bright for \textit{Gaia}, Tycho2 photometry ($B_T$, $V_T$; \citealt{hog2000}) was used instead. The stellar models were constructed by combining the stellar evolutionary model MESA \citep{paxton2011} with ATLAS9 model atmospheres \citep{castelli2003}. The stellar $T_{\mathrm{eff}}$ was also derived from the best-fit model.

Stellar bolometric luminosities in Table \ref{tab:obs_targs} were calculated as the median of three different estimates\footnote{Bolometric corrections and stellar sequences were taken from \url{http://www.pas.rochester.edu/~emamajek/EEM_dwarf_UBVIJHK_colors_Teff.txt}}: a bolometric correction from the $V$ absolute magnitude and based on $T_{\mathrm{eff}}$ \citep{pecaut2013}, the value from a stellar sequence table based only on $T_{\mathrm{eff}}$, and the value from a stellar sequence table based only on $V$ absolute magnitude. The given uncertainties are the standard deviation of these three luminosity estimates.

Most targets in our full sample are closer than 100 pc and all are within 156 pc, based on distances from the Gaia DR2 catalog unless otherwise noted \citep{gaia_mission, gaia_dr2}. The sample contains primarily stars of A (67), F (61), and G (31) spectral types and also includes late-B (12), some K (10), and a few M stars (3). The youngest of these may qualify as pre-main-sequence stars and thus still be evolving toward earlier spectral types. All targets have \Lir$>10^{-5}$, and approximately two-thirds have \Lir$>10^{-4}$.

A number of stellar moving groups are well represented in our sample, including $\beta$ Pictoris, Columba, Scorpius--Centaurus (Sco--Cen), and Tucana--Horologium. Sco--Cen, in particular, contributed 55 stars to our full sample. We ended up observing 18 Sco--Cen stars, which produced 15 disk detections---over half of the survey's total. This abundance of disks in Sco--Cen, all located 100--140 pc away, makes it appear as if detection frequency increases with distance (Figure \ref{fig:targ_histograms}); however, we see from Figure \ref{fig:targ_age_dist} that the age of the system is actually the more influential factor in detection, primarily because age is anticorrelated with dust content and \Lir.

We also include \textit{Herschel} PACS \citep{poglitsch2010} disk detection statuses in Table \ref{tab:obs_targs}, which we discuss in Section \ref{sect:discuss_detection_factors}. We reduced PACS photometer scanning mode data that was publicly available in the Herschel Science Archive circa 2015 July, performing aperture photometry on background-subtracted ``Level-2.5'' high-pass filter maps at 70, 100, and $160~\micron$. Source apertures had radii of $12\arcsec$, $12\arcsec$, and $22\arcsec$ at the three respective wavelengths. The map noise level was measured as the standard deviation of fluxes in six apertures (same size as the source aperture) distributed $50\arcsec$--$60\arcsec$ from the source. Source fluxes $\geq3\sigma$ in a given band are considered disk detections. For the targets that were added to the sample after 2015 July, we collected their detection statuses from publications cited parenthetically in Table \ref{tab:obs_targs}.

\subsection{Non-Debris Disk Targets} \label{sect:non-debris_targets}

While the focus of our survey and this paper is debris disks, 3 targets in our 104 star sample host disks that may be better classified as protoplanetary or transitional disks due to their high IR-excess magnitudes and gas contents. \object{HD 100546} (first by \citealt{pantin2000}) and \object{HD 141569} (first by \citealt{augereau1999a} and \citealt{weinberger1999}) were resolved in scattered light before GPIES began and were selected as GPI commissioning targets to help characterize the polarimetric mode performance. \object{AK Sco} was added as a target midsurvey after it was resolved in thermal emission with ALMA \citep{czekala2015a, lieman-sifry2016} and in scattered light with VLT/SPHERE \citep{janson2016}. In all three cases, we had the opportunity to obtain the first near-IR polarized intensity images of the disks and thus considered them worthwhile targets.

We generally exclude these three targets from our analyses (unless explicitly stated otherwise) because they likely represent an earlier evolutionary phase and are statistical outliers with respect to many disk parameters. Nevertheless, we present new morphological results from the images of AK~Sco and HD~100546 in Section \ref{sect:proto} in an effort to make this paper a comprehensive summary of GPIES disk survey results. A brief overview of previous HD~141569 results is also included there for completeness, but we direct the reader to \citet{bruzzone2019} for in-depth analysis of that disk.

\clearpage % TEMP for formatting

%%% Rotate to landscape for submission (probably)
% Campaign target table.
\startlongtable
%\begin{longrotatetable}
\begin{deluxetable*}{lrcccrcccccc}
%\tablecolumns{6}
%\tablewidth{0pt}
\tabletypesize{\scriptsize}
\tablecaption{\label{tab:obs_targs}GPIES-observed Disk Targets}
\tablehead{
% Header Line 1
\colhead{Name} & \colhead{$d$} & \colhead{Age} & \colhead{Age} & \colhead{$T_{\mathrm{eff}}$} & \colhead{$M_\star$} & \colhead{$L_\star$} & \colhead{\Lir} & \colhead{GPIES} & \colhead{Scat-Light} & \colhead{Herschel} & \colhead{Moving}  \vspace{-0.3cm} \\
% Header Line 2
 & \colhead{(pc)} & \colhead{(Myr)} & \colhead{ref} & \colhead{(K)} & \colhead{($\mathrm{M_\odot}$)} & \colhead{($\mathrm{L_\odot}$)} & \colhead{($10^{-4}$)} & \colhead{Metric} & \colhead{Resolved?} & \colhead{Det?} & \colhead{Group}
}
\startdata
 49 Cet         & $57.07\pm0.33$  & 45--55     & (1)   &        8900 & $2.02\substack{+0.03 \\ -0.05}$ & $16.65\pm2.89$  & 10.73                    & 0.0891   & (2)                  & 100,160         & Argus (1)      \\
 73 Her         & $42.75\pm0.12$  & 189--701   & (3)   &        7600 & $1.60\substack{+0.05 \\ -0.05}$ & $7.76\pm0.74$   & 3.6                      & 0.0384   & N                    & --              & --             \\
 AK Sco$^\dagger$         & $140.59\pm1.23$ & 14--18     & (4)   &        5460 & $1.30\substack{+0.07 \\ -0.05}$ & $4.68\pm2.06$   & 1098.29                  & 0.9026   & (5)                  & --              & UCL (6)        \\
 AU Mic         & $9.72\pm0.00$   & 23--29     & (7)   &        3500 & $0.64\substack{+0.02 \\ -0.03}$ & $0.06\pm0.03$   & 3.33                     & 0.0003   & (8)                  & 70,160          & bet Pic (9)    \\
 b Leo          & $38.87\pm0.27$  & 245--608   & (3)   &        9000 & $2.04\substack{+0.11 \\ -0.10}$ & $21.38\pm1.02$  & 0.39                     & 0.0030   & N                    & Null            & --             \\
 $\beta$ Pic    & $19.44\pm0.05$  & 23--29     & (7)   &        8200 & $1.73\substack{+0.00 \\ -0.02}$ & $9.33\pm3.13$   & 28.1                     & 0.1715   & (10)                 & 70,160          & bet Pic (9)    \\
 CE Ant         & $34.03\pm0.08$  & 7--13      & (9)   &        3420 & $0.31\substack{+0.06 \\ -0.06}$ & $0.07\pm0.07$   & 6.49                     & 0.0000   & (11)                 & 70,100,160 (12) & TWA (9)        \\
 $\epsilon$ Eri & $3.22\pm0.00$   & 165--835   & (3)   &        5600 & $0.86\substack{+0.01 \\ -0.01}$ & $0.41\pm0.18$   & 0.79                     & 0.0005   & N                    & 70,160          & --             \\
 $\eta$ Cha     & $94.97\pm1.44$  & 8--14      & (9)   &       11850 & $3.20\substack{+0.06 \\ -0.07}$ & $99.81\pm52.10$ & 0.86                     & 0.0010   & N                    & --              & eta Cha (9)    \\
 $\eta$ Tel A   & $48.22\pm0.49$  & 23--29     & (7)   &        9700 & $2.15\substack{+0.05 \\ -0.06}$ & $22.95\pm6.50$  & 1.4                      & 0.0126   & N                    & 100,160         & bet Pic (9)    \\
 Fomalhaut      & $7.70\pm0.03$   & 564--916   & (3)   &        8900 & $1.80\substack{+0.07 \\ -0.06}$ & $19.50\pm0.89$  & 0.64                     & 0.1227   & (13)                 & --              & --             \\
 g Lup          & $17.44\pm0.05$  & 66--334    & (3)   &        7000 & $1.42\substack{+0.01 \\ -0.01}$ & $3.71\pm1.09$   & 0.7                      & 0.0132   & (14)                 & 70,160          & --             \\
 $\gamma$ Dor   & $20.46\pm0.15$  & 535--1207  & (3)   &        7100 & $1.56\substack{+0.06 \\ -0.06}$ & $5.89\pm0.39$   & 0.26                     & 0.0038   & N                    & 70,160          & --             \\
 $\gamma$ Oph   & $31.52\pm0.21$  & 435--602   & (3)   &        9000 & $2.14\substack{+0.07 \\ -0.06}$ & $29.27\pm5.46$  & 1.11                     & 0.0099   & N                    & 70,100,160      & --             \\
 HD 377         & $38.52\pm0.09$  & 82--417    & (3)   &        5890 & $1.13\substack{+0.01 \\ -0.00}$ & $1.35\pm0.04$   & 4.01                     & 0.1922   & (11)                 & --              & --             \\
 HD 1466        & $42.97\pm0.05$  & 41--49     & (9)   &        6200 & $1.19\substack{+0.01 \\ -0.00}$ & $1.60\pm0.28$   & 1.17                     & 0.0007   & N                    & 70              & TucHor (9)     \\
 HD 3888        & $44.18\pm0.06$  & 20--120    & (3)   &        6200 & $1.67\substack{+0.05 \\ -0.03}$ & $1.82\pm0.11$   & 0.87                     & 0.0004   & N                    & --              & --             \\
 HD 7112        & $50.68\pm0.08$  & 91--464    & (3)   &        5800 & $0.99\substack{+0.00 \\ -0.01}$ & $0.78\pm0.12$   & 1.99                     & 0.0002   & N                    & --              & --             \\
 HD 8813        & $46.35\pm0.06$  & 50--200    & (3)   &        5800 & $1.04\substack{+0.00 \\ -0.01}$ & $0.84\pm0.10$   & 1.06                     & 0.0002   & N                    & --              & --             \\
 HD 10472       & $71.15\pm0.15$  & 41--49     & (9)   &        6900 & $1.44\substack{+0.01 \\ -0.00}$ & $3.47\pm0.75$   & 4.05                     & 0.0551   & N                    & --              & TucHor (15)    \\
 HD 13183       & $49.53\pm0.08$  & 41--49     & (9)   &        5700 & $1.51\substack{+0.08 \\ -0.08}$ & $0.78\pm0.09$   & 1.6                      & 0.0001   & N                    & Null            & TucHor (9)     \\
 HD 13246       & $45.61\pm0.06$  & 41--49     & (9)   &        6200 & $1.23\substack{+0.00 \\ -0.01}$ & $1.82\pm0.14$   & 1.98                     & 0.0028   & N                    & 70              & TucHor (9)     \\
 HD 15115       & $49.00\pm0.10$  & 38--48     & (9)   &        6900 & $1.45\substack{+0.01 \\ -0.00}$ & $3.55\pm0.71$   & 4.76                     & 0.0945   & (16)                 & 70,160          & Columba (9)    \\
 HD 15279       & $50.39\pm0.07$  & 43--220    & (3)   &        5910 & $1.09\substack{+0.00 \\ -0.01}$ & $1.09\pm0.16$   & 0.83                     & 0.0005   & N                    & --              & --             \\
 HD 16743       & $57.93\pm0.09$  & 66--334    & (3)   &        7000 & $1.57\substack{+0.00 \\ -0.01}$ & $5.50\pm0.36$   & 4.47                     & 0.0662   & N                    & 100,160         & --             \\
 HD 23208       & $56.80\pm0.13$  & 9--50      & (3)   &        5400 & $0.95\substack{+0.01 \\ -0.01}$ & $0.68\pm0.05$   & 2.71                     & 0.0005   & N                    & 70              & --             \\
 HD 24636       & $57.05\pm0.07$  & 41--49     & (9)   &        6900 & $1.44\substack{+0.01 \\ -0.01}$ & $3.49\pm0.72$   & 0.99                     & 0.0031   & N                    & 70              & TucHor (9)     \\
 HD 30447       & $80.54\pm0.25$  & 38--48     & (9)   &        6900 & $1.45\substack{+0.00 \\ -0.01}$ & $3.51\pm0.72$   & 9.11                     & 0.1270   & (17)                 & --              & Columba (9)    \\
 HD 31392       & $25.77\pm0.02$  & 66--334    & (3)   &        5500 & $0.99\substack{+0.00 \\ -0.01}$ & $0.56\pm0.06$   & 1.84                     & 0.0786   & N                    & 100             & --             \\
 HD 32195       & $62.79\pm0.14$  & 41--49     & (9)   &        6200 & $1.24\substack{+0.01 \\ -0.00}$ & $1.82\pm0.12$   & 0.93                     & 0.0006   & N                    & --              & TucHor (9)     \\
 HD 32297       & $132.79\pm1.06$ & 15--45     & (3)   &        7700 & $1.69\substack{+0.02 \\ -0.02}$ & $8.12\pm1.68$   & 60.54                    & 0.1016   & (18)                 & 100,160         & --             \\
 HD 32372       & $77.97\pm0.15$  & 38--48     & (9)   &        5680 & $1.12\substack{+0.01 \\ -0.01}$ & $0.91\pm0.03$   & 2.45                     & 0.0001   & N                    & Null (19)       & Columba (9)    \\
 HD 35841       & $103.68\pm0.30$ & 38--48     & (9)   &        6500 & $1.30\substack{+0.01 \\ -0.01}$ & $2.35\pm0.54$   & 3.25                     & 0.0315   & (17)                 & --              & Columba (20)   \\
 HD 37484       & $59.11\pm0.08$  & 38--48     & (9)   &        6900 & $1.44\substack{+0.00 \\ -0.01}$ & $3.47\pm0.77$   & 3.49                     & 0.0197   & N                    & --              & Columba (9)    \\
 HD 50554       & $31.19\pm0.06$  & 1890--4620 & (21)  &        6100 & $1.02\substack{+0.04 \\ -0.02}$ & $1.58\pm0.18$   & 0.45                     & 0.0139   & N                    & 70,100,160      & --             \\
 HD 53143       & $18.36\pm0.01$  & 45--55     & (22)  &        5500 & $1.00\substack{+0.01 \\ -0.07}$ & $0.57\pm0.05$   & 1.99                     & 0.0732   & (14)                 & 70,160          & IC2391 (23)    \\
 HD 57969       & $72.20\pm0.17$  & 78--409    & (3)   &        8500 & $1.71\substack{+0.04 \\ -0.05}$ & $10.69\pm4.51$  & 2.45                     & 0.0042   & N                    & 70              & --             \\
 HD 61005       & $36.49\pm0.04$  & 45--55     & (1)   &        5600 & $0.98\substack{+0.02 \\ -0.07}$ & $0.68\pm0.07$   & 27.91                    & 0.8611   & (24)                 & 70,100,160      & Argus (1)      \\
 HD 72687       & $45.44\pm0.09$  & 50--200    & (3)   &        5800 & $1.05\substack{+0.01 \\ -0.00}$ & $0.89\pm0.06$   & 0.97                     & 0.0610   & N                    & 70              & --             \\
 HD 80846       & $71.90\pm0.32$  & 50--200    & (3)   &        6100 & $1.23\substack{+0.01 \\ -0.01}$ & $1.82\pm0.04$   & 0.57                     & 0.0005   & N                    & --              & --             \\
 HD 82943       & $27.61\pm0.03$  & 165--835   & (3)   &        6000 & $1.23\substack{+0.00 \\ -0.01}$ & $1.62\pm0.18$   & 1.1                      & 0.0538   & N                    & 70,160          & --             \\
 HD 84040       & $71.61\pm0.24$  & 20--120    & (3)   &        6200 & $1.20\substack{+0.01 \\ -0.00}$ & $1.82\pm0.15$   & 0.71                     & 0.0001   & N                    & --              & --             \\
 HD 84075       & $64.10\pm0.09$  & 45--55     & (3)   &        6000 & $1.16\substack{+0.01 \\ -0.01}$ & $1.36\pm0.22$   & 1.86                     & 0.0796   & N                    & --              & --             \\
 HD 89452       & $36.26\pm0.08$  & 50--200    & (3)   &        5000 & $0.89\substack{+0.01 \\ -0.00}$ & $0.31\pm0.00$   & 3.64                     & 0.1711   & N                    & 70,160          & --             \\
 HD 95086       & $86.44\pm0.24$  & 12--18     & (4)   &        7600 & $1.61\substack{+0.02 \\ -0.01}$ & $6.74\pm1.46$   & 8.42                     & 0.0745   & N                    & 70,160          & LCC (6)        \\
 HD 100546$^\dagger$      & $110.02\pm0.62$ & 12--18     & (4)   &        7400 & $2.21\substack{+0.10 \\ -0.09}$ & $22.44\pm7.08$  & 7268.5                   & 0.1004   & (25)                 & 100,160         & LCC (6)        \\
 HD 106906      & $103.33\pm0.46$ & 12--18     & (4)   &        6500 & $2.70\substack{+0.12 \\ -0.11}$ & $5.89\pm1.15$   & 50.43                    & 0.8166   & N                    & 100             & LCC (6)        \\
 HD 107146      & $27.47\pm0.03$  & 50--200    & (3)   &        5900 & $1.08\substack{+0.00 \\ -0.01}$ & $0.99\pm0.19$   & 10.07                    & 0.4690   & (26)                 & 70,160          & --             \\
 HD 108857      & $104.52\pm0.93$ & 12--18     & (4)   &        6000 & $1.39\substack{+0.01 \\ -0.01}$ & $3.47\pm0.85$   & 6.94                     & 0.0034   & N                    & --              & LCC (6)        \\
 HD 110058      & $129.98\pm1.33$ & 12--18     & (4)   &        8000 & $1.70\substack{+0.03 \\ -0.02}$ & $9.33\pm2.13$   & 26.18                    & 0.0205   & (27)                 & 100,160         & LCC (6)        \\
 HD 111161      & $109.43\pm0.48$ & 12--18     & (4)   &        7800 & $1.72\substack{+0.02 \\ -0.03}$ & $9.33\pm1.17$   & 42.3                     & 0.2955   & N                    & --              & LCC (28)       \\
 HD 111520      & $108.94\pm0.65$ & 12--18     & (4)   &        6500 & $1.26\substack{+0.09 \\ -0.07}$ & $2.69\pm0.37$   & 10.28                    & 0.0105   & (29)                 & 70,160          & LCC (6)        \\
 HD 114082      & $95.65\pm0.45$  & 12--18     & (4)   &        7000 & $1.42\substack{+0.08 \\ -0.11}$ & $4.74\pm0.56$   & 36.32                    & 0.0627   & (30)                 & 100,160         & LCC (6)        \\
 HD 115600      & $109.62\pm0.49$ & 12--18     & (4)   &        7000 & $1.54\substack{+0.02 \\ -0.10}$ & $5.27\pm0.37$   & 22.61                    & 0.0478   & (31)                 & 100             & LCC (6)        \\
 HD 117214      & $107.61\pm0.50$ & 12--18     & (4)   &        6500 & $1.47\substack{+0.02 \\ -0.01}$ & $5.01\pm0.90$   & 26.74                    & 0.0969   & N                    & --              & LCC (6)        \\
 HD 129590      & $136.04\pm1.26$ & 14--18     & (4)   &        5910 & $1.40\substack{+0.02 \\ -0.01}$ & $3.35\pm0.96$   & 69.57                    & 0.4404   & (32)                 & --              & UCL (6)        \\
 HD 131835      & $133.65\pm3.55$ & 14--18     & (4)   &        8100 & $1.77\substack{+0.05 \\ -0.04}$ & $10.41\pm2.21$  & 30.88                    & 0.0531   & N                    & 70,100,160      & UCL (6)        \\
 HD 138813      & $137.41\pm1.07$ & 7--13      & (4)   &        8640 & $2.15\substack{+0.07 \\ -0.09}$ & $19.50\pm0.21$  & 13.39                    & 0.0403   & N                    & 70,100,160 (33) & US (6)         \\
 HD 141569$^\dagger$      & $110.63\pm0.54$ & 2--8       & (3)   &        8400 & $2.04\substack{+0.04 \\ -0.07}$ & $15.27\pm0.57$  & 127.42                   & 0.2915   & (34)                 & 100,160         & --             \\
 HD 142315      & $145.34\pm1.11$ & 7--13      & (4)   &        8800 & $4.17\substack{+0.10 \\ -0.07}$ & $34.67\pm6.48$  & 6.4                      & 0.0254   & N                    & 100,160         & US (6)         \\
 HD 143675      & $139.20\pm1.07$ & 14--18     & (4)   &        7900 & $1.78\substack{+0.03 \\ -0.03}$ & $9.67\pm2.03$   & 5.58                     & 0.0008   & N                    & --              & UCL (6)        \\
 HD 145560      & $120.44\pm0.96$ & 14--18     & (4)   &        6500 & $1.29\substack{+0.14 \\ -0.05}$ & $3.47\pm0.14$   & 12.67                    & 0.0117   & N                    & --              & UCL (6)        \\
 HD 146897      & $131.50\pm0.93$ & 7--13      & (4)   &        6200 & $1.28\substack{+0.02 \\ -0.01}$ & $3.40\pm0.66$   & 101.93                   & 0.3294   & (35)                 & --              & US (6)         \\
 HD 156623      & $111.75\pm0.96$ & 14--18     & (4)   &        8350 & $1.90\substack{+0.04 \\ -0.05}$ & $13.06\pm1.80$  & 43.32                    & 0.0575   & N                    & --              & UCL (28)       \\
 HD 157587      & $100.51\pm0.60$ & 165--835   & (3)   &        6300 & $1.44\substack{+0.01 \\ -0.01}$ & $2.69\pm0.23$   & 32.02                    & 0.3469   & (29)                 & --              & --             \\
 HD 164249 A    & $49.61\pm0.12$  & 23--29     & (7)   &        6600 & $1.38\substack{+0.01 \\ -0.00}$ & $3.04\pm0.59$   & 10.31                    & 0.2826   & N                    & 100,160         & bet Pic (9)    \\
 HD 181327      & $48.21\pm0.13$  & 23--29     & (7)   &        6400 & $1.39\substack{+0.01 \\ -0.01}$ & $2.69\pm0.02$   & 13.47                    & 0.1301   & (36)                 & 100,160         & bet Pic (9)    \\
 HD 191089      & $50.13\pm0.11$  & 23--29     & (7)   &        6400 & $1.35\substack{+0.01 \\ -0.01}$ & $2.54\pm0.17$   & 14.98                    & 0.1169   & (17)                 & 100,160         & bet Pic (9)    \\
 HD 202917      & $46.85\pm0.09$  & 41--49     & (9)   &        5700 & $1.02\substack{+0.01 \\ -0.08}$ & $0.69\pm0.14$   & 2.85                     & 0.0366   & (17)                 & 70,100          & TucHor (9)     \\
 HD 205674      & $56.40\pm0.23$  & 100--500   & (3)   &        7000 & $1.41\substack{+0.01 \\ -0.00}$ & $4.07\pm0.92$   & 2.43                     & 0.0348   & N                    & 100,160         & --             \\
 HD 206893      & $40.81\pm0.11$  & 66--334    & (3)   &        6600 & $1.35\substack{+0.01 \\ -0.01}$ & $2.51\pm0.79$   & 2.26                     & 0.0687   & N                    & 70,160          & --             \\
 HD 209253      & $31.42\pm0.08$  & 100--500   & (3)   &        6300 & $1.22\substack{+0.01 \\ -0.01}$ & $1.82\pm0.43$   & 0.92                     & 0.0283   & N                    & --              & --             \\
 HD 221853      & $65.42\pm0.21$  & 50--200    & (3)   &        6900 & $1.46\substack{+0.01 \\ -0.01}$ & $3.64\pm0.69$   & 6.77                     & 0.0971   & N                    & 100,160         & --             \\
 HIP 25434      & $92.41\pm0.26$  & 38--48     & (9)   &        6320 & $1.23\substack{+0.01 \\ -0.01}$ & $1.94\pm0.39$   & 2.69                     & 0.0001   & N                    & --              & Columba (37)   \\
 HR 9           & $39.96\pm0.10$  & 23--29     & (7)   &        6900 & $1.49\substack{+0.02 \\ -0.01}$ & $4.07\pm0.49$   & 1.39                     & 0.0034   & N                    & 100             & bet Pic (9)    \\
 HR 506         & $17.34\pm0.02$  & 700--2830  & (21)  &        6100 & $1.14\substack{+0.01 \\ -0.02}$ & $1.57\pm0.18$   & 3.55                     & 0.0893   & (38)                 & 70,100,160      & --             \\
 HR 520         & $62.03\pm0.65$  & 448--603   & (3)   &        9200 & $2.16\substack{+0.07 \\ -0.07}$ & $34.67\pm4.43$  & 0.68                     & 0.0067   & N                    & 70,100,160      & --             \\
 HR 826         & $48.20\pm0.08$  & 100--500   & (3)   &        6900 & $1.53\substack{+0.01 \\ -0.00}$ & $4.32\pm0.40$   & 1.96                     & 0.0432   & N                    & --              & --             \\
 HR 1082        & $69.64\pm0.20$  & 38--48     & (9)   &        8400 & $1.80\substack{+0.03 \\ -0.03}$ & $11.75\pm2.00$  & 4.55                     & 0.0470   & N                    & 70,100,160      & Columba (9)    \\
 HR 1139        & $41.81\pm0.22$  & 279--778   & (3)   &        7800 & $1.62\substack{+0.05 \\ -0.05}$ & $8.11\pm1.68$   & 2.01                     & 0.0002   & N                    & Null            & --             \\
 HR 1254        & $34.85\pm0.16$  & 198--1002  & (3)   &        7000 & $1.64\substack{+0.02 \\ -0.02}$ & $5.89\pm0.30$   & 0.81                     & 0.0111   & N                    & --              & --             \\
 HR 1919        & $70.46\pm0.39$  & 38--48     & (9)   &        8900 & $1.99\substack{+0.02 \\ -0.04}$ & $16.46\pm2.91$  & 0.78                     & 0.0010   & N                    & 70,100,160      & Columba (39)   \\
 HR 2562        & $34.04\pm0.05$  & 100--500   & (3)   &        6800 & $1.43\substack{+0.01 \\ -0.01}$ & $3.47\pm0.81$   & 0.71                     & 0.0297   & N                    & 70,100,160      & --             \\
 HR 3300        & $73.25\pm0.38$  & 41--49     & (9)   &        9400 & $2.21\substack{+0.02 \\ -0.05}$ & $22.56\pm2.53$  & 0.44                     & 0.0041   & N                    & Null            & Carina (40)    \\
 HR 3341        & $69.33\pm0.27$  & 130--200   & (9)   &        8800 & $1.96\substack{+0.03 \\ -0.05}$ & $16.75\pm2.42$  & 1.32                     & 0.0127   & N                    & 100             & AB Dor (40)    \\
 HR 4796 A      & $72.78\pm1.75$  & 7--13      & (9)   &        9600 & $2.23\substack{+0.04 \\ -0.05}$ & $26.44\pm5.48$  & 48.88                    & 0.2415   & (41)                 & 100,160         & TWA (9)        \\
 HR 5751        & $55.32\pm0.15$  & 14--18     & (4)   &        7800 & $1.69\substack{+0.03 \\ -0.02}$ & $8.02\pm1.70$   & 0.79                     & 0.0001   & N                    & --              & UCL (28)       \\
 HR 6948        & $37.05\pm0.06$  & 50--200    & (3)   &        6600 & $1.46\substack{+0.01 \\ -0.01}$ & $3.14\pm0.58$   & 3.14                     & 0.0966   & N                    & 70,100,160      & --             \\
 HR 7012        & $28.55\pm0.15$  & 23--29     & (7)   &        7700 & $1.70\substack{+0.01 \\ -0.02}$ & $8.13\pm1.67$   & 8.39                     & 0.0021   & N                    & 100,160         & bet Pic (9)    \\
 HR 7380        & $69.93\pm1.76$  & 23--29     & (7)   &        9700 & $2.34\substack{+0.05 \\ -0.05}$ & $27.78\pm3.83$  & 1.91                     & 0.0180   & N                    & 70,100,160      & bet Pic (40)   \\
 HR 8323        & $15.56\pm0.02$  & 1050--3150 & (42)  &        6100 & $1.12\substack{+0.05 \\ -0.08}$ & $1.19\pm0.34$   & 1.27                     & 0.0270   & (43)                 & --              & --             \\
 HR 8799        & $41.29\pm0.15$  & 38--48     & (9)   &        7300 & $1.55\substack{+0.00 \\ -0.01}$ & $5.56\pm1.19$   & 2.63                     & 0.0387   & N                    & 70,100,160      & Columba (9)    \\
 $\nu$ Hor      & $50.45\pm0.46$  & 416--713   & (3)   &        8700 & $1.89\substack{+0.08 \\ -0.07}$ & $18.05\pm2.27$  & 0.42                     & 0.0041   & N                    & 70,100,160      & --             \\
 NZ Lup         & $60.34\pm0.18$  & 14--18     & (3)   &        6000 & $1.29\substack{+0.01 \\ -0.01}$ & $2.29\pm0.26$   & 1.34                     & 0.0151   & (17)                 & --              & --             \\
 $o^1$ Cnc      & $48.40\pm0.41$  & 597--883   & (3)   &        8200 & $1.82\substack{+0.06 \\ -0.06}$ & $15.85\pm0.19$  & 0.88                     & 0.0097   & N                    & 100,160         & --             \\
 $o^2$ Cnc      & $48.80\pm0.23$  & 589--980   & (3)   &        7700 & $1.67\substack{+0.06 \\ -0.05}$ & $10.37\pm0.88$  & 1.65                     & 0.0220   & N                    & 100,160         & --             \\
 $\phi^1$ Pav   & $27.79\pm0.19$  & 677--1114  & (3)   &        7300 & $1.62\substack{+0.05 \\ -0.05}$ & $7.76\pm0.17$   & 0.96                     & 0.0154   & N                    & 70,160          & --             \\
 $\pi^1$ Ori    & $35.66\pm0.32$  & 265--579   & (3)   &        9000 & $1.85\substack{+0.06 \\ -0.06}$ & $16.36\pm2.92$  & 0.74                     & 0.0068   & N                    & 100,160         & --             \\
 $\rho$ Vir     & $36.27\pm0.28$  & 75--347    & (3)   &        9000 & $1.83\substack{+0.04 \\ -0.05}$ & $13.08\pm4.34$  & 0.7                      & 0.0061   & N                    & 70,100,160      & --             \\
 $\tau$ Cet     & $3.65\pm0.00$   & 2900--8700 & (42)  &        5750 & $0.85\substack{+0.02 \\ -0.01}$ & $0.56\pm0.23$   & 0.99                     & 0.0410   & N                    & 70,160 (44)     & --             \\
 TWA 25         & $53.11\pm0.19$  & 7--13      & (9)   &        3550 & $0.62\substack{+0.08 \\ -0.07}$ & $0.13\pm0.10$   & --                       & --       & (11)                 & Null (12)       & TWA (9)        \\
 V419 Hya       & $21.54\pm0.02$  & 130--200   & (9)   &        5300 & $0.90\substack{+0.00 \\ -0.01}$ & $0.41\pm0.04$   & 4.85                     & 0.1521   & (45)                 & 70,160          & AB Dor (46)    \\
 V435 Car       & $71.43\pm0.17$  & 206--675   & (3)   &        7700 & $1.64\substack{+0.05 \\ -0.05}$ & $8.29\pm1.64$   & 5.34                     & 0.0647   & N                    & 100,160         & --             \\
 $\zeta$ Lep    & $21.61\pm0.07$  & 23--29     & (7)   &        8600 & $2.06\substack{+0.03 \\ -0.05}$ & $15.72\pm2.14$  & 1.91                     & 0.0047   & N                    & 100             & bet Pic (23)   \\
%\targtable
\enddata
%\vspace{-0.8cm} 
\tablecomments{GPIES-observed disk targets sorted by name, with a $^\dagger$ denoting the three protoplanetary/transitional disks. Values for $T_{\mathrm{eff}}$, $M_\star$, $L_\star$, \Lir, and some ages were newly estimated for GPIES as described here and in \citet{nielsen2019}; see Sections \ref{sect:targ_selection} and \ref{sect:final_sample} for details.\ Column descriptions: star name; distance with $1\sigma$ uncertainties; stellar age; reference for stellar age; stellar mass with 68\% confidence range; stellar bolometric luminosity with $1\sigma$ uncertainties; stellar effective temperature; IR-excess magnitude; GPIES detectability metric value; whether the disk was previously resolved in scattered light when selected as a GPIES target, listing either the discovery reference or an ``N'' for no previous detection; wavelengths (in $\micron$) at which the disk was detected at ${\geq}3\sigma$ with \textit{Herschel} PACS (``Null'' for no detection, ``--'' for no data); moving group membership. References are in parentheses.}

\tablerefs{Distances are from Gaia DR2 \citep{gaia_mission, gaia_dr2}. Other references: (1) \citet{zuckerman2019}, (2) \citet{choquet2017}, (3) estimated for the GPIES campaign and described in \citealt{nielsen2019}, (4) \citet{pecaut2016}, (5) \citet{janson2016}, (6) \citet{dezeeuw1999}, (7) \citet{nielsen2016}, (8) \citet{kalas2004}, (9) \citet{bell2015}, (10) \citet{smith1984}, (11) \citet{choquet2016}, (12) \citet{riviere-marichalar:2013}, (13) \citet{kalas2005_fomb}, (14) \citet{kalas2006}, (15) \citet{zuckerman2004}, (16) \citet{kalas2007a}, (17) \citet{soummer2014}, (18) \citet{schneider2005}, (19) \citet{moor2016}, (20) \citet{torres2008}, (21) \citet{aguilera-gomez2018}, (22) \citet{barrado2004}, (23) \citet{nakajima2012}, (24) \citet{hines2007}, (25) \citet{pantin2000}, (26) \citet{ardila2004}, (27) \citet{kasper2015}, (28) \citet{rizzuto2012}, (29) \citet{padgett2016}, (30) \citet{wahhaj2016}, (31) \citet{currie2015a}, (32) \citet{matthews2017}, (33) \citet{carpenter:2009, mathews2013}, (34) \citet{weinberger1999, augereau1999a}, (35) \citet{thalmann2013}, (36) \citet{schneider2006}, (37) \citet{moor2013a}, (38) \citet{stapelfeldt2007_lyot}, (39) \citet{zuckerman2012_49cet}, (40) ${>}90$\% membership probability from the BANYAN $\Sigma$ tool \citep{gagne2018a}, (41) \citet{schneider1999}, (42) \citet{mamajek2008}, (43) \citet{krist2010}, (44) \citet{lawler2014}, (45) \citet{golimowski2011}, (46) \citet{lopez-santiago2006}.}

\end{deluxetable*}
%\end{longrotatetable}

\section{Observations}

\subsection{Observed Sample} \label{sect:obs_sample}

In total, we observed \Nobs targets: 96 during the survey, plus an additional 8 that were observed during GPI commissioning. We include this commissioning data in our analyses because those targets would have been observed in the regular survey had we not already had high-quality data in hand. All observed targets are listed in Table \ref{tab:obs_targs} in order of increasing right ascension (RA). This table also contains the basic stellar properties used elsewhere in our analysis and some information known about the disks at the start of GPIES observations, such as whether a disk had already been resolved in scattered light at the time that it was added to our target list.

We list basic information for every data set by target in Table \ref{tab:datasets}. This includes each data set's observing mode (described in Section \ref{sect:obs_modes}), exposure time per individual frame, total integration time, total parallactic angle rotation, whether a disk was detected in that specific data set (``P'' for a polarized detection and/or ``I'' for a total intensity detection), and the observation date. We also repeat the IR-excess magnitude from Table \ref{tab:obs_targs} for reference and provide the target star's apparent magnitude in the Cousins $I$ and 2MASS $H$ filter bands for reference. These are synthetic magnitudes calculated by fitting a stellar atmosphere model to real photometric measurements from the literature and then convolving this model with the filter transmission profiles (see \citealt{cotten2016} for the detailed procedure). Per normal GPI operations, the AO high-order wavefront sensor was read out at a rate of 1 kHz for stars with $I<8$ mag and 500 Hz for stars with $I\geq8$ mag (exchanging slightly degraded contrast for a higher signal-to-noise ratio (S/N) wavefront measurement on faint stars; \citealt{bailey2016}).

% AVOID ORPHANED SECTION HEADING
\vspace{40pt}

\subsection{Observing Modes}
\label{sect:obs_modes}

The GPIES disk survey was conducted using GPI's polarimetric mode \citep{perrin2010, perrin2015_4796}. After the coronagraphic focal plane, a rotating half-wave plate modulates the polarization state (i.e., polarization direction) of the incident linearly polarized light. It then employs a polarizing Wollaston prism beamsplitter placed after the IFS lenslet array to divide the light into two orthogonal polarization states so that each lenslet produces a pair of polarization spots on the detector. Observing sequences typically consist of one or more sets of four images with the half-wave plate at positions of [$0\fdg0$, $22\fdg5$, $45\fdg0$, $67\fdg5$]. This enables measurements of the Stokes \I, \Q, and \U vectors in each spatial element (``spaxel''). Details of our data reduction methods are provided in Section \ref{sect:reduction}.

For most analyses, we convert the regular Stokes vectors \Q and \U into their ``radial'' components, \Qr and \Ur, via

\begin{equation} \label{eq:radial_stokes}
\begin{aligned}
    \mathrm{\mathcal{Q_{\phi}}} &= \mathrm{\mathcal{Q}} \cos{2\phi} + \mathrm{\mathcal{U}} \sin{2\phi} \\
    \mathrm{\mathcal{U_{\phi}}} &= -\mathrm{\mathcal{Q}} \sin{2\phi} + \mathrm{\mathcal{U}} \cos{2\phi},
\end{aligned}
\end{equation}

where $\phi$ is the azimuthal angle of a given pixel around the star as measured counterclockwise from $-\pi$ to $\pi$ radians starting at the $-X$-axis \citep{schmid2006, millar-blanchaer2015}. Afterwards, \Qr contains all of the linear polarized intensity that is aligned perpendicularly or parallel to the vector from a given pixel to the central star. The \Ur image contains the polarized intensity aligned $\pm45\degr$ to that same vector. In the context of optically thin single scattering in debris disks, which we assume here, the disk signal is expected to be found only in the \Qr frame, and the \Ur frame should only contain noise (see \ref{appendix-a} for more discussion).

Pol-mode observations of disk targets were divided into two categories based on total integration time: short ``snapshot'' and longer ``deep'' observations, which we discuss in detail in Sections \ref{sect:snapshots} and \ref{sect:deeps}. The pol-mode data allow us to retrieve both polarized intensity and total intensity (i.e. polarization-agnostic) disk signals.

Spec-mode imaging replaces the Wollaston prism with a dispersing prism to produce low-resolution micro-spectra ($R\approx50$) on the detector. This only measures the total intensity of the disk (i.e., Stokes \I), but it is the primary mode used for the GPIES planet-search campaign. Given the substantial overlap between the planet and disk target lists, 91 of our observed disk targets were also observed in spec-mode.

Regardless of the light dispersion mode, GPI always operates in angular differential imaging (ADI) mode so that the sky appears to constantly rotate on the detector \citep{marois2006}. Therefore, astrophysical sources in our images rotate over time while the stellar PSF remains approximately fixed. We take advantage of this angular diversity to subtract the stellar signal from the data and increase the final contrast in the total intensity images obtained from both pol-mode and spec-mode data sets. We discuss our data reduction methods in more detail in Section \ref{sect:spec_mode}.

All GPIES images also include four ``satellite'' spots---fiducial images of the occulted star produced by a periodic grid superimposed on GPI's pupil apodizer such that they have known locations and fixed flux ratios relative to that star \citep{marois2006, sivaramakrishnan2006}. The satellite spots are used to precisely determine the star's position in every image and also to photometrically calibrate the data (see Section \ref{sect:pol_reduction} for details).

\subsubsection{Snapshot Polarimetric Observations} \label{sect:snapshots}

The majority (83) of our polarimetric observations are relatively short ``snapshots'' in order to maximize the number of targets observed. Overall, 12 of these snapshots produced detections. The snapshot data for each target totaled less than 20 minutes of integration time, or about 30 minutes of wall-clock time when considering telescope and instrument overheads. The median integration time was 15.5 minutes. Our operating assumption was that most disks detectable with GPI in any amount of integration time reasonable for a broad survey would be detected at some level in a snapshot. Disks with high S/N in a snapshot were immediately ready for analysis, and disks with low S/N could be followed up with a deep observation. ``Quicklook'' reductions of the polarized intensity produced automatically on remote computers during observations were helpful in quickly identifying detections in real time, sometimes allowing us to seamlessly extend a snapshot into a deep observation and avoid revisiting a target.

At the start of GPIES, snapshots consisted of eight frames that encompassed two sets of four half-wave plate rotations. However, we increased the number of frames to 16 (four sets of rotations) after the first ${\sim}$6 months because detector persistence was introducing additional noise into the first frames of a pol-mode data set \citep{Millar-Blanchaer2016}. This persistence occurs when electrons previously freed by incident light get trapped in the HAWAII-2RG detector's crystal lattice and then become indistinguishable from newly released electrons in subsequent readouts. Due to the high incident flux and tendency to bias the difference between pairs of polarization spots, persistence is particularly strong shortly after observing a bright star in spec-mode, sometimes remaining significant over ${\sim}20$ minutes later. To mitigate this effect, we made efforts to place pol-mode observations before spec-mode observations on nightly schedules.

Due to the inherently brief observations of the snapshots and our focus on polarized intensity detections, accruing field rotation was not a priority; the median rotation was only $6\fdg9$. Consequently, we had limited sensitivity to disks in total intensity in the snapshots. Deep pol and spec data sets were typically longer and encompassed more rotation, leading to more effective stellar PSF subtraction with less suppression of the disk light, hence better sensitivity to total intensity disk signals.

\subsubsection{Deep Polarimetric Observations} \label{sect:deeps}

Deep pol-mode observations were identical to snapshots except for longer total integration times, defined as 20 minutes or more. The median integration time was 35.8 minutes across the 36 deep data sets acquired. Of these, 23 produced detections (all in \Qr, with 18 also in total intensity). Deep observations were primarily assigned for follow-up of targets with disks detected in GPIES snapshots or for targets previously detected with another instrument in scattered light. These selection biases are largely responsible for the higher detection rate of deep observations compared to snapshots.

As expected, deep pol-mode detections typically had higher \Qr S/N than snapshots and were therefore used for the main analysis where they existed. With the greater time on source, the deep observations also generally encompassed more field rotation (median $21\fdg9$) than snapshots and tended to yield higher-S/N total intensity detections as well, or a first detection where the snapshot produced none.

As a point of reference, our longest observation was 155.2 minutes, resulting in a non-detection of HR 2562's disk. This data set represents an unusual case, though, with more than twice the integration time of the next longest. HR 2562 hosts a directly imaged brown dwarf discovered by the GPIES planet search \citep{konopacky2016}, and our goal was to better resolve the debris disk that had been previously resolved on a larger scale with \textit{Herschel} PACS \citep{moor2011a, moor2015} but had not yet been seen in scattered light. This would answer open questions about whether the brown dwarf orbits within the disk's inner hole and whether that orbit is coplanar with the disk \citep{maire2018}. The length of the sequence was based on the disk's non-detection in a 12 minute snapshot, non-detection in archival NICMOS data \citep{choquet2015}, and weak IR excess (\Lir of $\num{7e-5}$ from our model fitting and $\num{1e-5}$ from \citealt{moor2015}). Despite the long integration, we did not detect the disk.

\subsubsection{Spectroscopic Observations}

In tandem with the polarimetric observations, 91 disk targets were also observed in spec-mode as part of the GPIES planet-search campaign. These observations provided additional total intensity data but no polarization information. The spec- and pol-mode observations were typically made consecutively to minimize time spent on target acquisition, although scheduling constraints, delayed follow-up, and variable weather occasionally separated the observations by periods from days to years.

The median spec-mode data set comprised 37.8 minutes of integration (across 38 exposures) and $29\fdg6$ of field rotation. This is 35\% more rotation than the median deep pol observation, mainly because the planet search prioritized observing targets near transit to aid ADI PSF subtraction for faint point-source detection. Partly for this reason, five disks were detected in total intensity only in their spec data. Two of those (HD 15115 and NZ Lup) represent the only GPIES detections of the disks in either mode, emphasizing the complementarity of the spec data.

We do not investigate the spectral properties of the disks in this work, although the information is available for future studies. For many disks, the low surface brightness makes extracting the signal in individual wavelength channels challenging. When it can be measured, that brightness tends to be approximately constant relative to the stellar flux as a function of wavelength within the $H$ band.

% Campaign target table.
\startlongtable
\begin{deluxetable*}{lccccrrrcl}
%\tablecolumns{6}
%\tablewidth{0pt}
\tablecaption{GPIES Disk Observations by Target}
\tablehead{
% Header Line 1
\colhead{Name} & \colhead{\Lir} & \colhead{$I$} & \colhead{$H$} & \colhead{Mode} & \colhead{$t_{\mathrm{exp}}$} & \colhead{$t_{\mathrm{int}}$} & \colhead{$\Delta$PA} & \colhead{Detected?} & \colhead{Date} \vspace{-0.3cm} \\
% \colhead{$N_{frame}$}
% Header Line 2
 & \colhead{($10^{-4}$)} & \colhead{(mag)} & \colhead{(mag)} & & \colhead{(s)} & \colhead{(s)} & \colhead{($\degr$)} & &
}
\startdata
49 Cet & 10.7 & 5.5 & 5.5 & Deep & 59.65 & 2385.86 & 27.8 & -- & 141111 \\
... & ... & ... & ... & Spec & 59.65 & 1013.99 & 17.3 & -- & 141111 \\
73 Her & 3.6 & 5.5 & 5.1 & Snap & 29.10 & 931.07 & 5.1 & -- & 170809 \\
... & ... & ... & ... & Spec & 59.65 & 2326.21 & 13.0 & -- & 170809 \\
AK Sco$^\dagger$ & 1098.3 & 8.0 & 7.1 & Deep & 59.65 & 2505.15 & 54.9 & P,I & 180811 \\
... & ... & ... & ... & Spec & 59.65 & 1849.04 & 65.6 & I & 180809 \\
AU Mic & 3.3 & 6.6 & 4.8 & Deep & 59.65 & 2624.44 & 166.9 & P,I & 140515* \\
... & ... & ... & ... & Spec & 59.65 & 2863.03 & 8.5 & -- & 140912* \\
b Leo & 0.4 & 4.4 & 4.3 & Snap & 4.36 & 523.72 & 5.5 & -- & 180326 \\
... & ... & ... & ... & Spec & 59.65 & 2266.56 & 12.8 & -- & 160427 \\
$\beta$ Pic & 28.1 & 3.7 & 3.5 & Deep & 5.82 & 3258.73 & 91.5 & P,I & 131212* \\
... & ... & ... & ... & Spec & 59.65 & 4115.60 & 38.5 & -- & 180924 \\
CE Ant & 6.5 & 9.1 & 7.1 & Snap & 59.65 & 1192.93 & 12.3 & P & 180204 \\
... & ... & ... & ... & Deep & 119.29 & 3817.37 & 12.8 & P & 180405 \\
... & ... & ... & ... & Spec & 59.65 & 2266.56 & 112.4 & -- & 180204 \\
$\epsilon$ Eri & 0.8 & 2.5 & 1.8 & Snap & 1.45 & 132.39 & 6.9 & -- & 141110 \\
... & ... & ... & ... & Snap & 1.45 & 907.79 & 15.4 & -- & 180130 \\
... & ... & ... & ... & Spec & 14.55 & 1978.51 & 27.1 & -- & 141110 \\
$\eta$ Cha & 0.9 & 5.5 & 5.7 & Snap & 43.64 & 698.30 & 3.9 & -- & 151218 \\
... & ... & ... & ... & Spec & 59.65 & 2266.56 & 13.4 & -- & 180105 \\
$\eta$ Tel A & 1.4 & 5.0 & 5.0 & Snap & 29.10 & 465.53 & 4.6 & -- & 150501 \\
... & ... & ... & ... & Spec & 59.65 & 2505.15 & 26.7 & -- & 150501 \\
Fomalhaut & 0.6 & 1.0 & 0.9 & Snap & 1.45 & 209.49 & 0.3 & -- & 150829 \\
... & ... & ... & ... & Spec & 2.91 & 1222.02 & 166.1 & -- & 150830 \\
g Lup & 0.7 & 4.1 & 3.7 & Snap & 4.36 & 628.47 & 17.2 & -- & 160322 \\
... & ... & ... & ... & Spec & 52.37 & 2356.76 & 43.7 & -- & 160322 \\
$\gamma$ Dor & 0.3 & 3.8 & 3.5 & Snap & 4.36 & 279.32 & 4.0 & -- & 141110 \\
... & ... & ... & ... & Spec & 59.65 & 2326.21 & 28.7 & -- & 141110 \\
$\gamma$ Oph & 1.1 & 3.7 & 3.6 & Snap & 4.36 & 418.98 & 3.4 & -- & 150702 \\
... & ... & ... & ... & Spec & 59.65 & 2505.15 & 16.9 & -- & 150702 \\
HD 377 & 4.0 & 6.9 & 6.1 & Deep & 59.65 & 2147.27 & 12.2 & -- & 171128 \\
... & ... & ... & ... & Spec & 59.65 & 2863.03 & 22.0 & -- & 171128 \\
HD 1466 & 1.2 & 6.8 & 6.2 & Snap & 59.65 & 954.34 & 7.3 & -- & 181122 \\
... & ... & ... & ... & Spec & 59.65 & 2564.79 & 21.0 & -- & 150703 \\
HD 3888 & 0.9 & 6.8 & 6.2 & Snap & 59.65 & 954.34 & 7.6 & -- & 181221 \\
... & ... & ... & ... & Spec & 59.65 & 2206.92 & 25.6 & -- & 150831 \\
HD 7112 & 2.0 & 8.0 & 7.2 & Snap & 59.65 & 894.70 & 6.1 & -- & 181121 \\
... & ... & ... & ... & Spec & 59.65 & 2266.56 & 16.8 & -- & 171111 \\
HD 8813 & 1.1 & 7.7 & 6.8 & Snap & 59.65 & 954.34 & 5.8 & -- & 181122 \\
... & ... & ... & ... & Spec & 59.65 & 2087.62 & 12.8 & -- & 171110 \\
HD 10472 & 4.1 & 7.1 & 6.7 & Snap & 59.65 & 715.76 & 6.2 & -- & 151106 \\
... & ... & ... & ... & Deep & 88.74 & 2307.30 & 21.1 & -- & 151221 \\
... & ... & ... & ... & Spec & 59.65 & 2266.56 & 19.5 & -- & 151106 \\
HD 13183 & 1.6 & 7.9 & 7.0 & Snap & 59.65 & 954.34 & 8.7 & -- & 181121 \\
... & ... & ... & ... & Spec & 59.65 & 2266.56 & 25.3 & -- & 181122 \\
HD 13246 & 2.0 & 6.9 & 6.3 & Snap & 59.65 & 954.34 & 8.4 & -- & 180923 \\
... & ... & ... & ... & Spec & 59.65 & 2326.21 & 27.4 & -- & 141218 \\
HD 15115 & 4.8 & 6.3 & 5.8 & Deep & 59.65 & 1670.10 & 12.5 & -- & 131212* \\
... & ... & ... & ... & Spec & 59.65 & 3757.72 & 35.2 & I & 141216 \\
HD 15279 & 0.8 & 7.6 & 6.9 & Snap & 59.65 & 954.34 & 10.5 & -- & 181221 \\
... & ... & ... & ... & Spec & 59.65 & 2206.92 & 25.0 & -- & 170806 \\
HD 16743 & 4.5 & 6.3 & 5.9 & Snap & 59.65 & 954.34 & 10.7 & -- & 161118 \\
... & ... & ... & ... & Spec & 59.65 & 2266.56 & 26.1 & -- & 161118 \\
HD 23208 & 2.7 & 8.3 & 7.3 & Snap & 59.65 & 954.34 & 21.5 & -- & 171111 \\
... & ... & ... & ... & Spec & 59.65 & 2266.56 & 45.2 & -- & 171111 \\
HD 24636 & 1.0 & 6.7 & 6.2 & Snap & 59.65 & 954.34 & 6.0 & -- & 181122 \\
... & ... & ... & ... & Spec & 59.65 & 2266.56 & 14.8 & -- & 180924 \\
HD 30447 & 9.1 & 7.4 & 6.9 & Snap & 59.65 & 1192.93 & 19.9 & P & 160920 \\
... & ... & ... & ... & Deep & 59.65 & 3101.61 & 125.8 & P,I & 160922 \\
... & ... & ... & ... & Spec & 59.65 & 2266.56 & 91.7 & I & 160920 \\
HD 31392 & 1.8 & 6.8 & 5.9 & Snap & 29.10 & 465.53 & 3.6 & -- & 141216 \\
... & ... & ... & ... & Spec & 59.65 & 2445.50 & 63.3 & -- & 141216 \\
HD 32195 & 0.9 & 7.5 & 7.0 & Snap & 59.65 & 954.34 & 5.5 & -- & 181122 \\
... & ... & ... & ... & Spec & 59.65 & 2266.56 & 17.9 & -- & 161220 \\
HD 32297 & 60.5 & 7.9 & 7.6 & Deep & 59.65 & 2147.27 & 19.1 & P,I & 141218 \\
... & ... & ... & ... & Spec & 59.65 & 2206.92 & 16.7 & I & 161220 \\
HD 32372 & 2.4 & 8.6 & 7.8 & Snap & 59.65 & 1192.93 & 16.4 & -- & 181121 \\
... & ... & ... & ... & Spec & 59.65 & 2266.56 & 46.5 & -- & 161221 \\
HD 35841 & 3.3 & 8.6 & 7.8 & Snap & 59.65 & 954.34 & 2.6 & P & 160228 \\
... & ... & ... & ... & Deep & 88.74 & 2484.78 & 3.7 & P,I & 160318 \\
... & ... & ... & ... & Spec & 59.65 & 2863.03 & 46.9 & I & 160228 \\
HD 37484 & 3.5 & 6.8 & 6.3 & Snap & 59.65 & 954.34 & 4.8 & -- & 171110 \\
... & ... & ... & ... & Spec & 59.65 & 1849.04 & 141.4 & -- & 171110 \\
HD 50554 & 0.5 & 6.3 & 5.5 & Deep & 59.65 & 2863.03 & 19.6 & -- & 180201 \\
HD 53143 & 2.0 & 6.1 & 5.1 & Snap & 29.10 & 698.30 & 5.9 & -- & 151219 \\
... & ... & ... & ... & Spec & 59.65 & 2206.92 & 20.3 & -- & 151219 \\
HD 57969 & 2.4 & 6.5 & 6.3 & Deep & 59.65 & 1371.87 & 13.2 & -- & 171128 \\
... & ... & ... & ... & Deep & 59.65 & 1431.51 & 14.3 & -- & 171230 \\
... & ... & ... & ... & Spec & 59.65 & 3161.26 & 30.1 & -- & 180130 \\
HD 61005 & 27.9 & 7.4 & 6.5 & Deep & 59.65 & 2087.62 & 140.1 & P,I & 140324* \\
... & ... & ... & ... & Spec & 59.65 & 2445.50 & 112.3 & I & 150130 \\
HD 72687 & 1.0 & 7.5 & 6.7 & Snap & 59.65 & 1133.28 & 1.5 & -- & 171228 \\
... & ... & ... & ... & Spec & 59.65 & 2147.27 & 172.3 & -- & 160229 \\
HD 80846 & 0.6 & 7.8 & 7.2 & Snap & 59.65 & 954.34 & 7.7 & -- & 171228 \\
... & ... & ... & ... & Spec & 59.65 & 1849.04 & 37.8 & -- & 160225 \\
HD 82943 & 1.1 & 6.1 & 5.2 & Snap & 29.10 & 872.87 & 13.6 & -- & 160226 \\
... & ... & ... & ... & Spec & 59.65 & 3817.37 & 47.4 & -- & 171229 \\
HD 84040 & 0.7 & 7.9 & 7.3 & Snap & 59.65 & 1192.93 & 18.9 & -- & 180309 \\
... & ... & ... & ... & Spec & 59.65 & 2147.27 & 107.4 & -- & 160229 \\
HD 84075 & 1.9 & 7.9 & 7.2 & Snap & 59.65 & 954.34 & 5.5 & -- & 180106 \\
... & ... & ... & ... & Spec & 59.65 & 3519.14 & 22.9 & -- & 180128 \\
HD 89452 & 3.6 & 8.0 & 6.9 & Snap & 59.65 & 954.34 & 6.8 & -- & 180105 \\
... & ... & ... & ... & Spec & 59.65 & 2266.56 & 22.7 & -- & 170213 \\
HD 95086 & 8.4 & 7.1 & 6.8 & Snap & 59.65 & 536.82 & 8.9 & -- & 150408 \\
... & ... & ... & ... & Snap & 59.65 & 954.34 & 6.9 & -- & 160229 \\
... & ... & ... & ... & Spec & 59.65 & 4652.42 & 47.6 & -- & 160306 \\
HD 100546$^\dagger$ & 7268.5 & 6.7 & 6.0 & Snap & 29.10 & 232.77 & 1.2 & P & 131212* \\
... & ... & ... & ... & Spec & 59.65 & 7157.57 & 51.6 & I & 160227 \\
HD 106906 & 50.4 & 7.3 & 6.8 & Snap & 88.74 & 709.94 & 7.1 & P,I & 150504 \\
... & ... & ... & ... & Deep & 59.65 & 2564.79 & 20.3 & P,I & 150701 \\
... & ... & ... & ... & Spec & 59.65 & 2505.15 & 25.3 & I & 150504 \\
HD 107146 & 10.1 & 6.4 & 5.6 & Snap & 59.65 & 954.34 & 6.7 & -- & 160225 \\
... & ... & ... & ... & Spec & 59.65 & 2266.56 & 21.9 & -- & 160225 \\
HD 108857 & 6.9 & 8.0 & 7.2 & Snap & 59.65 & 775.40 & 3.9 & -- & 181221 \\
... & ... & ... & ... & Spec & 59.65 & 2266.56 & 22.6 & -- & 160427 \\
HD 110058 & 26.2 & 7.8 & 7.5 & Snap & 59.65 & 1073.64 & 14.1 & -- & 160126 \\
... & ... & ... & ... & Deep & 59.65 & 2147.27 & 25.2 & P,I & 160319 \\
... & ... & ... & ... & Spec & 59.65 & 2147.27 & 29.6 & I & 160319 \\
HD 111161 & 42.3 & 7.3 & 7.2 & Snap & 59.65 & 954.34 & 6.8 & P & 180204 \\
... & ... & ... & ... & Deep & 59.65 & 4533.13 & 38.0 & P & 180310 \\
... & ... & ... & ... & Spec & 88.74 & 2484.78 & 16.9 & -- & 180204 \\
HD 111520 & 10.3 & 8.3 & 7.7 & Snap & 29.10 & 581.92 & 7.4 & P,I & 150702 \\
... & ... & ... & ... & Deep & 88.74 & 2839.75 & 28.3 & P,I & 160318 \\
... & ... & ... & ... & Spec & 59.65 & 2445.50 & 34.8 & I & 150702 \\
HD 114082 & 36.3 & 7.7 & 7.2 & Deep & 59.65 & 2087.62 & 12.3 & P & 170807 \\
... & ... & ... & ... & Spec & 59.65 & 2803.38 & 25.8 & I & 180129 \\
HD 115600 & 22.6 & 7.8 & 7.3 & Deep & 59.65 & 2624.44 & 24.0 & P,I & 150703 \\
... & ... & ... & ... & Spec & 59.65 & 2266.56 & 19.9 & I & 180310 \\
HD 117214 & 26.7 & 7.5 & 6.9 & Deep & 59.65 & 1908.68 & 18.5 & P,I & 180311 \\
... & ... & ... & ... & Spec & 59.65 & 2206.92 & 19.8 & I & 180311 \\
HD 129590 & 69.6 & 8.5 & 7.8 & Deep & 59.65 & 2147.27 & 17.9 & P,I & 170809 \\
HD 131835 & 30.9 & 7.7 & 7.5 & Deep & 59.65 & 1908.68 & 74.2 & P & 150501 \\
... & ... & ... & ... & Spec & 59.65 & 2445.50 & 94.8 & -- & 150504 \\
HD 138813 & 13.4 & 7.2 & 7.2 & Snap & 59.65 & 536.82 & 25.8 & -- & 180311 \\
HD 141569$^\dagger$ & 127.4 & 7.0 & 6.8 & Snap & 59.65 & 238.59 & 1.6 & P & 140321* \\
... & ... & ... & ... & Deep & 59.65 & 3041.97 & 47.3 & P,I & 140322* \\
... & ... & ... & ... & Spec & 59.65 & 2445.50 & 24.3 & I & 150404 \\
HD 142315 & 6.4 & 6.8 & 6.7 & Snap & 59.65 & 954.34 & 0.2 & -- & 180922 \\
... & ... & ... & ... & Spec & 59.65 & 1670.10 & 47.0 & -- & 150501 \\
HD 143675 & 5.6 & 7.8 & 7.6 & Snap & 59.65 & 954.34 & 21.0 & P & 180408 \\
... & ... & ... & ... & Spec & 59.65 & 3101.61 & 94.3 & I & 180408 \\
HD 145560 & 12.7 & 8.4 & 7.8 & Deep & 59.65 & 1670.10 & 17.6 & P & 180812 \\
... & ... & ... & ... & Spec & 59.65 & 2266.56 & 36.1 & -- & 180812 \\
HD 146897 & 101.9 & 8.6 & 7.8 & Deep & 88.74 & 1774.84 & 28.9 & P,I & 160321 \\
... & ... & ... & ... & Spec & 59.65 & 2266.56 & 59.5 & -- & 180815 \\
HD 156623 & 43.3 & 7.1 & 7.0 & Snap & 59.65 & 954.34 & 11.2 & P & 170421 \\
... & ... & ... & ... & Deep & 88.74 & 2129.81 & 28.2 & P & 190427 \\
... & ... & ... & ... & Spec & 59.65 & 2266.56 & 38.3 & -- & 170421 \\
HD 157587 & 32.0 & 7.9 & 7.3 & Deep & 88.74 & 2484.78 & 49.9 & P & 150829 \\
HD 164249 A & 10.3 & 6.5 & 5.9 & Snap & 29.10 & 465.53 & 5.3 & -- & 150501 \\
... & ... & ... & ... & Spec & 59.65 & 2505.15 & 29.6 & -- & 150501 \\
HD 181327 & 13.5 & 6.5 & 6.0 & Snap & 59.65 & 1133.28 & 13.2 & -- & 140512* \\
... & ... & ... & ... & Spec & 59.65 & 2505.15 & 26.8 & -- & 150703 \\
HD 191089 & 15.0 & 6.6 & 6.1 & Snap & 59.65 & 775.40 & 0.6 & P & 150831 \\
... & ... & ... & ... & Deep & 88.74 & 2484.78 & 101.3 & P & 150901 \\
... & ... & ... & ... & Spec & 59.65 & 2206.92 & 7.2 & -- & 150831 \\
HD 202917 & 2.9 & 7.9 & 7.0 & Snap & 29.10 & 698.30 & 7.1 & -- & 150704 \\
... & ... & ... & ... & Spec & 59.65 & 2385.86 & 28.2 & -- & 150704 \\
HD 205674 & 2.4 & 6.7 & 6.2 & Snap & 59.65 & 954.34 & 20.3 & -- & 180921 \\
HD 206893 & 2.3 & 6.2 & 5.7 & Snap & 59.65 & 954.34 & 10.3 & -- & 160922 \\
... & ... & ... & ... & Spec & 59.65 & 2266.56 & 32.7 & -- & 160922 \\
HD 209253 & 0.9 & 6.0 & 5.5 & Deep & 59.65 & 1431.51 & 2.9 & -- & 181119 \\
HD 221853 & 6.8 & 6.9 & 6.4 & Snap & 59.65 & 715.76 & 4.9 & -- & 161119 \\
... & ... & ... & ... & Spec & 59.65 & 1312.22 & 8.7 & -- & 151106 \\
HIP 25434 & 2.7 & 8.3 & 7.8 & Snap & 59.65 & 835.05 & 11.7 & -- & 181121 \\
... & ... & ... & ... & Spec & 59.65 & 2266.56 & 46.5 & -- & 151221 \\
HR 9 & 1.4 & 5.7 & 5.2 & Snap & 29.10 & 931.07 & 16.6 & -- & 181122 \\
... & ... & ... & ... & Spec & 59.65 & 2266.56 & 68.3 & -- & 181122 \\
HR 506 & 3.5 & 5.0 & 4.3 & Deep & 5.82 & 1489.70 & 22.7 & -- & 151106 \\
HR 520 & 0.7 & 5.0 & 5.0 & Snap & 14.55 & 931.07 & 11.2 & -- & 170906 \\
... & ... & ... & ... & Spec & 59.65 & 2206.92 & 22.2 & -- & 170906 \\
HR 826 & 2.0 & 6.1 & 5.5 & Snap & 59.65 & 477.17 & 5.8 & -- & 141109 \\
... & ... & ... & ... & Spec & 59.65 & 1849.04 & 69.8 & -- & 141109 \\
HR 1082 & 4.6 & 6.2 & 6.1 & Snap & 59.65 & 954.34 & 38.0 & -- & 161117 \\
... & ... & ... & ... & Spec & 59.65 & 2326.21 & 79.3 & -- & 181221 \\
HR 1139 & 2.0 & 5.3 & 5.1 & Snap & 29.10 & 931.07 & 11.8 & -- & 181120 \\
... & ... & ... & ... & Spec & 59.65 & 2266.56 & 22.1 & -- & 160918 \\
HR 1254 & 0.8 & 5.1 & 4.6 & Snap & 11.64 & 744.85 & 4.1 & -- & 170906 \\
HR 1919 & 0.8 & 6.0 & 5.9 & Snap & 59.65 & 1133.28 & 10.3 & -- & 171111 \\
... & ... & ... & ... & Spec & 59.65 & 4533.13 & 58.0 & -- & 180204 \\
HR 2562 & 0.7 & 5.6 & 5.1 & Snap & 29.10 & 698.30 & 6.4 & -- & 160125 \\
... & ... & ... & ... & Deep & 29.10 & 9310.66 & 77.3 & -- & 180311 \\
... & ... & ... & ... & Spec & 59.65 & 3459.49 & 30.7 & -- & 171129 \\
HR 3300 & 0.4 & 5.8 & 5.8 & Deep & 59.65 & 1431.51 & 14.7 & -- & 171128 \\
... & ... & ... & ... & Spec & 59.65 & 2922.67 & 29.9 & -- & 180310 \\
HR 3341 & 1.3 & 6.0 & 5.9 & Deep & 59.65 & 1431.51 & 16.1 & -- & 171128 \\
... & ... & ... & ... & Spec & 59.65 & 2266.56 & 23.2 & -- & 151221 \\
HR 4796 A & 48.9 & 5.8 & 5.8 & Snap & 29.10 & 640.11 & 2.1 & P,I & 131212* \\
... & ... & ... & ... & Spec & 59.65 & 2206.92 & 53.0 & I & 160318 \\
HR 5751 & 0.8 & 6.0 & 5.7 & Snap & 59.65 & 954.34 & 12.9 & -- & 160318 \\
... & ... & ... & ... & Spec & 59.65 & 6322.52 & 116.2 & -- & 160318 \\
HR 6948 & 3.1 & 5.7 & 5.3 & Snap & 29.10 & 698.30 & 6.9 & -- & 150701 \\
... & ... & ... & ... & Spec & 59.65 & 2505.15 & 51.0 & -- & 150701 \\
HR 7012 & 8.4 & 4.6 & 4.2 & Snap & 4.36 & 1117.28 & 19.3 & P & 180921 \\
... & ... & ... & ... & Spec & 59.65 & 2505.15 & 20.8 & -- & 150408 \\
HR 7380 & 1.9 & 5.6 & 5.6 & Snap & 59.65 & 954.34 & 0.3 & -- & 180923 \\
... & ... & ... & ... & Spec & 59.65 & 2206.92 & 165.8 & -- & 180923 \\
HR 8323 & 1.3 & 4.9 & 4.3 & Snap & 4.36 & 663.38 & 8.2 & -- & 180812 \\
HR 8799 & 2.6 & 5.7 & 5.3 & Snap & 29.10 & 872.87 & 8.3 & -- & 160919 \\
... & ... & ... & ... & Spec & 59.65 & 3578.78 & 20.9 & -- & 160919 \\
$\nu$ Hor & 0.4 & 5.1 & 5.1 & Snap & 29.10 & 931.07 & 8.5 & -- & 161119 \\
... & ... & ... & ... & Spec & 59.65 & 2206.92 & 17.4 & -- & 161119 \\
NZ Lup & 1.3 & 7.1 & 6.4 & Snap & 29.10 & 698.30 & 18.0 & -- & 150408 \\
... & ... & ... & ... & Spec & 59.65 & 2505.15 & 47.6 & I & 150408 \\
$o^1$ Cnc & 0.9 & 5.1 & 4.9 & Snap & 29.10 & 931.07 & 6.6 & -- & 171228 \\
$o^2$ Cnc & 1.7 & 5.5 & 5.2 & Snap & 29.10 & 931.07 & 7.7 & -- & 160430 \\
$\phi^1$ Pav & 1.0 & 4.4 & 4.1 & Snap & 4.36 & 698.30 & 6.4 & -- & 181119 \\
$\pi^1$ Ori & 0.7 & 4.5 & 4.5 & Snap & 11.64 & 931.07 & 6.2 & -- & 181119 \\
... & ... & ... & ... & Spec & 59.65 & 1431.51 & 9.9 & -- & 151221 \\
$\rho$ Vir & 0.7 & 4.8 & 4.8 & Snap & 11.64 & 744.85 & 6.0 & -- & 160226 \\
... & ... & ... & ... & Spec & 59.65 & 2206.92 & 16.3 & -- & 160226 \\
$\tau$ Cet & 1.0 & 2.7 & 1.9 & Snap & 1.45 & 654.66 & 29.6 & -- & 161219 \\
TWA 25 & -- & 9.3 & 7.5 & Deep & 59.65 & 2147.27 & 20.3 & -- & 170213 \\
... & ... & ... & ... & Spec & 59.65 & 2266.56 & 53.7 & -- & 170213 \\
V419 Hya & 4.9 & 6.8 & 5.8 & Snap & 59.65 & 894.70 & 0.7 & -- & 160226 \\
... & ... & ... & ... & Spec & 59.65 & 2147.27 & 155.2 & -- & 160226 \\
V435 Car & 5.3 & 6.5 & 6.2 & Snap & 29.10 & 465.53 & 5.4 & -- & 141217 \\
... & ... & ... & ... & Spec & 59.65 & 2385.86 & 29.9 & -- & 141217 \\
$\zeta$ Lep & 1.9 & 3.4 & 3.3 & Snap & 2.91 & 698.30 & 5.2 & -- & 160229 \\
... & ... & ... & ... & Spec & 59.65 & 4712.06 & 51.8 & -- & 180131 \\
\enddata
%\vspace{-0.8cm} 
\tablecomments{Targets are sorted by name, with $^\dagger$ denoting the three protoplanetary/transitional disks. Other column headings: IR-excess magnitude repeated from Table \ref{tab:obs_targs} for reference; star's synthetic apparent magnitudes in the Cousins $I$ and 2MASS $H$ bands; $t_{\mathrm{exp}}$ = exposure time per frame; $t_{\mathrm{int}}$ = total integration time per data set; $\Delta$PA = total parallactic angle rotation per data set; detection status with ``P'' for polarized intensity, ``I'' for total intensity, and ``--'' for no detection; and date of observation as YYMMDD, with * denoting GPI commissioning data.\label{tab:datasets}}
\end{deluxetable*}

\section{Data Reduction}
\label{sect:reduction}
All of the data were reduced using the standard GPI Data Reduction Pipeline (DRP; version 1.4) procedures, previously documented in \citet{perrin2014_drp, perrin2016_drp} and \citet{Wang2018}. We summarize the basic steps in the following subsections and refer readers to the referenced publications for more details.

Generally, we consider the effective outer radius of our field of view in final reduced images to be $1\farcs4$ from the star. This is because our final images probe most PAs out to at least $1\farcs4$ after combining the differentially rotated individual frames of a given dataset (due to observation in ADI mode). Radial separations up to ${\sim}1\farcs8$ are visible over limited ranges of PAs (varying by data set) that are aligned with the corners of the individual frames' $2\farcs6\times2\farcs6$ fields of view. The minimum projected separation we can probe is $0\farcs123$, a limit set by the radius of the FPM. The GPI pixel scale is $14.166 \pm 0.007$ mas lenslet$^{-1}$ \citep{derosa2015b}.

Regarding units, the DRP default is to express intensity in GPI images in analog-to-digital units per coadd (ADU coadd$^{-1}$); however, we took additional steps (described throughout this section) to convert our images to stellar contrast units and/or surface brightness units of mJy arcsec$^{-2}$ for presentation and analysis.

\subsection{Polarization Mode}
\label{sect:pol_reduction}
Raw polarimetry data consist of a pair of PSFs for each lenslet in the IFS: one PSF for each of the two orthogonal polarization states. For nearly real-time analysis during observations, those raw data were transformed into final data products by the GPIES Data Cruncher \citep{Wang2018}, which is an automated data-processing architecture created for GPIES that implements the GPI DRP (among other data-archiving steps). Each data set was then inspected visually for disk signals. Those found to contain candidate disks, as well as any that were too noisy to be properly processed automatically, were subsequently re-reduced manually with the GPI DRP, tailoring to the individual noise properties of the data set. All disk detections presented here have been manually reduced to maximize disk S/N.

In this reduction process, each raw 2D image was dark subtracted, ``destriped'' to remove correlated detector noise and vibration-induced microphonics \citep{Ingraham2014}, and corrected for bad pixels. After a cross-correlation procedure to determine the exact location of each lenslet's two PSFs \citep{Draper2014}, the 2D data were assembled into a 3D ``polarization datacube'', where the first two dimensions held the spatial coordinates in the lenslet array and the third dimension corresponded to the orthogonal polarizations. Each image in the datacube was flat-fielded, another bad pixel correction step was applied, and then the 3D datacubes were corrected for field distortion from the instrument optics \citep{konopacky2014}. The position of the central star was then measured via the fiducial satellite spots \citep{wang2014} and their fluxes were recorded for later photometric calibration. In pol-mode images, we measured satellite spot flux as the integrated flux within a finite aperture\footnote{The aperture used to measure pol-mode satellite spot fluxes had a ``race track'' (i.e. rounded rectangle) shape with major and minor axis diameters of 18 and 8 pixels, respectively.} centered on each of the four spots in each of the two orthogonal polarization states (for a total of eight satellite spot fluxes per datacube), as described in \citet{hung2016}.

The entire set of 3D polarization datacubes then went through a double differencing cleaning procedure developed specifically for GPI ADI data \citep{perrin2015_4796} to account for biases between the two orthogonal polarization channels. The individual datacubes were smoothed with a Gaussian kernel with a full-width at half-maximum (FWHM) of 1 pixel to smooth out pixel-to-pixel noise without significantly affecting the final spatial resolution (GPI's PSF has a diffraction-limited FWHM of 3.8 pixels in the $H$ band). The mean stellar polarization (a term that includes both the intrinsic polarization of the host star and the instrumental polarization) was measured in an annulus near the FPM edge and then subtracted off from each pixel after scaling by that pixel's total intensity \citep{Millar-Blanchaer2016}. The default annulus was between 7 and 13 pixels from the star (i.e., from inside to outside the FPM) but its location varied by data set from 1--2 pixels to 13--15 pixels in manual reductions to minimize the stellar and instrumental polarization noise. This step assumes that there was no significant polarized intensity in that annular region from a given disk. Each frame was then rotated via interpolation to have north aligned with the $+Y$-axis of the image. The GPI DRP does not correct for the entire systematic offset of the instrument's north angle from true north, so all GPI-based PA measurements presented herein (including those from past publications) have been corrected by an additional offset angle. These additional offsets ranged from $0\fdg17 \pm 0\fdg14$ to $0\fdg45 \pm 0\fdg11$ depending on the observation date, using the revised offsets reported in Table 4 of \citet{derosa2020}, and the associated systematic uncertainties were added in quadrature with the measurement error to get the total uncertainty.

The ensemble of polarization datacubes was then converted to a single Stokes datacube containing vectors \{$\mathcal{I}$, $\mathcal{Q}$, $\mathcal{U}$, $\mathcal{V}$\} by inverting a ``measurement matrix'' that described how an incident Stokes vector was converted to the orthogonal polarization states measured in each polarization datacube \citep{perrin2015_4796}. Although we formally include Stokes $\mathcal{V}$ in our algebra, GPI is only sensitive to $\mathcal{V}$ where the half-wave plate deviates from perfect behavior. In practice, the $\mathcal{V}$ image was disregarded. Finally, the Stokes \Q and \U images were converted to their radial components, \Qr and \Ur, as described earlier. This was typically the end of the reduction. A quadrupole-like pattern occasionally remained in the \Qr and \Ur images, however, due to imperfect subtraction of the instrumental polarization. In this case, we fit a function of the form $B=B_0 \mathcal{I}_r \sin{2(\theta + \theta_0)}$ to the \Ur image by varying the scalar factor $B_0$ and offset angle $\theta_0$ to minimize the sum of the squared residuals. $\mathcal{I}_r$ is the azimuthally averaged total intensity as a function of radius. The best-fit function $B$ was subtracted from the \Ur image and also rotated $45\degr$ counterclockwise and subtracted from the \Qr image for a final result containing less instrumental polarization noise.

To photometrically calibrate the radial Stokes cubes, we followed the procedure described in \citet{hung2016} and briefly summarized here. For a given radial Stokes cube, we took the satellite spot flux measurements made previously for each constituent polarization datacube, summed those fluxes over the datacube's two polarization states, and then averaged the four resulting sums. We then averaged those fluxes over the polarization datacubes to get a mean satellite spot flux for the data set. Next, we converted the radial Stokes cube from intensity units of ADU coadd$^{-1}$ to stellar contrast units via the equation

\begin{equation} \label{eq:convert_contrast}
    \mathrm{img}(contrast) = \mathrm{img}({\textstyle\frac{ADU}{coadd}})\cdot \frac{\num{1.74e-4}}{F_{\mathrm{spot}}},
\end{equation}

where $F_{\mathrm{spot}}$ is the data set's mean satellite spot flux in ADU coadd$^{-1}$. The constant $\num{1.74e-4}$ is a revised measurement of GPI's peak satellite spot intensity to stellar flux ratio\footnote{The current ratio of $\num{1.74e-4}$ was revised downward from $\num{2.035e-4}$ and updated in the GPI DRP in 2019 February. For details, see the related GPI calibration information posted on the Gemini Observatory website at https://www.gemini.edu/sciops/instruments/gpi/instrument-performance/satellite-spots.}. Then we multiplied the radial Stokes image (in stellar contrast units at this point) by the stellar $H$-band flux in millijansky (converted from the magnitude given in Table \ref{tab:datasets}). Finally, we divided that product by the square of the GPI pixel scale ($14.166$ mas pixel$^{-1}$) to arrive at images in surface brightness units of mJy arcsec$^{-2}$. The list of resulting calibration factors for each disk's polarized intensity detection data set is given in \ref{appendix-b}. Some of these factors, and thus the calibrated disk surface brightnesses, differ from previously published values due to the revised satellite spot to star ratio (a ${\sim}15$\% difference) and the new data reductions used in this work (typically a ${\sim}1\sigma$ effect).

Total intensity images were also produced from the polarimetric datacubes; this is described in Section \ref{sect:spec_mode}.

\subsection{Polarimetric Noise Sources}

Noise in polarimetry images comes from a combination of random noise (i.e., photon and read noise) and systematic noise. At the smallest separations (${<}0\farcs3$), the images appear to be dominated by as-of-yet uncharacterized systematics. Outside of $0\farcs3$, most images are dominated by the photon noise of the residual speckles. At the largest separations, read noise dominates. The relative contribution of each noise source at a given angular separation differs and depends on the brightness of the target and the total number of frames in the data set.

Other systematic noise sources in polarimetric data include residual instrumental polarization, detector persistence, and sky/twilight polarization. As mentioned previously, residual instrumental polarization appears as a quadrupole pattern around the star in the \Qr and \Ur frames. It can be mitigated by carefully selecting the region where the instrumental polarization is measured in the pipeline and/or by subtracting an additional quadrupole term from the radial Stokes cube. In most cases, this is not a limiting systematic for disk detection, but it has the potential to bias the interpretation of disk morphology. Detector persistence is strongest in GPI polarimetry images when the previous observation was of a bright star in spectroscopic mode. It usually manifests in polarization datacubes as thick positive and negative bars when looking at the difference of the two orthogonal polarization states (see the thick vertical bars of the positive and negative background noise in the HD 191089 image in Figure~\ref{fig:det_gallery_Qr}). The exact appearance of the noise in final Stokes cubes depends on the strength and decay rate of the persistence, as well as the amount of field rotation in the full sequence \citep{Millar-Blanchaer2016}. Finally, in some circumstances, a disk was observed during dawn or dusk twilight when the sky polarization adds a signal that is constant spatially across a given frame but varies between frames (i.e., with time). This signal can be compensated for by measuring the mean normalized difference in a polarization datacube and subtracting that value from the entire cube. Because it is relatively easy to measure and subtract, this type of systematic is not a limiting factor for any of the disks presented here.

\subsection{Total Intensity from Spectral and Polarimetric Modes}
\label{sect:spec_mode}

We retrieve total intensity information from both spectral mode and polarimetric mode data. The data reduction steps for GPI spectral mode data are detailed in \citet{Wang2018}. Briefly, the GPI DRP takes raw 2D frames containing ${\sim}$35,000 microspectra and converts them into 3D spectral datacubes after dark subtraction and bad pixel correction. The microspectra positions are calibrated using an argon arc lamp image taken before the observing sequence \citep{Wolff2014}. Distortion and any remaining bad pixels are corrected in the datacubes \citep{konopacky2014}. The satellite spots are then located and measured in each frame of each spectral datacube for astrometric and spectrophotometric calibration \citep{wang2014}. 

To recover disks in the data, the open-source \texttt{pyKLIP} package is used to subtract the stellar PSF from datacubes in each observing sequence \citep{wang2015_pyklip}. \texttt{pyKLIP} uses Karhunen--Lo\`{e}ve image projection (KLIP) to model and subtract the stellar PSF \citep{soummer2012, pueyo2015}. We employ only ADI for PSF subtraction; debris disks generally show too little brightness variation across narrow IR bandpasses like the $H$ band to make spectral diversity an effective PSF-subtraction tool. The automatic reduction performed by the GPIES Data Cruncher uses frames at the same wavelength where the disk has moved by at least 1 pixel due to ADI to build the Karhunen--Lo\'eve (KL) modes and saves images where 1, 3, 10, 20, and 50 KL modes were used to model the stellar PSF. Having a wide variety allows us to gauge the effect that the number of KL modes has on the disk shape and surface brightness. For this work, we chose to present the final image for each disk that provided the best qualitative balance between preserving the (presumed) intrinsic disk signal and minimizing stellar PSF residuals.

An analogous process is used for extracting total intensity signals from the polarimetric datacubes. In this case, the total intensity of each cube is the sum of the two orthogonal polarization states, which is computed for each datacube in the data set and then fed into the \texttt{pyKLIP} algorithm in the same way as the spec-mode data.

Other methods exist to construct the model for stellar PSF subtraction but we do not employ them here. Using a reference library of diskless images (Reference Differential Imaging) avoids the self-subtraction biases discussed below, and our large survey has created a sizable library. However, we have found the substantial PSF diversity in our AO-corrected data to limit the effectiveness of speckle suppression and produce lower S/N results than KLIP. Masking the disk and interpolating the PSF from unmasked regions is another option that has produced promising results for a few GPIES disks \citep{kalas2015, perrin2015_4796, draper2016} but requires manual tuning of the mask shape for each disk and is ineffective for azimuthally wide disks that demand interpolation over a large area. Non-negative matrix factorization \citep{ren2018_nmf} is a relatively new technique that has been shown to preserve extended emission better than KLIP, but it is computationally expensive so we leave its application to our full sample for a future study.

We photometrically calibrated the total intensity data to present it in surface brightness units of mJy arcsec$^{-2}$. To do so, the final images were first converted from ADU coadd$^{-1}$ to stellar contrast units using Equation (\ref{eq:convert_contrast}) as described in Section \ref{sect:pol_reduction}. For each image derived from pol-mode data, the image was then multiplied by the target star's $H$-band flux ($F_\star$) in mJy (converted from magnitudes listed in Table \ref{tab:obs_targs}). For each image derived from spec-mode data, however, the image was instead multiplied by $F_\star$/13.5. The factor of 13.5 is the approximate ratio (with an estimated $1\sigma$ accuracy of $\pm10$\%) of the GPI $H$-band PSF's aperture-integrated flux to its peak flux, assuming a circular aperture with $\mathrm{radius} = 4$ pixels that is equivalent to the size of the aperture used for pol-mode satellite spot measurements. This scaling of the stellar flux was necessary because satellite spot fluxes from spec-mode data ($F_{\mathrm{spot}}$ in Eq. 1) were measured by the DRP as the peak flux of the satellite spot \citep{wang2014} rather than its aperture-integrated flux (as was done with pol-mode data). Finally, all images were divided by the square of the GPI pixel scale to arrive at units of mJy arcsec$^{-2}$. We stress, however, that this calibration does not correct for the biases introduced by the PSF-subtraction process, so our total intensity surface brightnesses as presented here are likely underestimates of the true values by varied degree.

\subsection{Spectral Noise Sources}
Noise in our final PSF-subtracted images comes in three main types: read noise, residual speckle noise, and data reduction systematics. In spectral mode, the light from each lenslet is dispersed into a spectrum across several pixels rather than concentrated into two spots (as in polarization mode). Thus, for faint disks at large separations from the star, read noise is a significant limitation.

Primarily, though, diffraction from the star limits our ability to detect disks. The most problematic manifestation of diffraction is in the form of speckles. These constitute either a smooth time-averaged halo due to uncorrected atmospheric turbulence, or individual point-source-like brightness fluctuations resulting from aberrations along the light path that are not sensed by the AO system and remain quasistatic over time. The chief purpose of PSF-subtraction routines like KLIP is to suppress this speckle noise, and they are effective; however, residual speckle noise still sets the noise floor close to the star. The speckle contribution decreases with separation until it reaches the read noise floor at some point. Additionally, when the wind is strong, it induces a broad spatial asymmetry with a ``butterfly'' shape in the smooth halo that is aligned with the wind direction \citep{Cantalloube2018, Madurowicz2018}. This wind butterfly is especially problematic because it rotates through the image sequence due to field rotation, so ADI-based PSF-subtraction techniques are largely ineffective at removing it. As a result, these broad diffraction features can obscure low-surface-brightness extended emission from disks.

In the process of subtracting the stellar PSF, systematic errors are introduced by the PSF-subtraction algorithm. The PSF model constructed by KLIP is built from a subset of the science frames themselves; hence, the PSF model contains disk signal in addition to the stellar signal. When this model is subtracted from the data, some of the disk signal is removed as well. The resulting self-subtraction and over-subtraction effects vary with position in the image and the parameterization of the KLIP subtraction \citep{marois2006, milli2012}. Generally, these effects both attenuate low-frequency astrophysical signals (similar to a high-pass filter) and distort higher frequency signals like those that define the disk morphology. Consequently, our final total intensity images present biased disk surface brightnesses and morphologies. One can partially correct for these biases by forward modeling the disk signal through the PSF-subtraction process \citep{esposito2014, pueyo2016}. However, this forward modeling relies on accurately modeling the underlying disk brightness distribution and is typically a linear approximation of KLIP's higher order effects. When either of these assumptions breaks down, our estimated disk parameters will be biased. We have applied corrections to some individual disks in previous publications \citep{esposito2016, esposito2018}, but this is a time-intensive task and we leave its application to our full sample for the future. In this work, we present the total intensity images and subsequent measurements without any corrections from forward modeling.

\subsection{Image Contrasts}
\label{sect:contrasts}

% Polarization datacube contrast histogram
\begin{figure}
\centering
\includegraphics[width=\columnwidth]{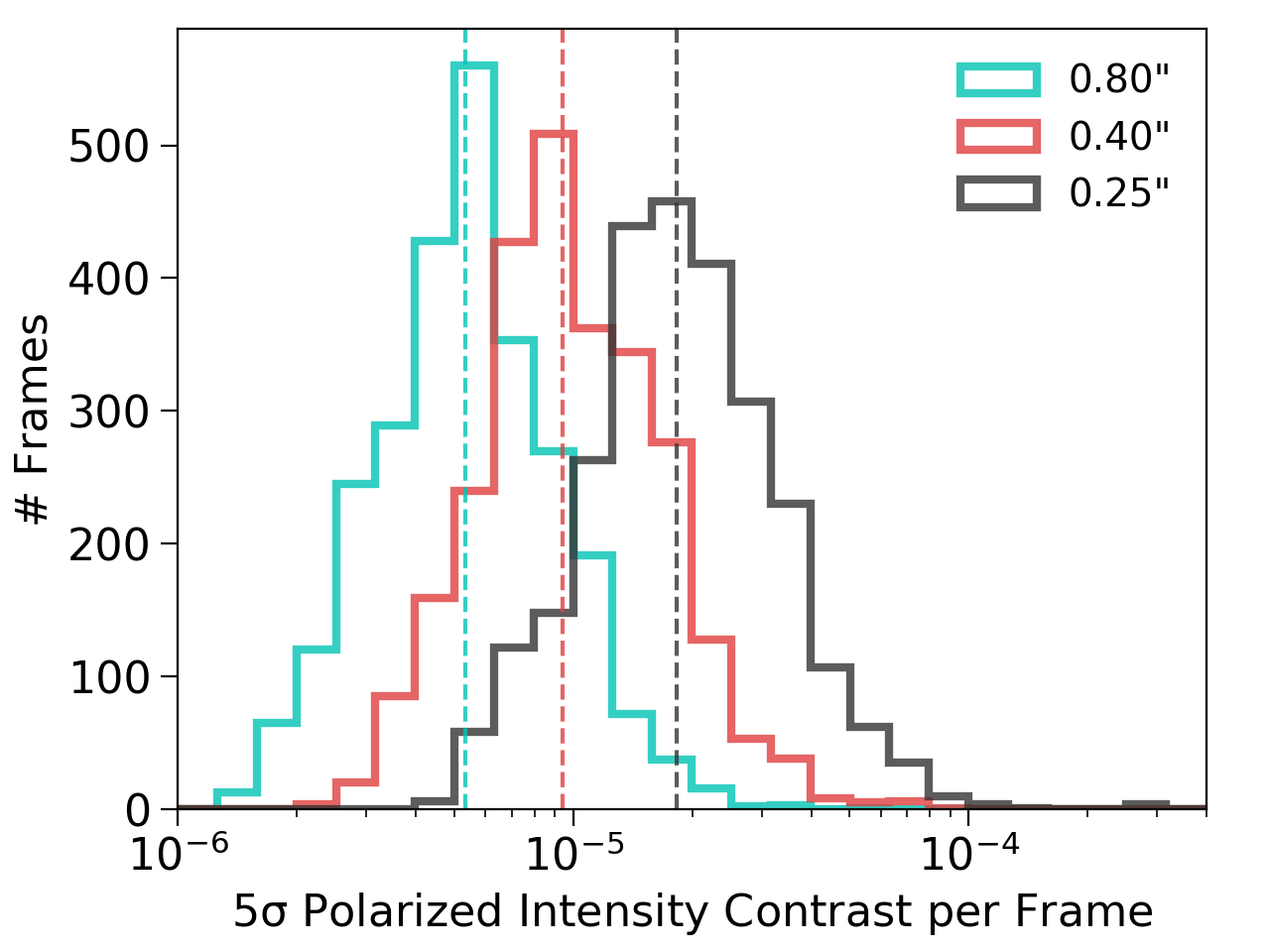}
\caption{$5\sigma$ polarized intensity contrasts for all individual polarization datacubes in the $H$ band. Separate histograms are drawn for contrasts measured at projected separations of $0\farcs25$, $0\farcs40$, and $0\farcs80$, with dashed lines marking the median contrast at each separation.}
\label{fig:contrast_podc}
\end{figure}

% Radial Stokes contrast whisker plot
\begin{figure*}[ht]
\centering
\includegraphics[width=5in]{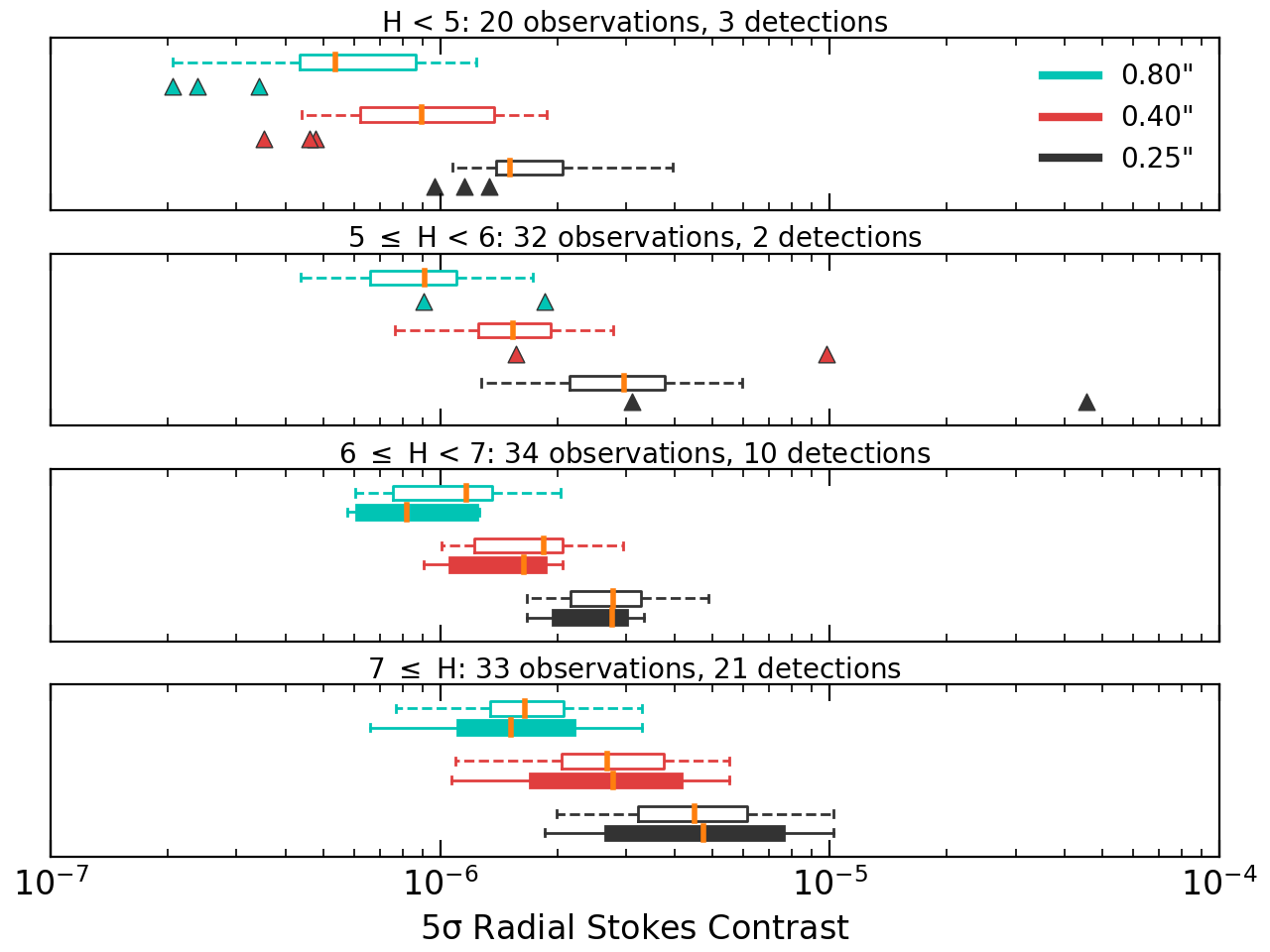}
\caption{Polarized intensity radial Stokes contrasts for each pol-mode dataset listed in Table \ref{tab:datasets}. Panels divide the stars by $H$-band magnitude. Each panel shows the $5\sigma$ radial Stokes contrast at $0\farcs25$, $0\farcs40$, and $0\farcs80$ for all targets in that stellar magnitude bin (unfilled box and dashed line) and separately for only detected disks (filled box and solid line). The box edges mark the 25th and 75th percentiles, with an orange line at the median, while the lines extending outside the boxes stop at the 5th and 95th percentiles. For the top two panels, we plot contrasts for the detected disks as individual triangles, due to their small number.}
\label{fig:contrast_rstokes}
\end{figure*}

One objective quantitative assessment of the sensitivity of each observation is the contrast achieved relative to the target star's flux. We measured this routinely for polarized intensity in individual polarization datacubes (from the difference of the two polarization states in single exposures) and for the final radial Stokes intensity (from the \Ur channel of combined Stokes cubes). In both cases, we followed the procedures described in \citet{Millar-Blanchaer2016}. Briefly, for a given image, we first converted it to contrast units via Equation (\ref{eq:convert_contrast}), with $F_{\mathrm{spot}}$ being either the average aperture-integrated satellite spot flux for a single polarization datacube (for the individual frame contrasts in Figure \ref{fig:contrast_podc}) or those same satellite spot fluxes averaged over all constituent datacubes of a radial Stokes cube (for the radial Stokes contrasts in Figure \ref{fig:contrast_rstokes}). We make our contrast values analogous to total intensity point-source contrasts by dividing the contrast image by the ratio of the GPI $H$-band PSF's aperture-integrated flux to its peak flux, which is approximately 13.5 (see Section \ref{sect:spec_mode}). We note that our value of 13.5 for this ratio is ${\sim}25$\% greater than that used in previous publications that approximated the PSF as a 2D Gaussian function, including \citet{Millar-Blanchaer2016} and \citet{millar-blanchaer2017}, so our contrasts are correspondingly ${\sim}25$\% deeper. Finally, we measure the standard deviations of pixel values in concentric 1-pixel-wide annuli around the star at each projected separation, and multiply those standard deviations by 5 to arrive at the $5\sigma$ contrast.

The distributions of polarized intensity contrasts among polarization datacubes at multiple projected separations are shown in Figure \ref{fig:contrast_podc}. As expected, contrast improves at wider separations, with variation according to stellar magnitude, weather conditions, atmospheric seeing, and AO performance.

The radial Stokes contrasts of every data set are shown in Figure \ref{fig:contrast_rstokes} with one contrast value per data set per separation and broken down by stellar $H$-band magnitude. Disk detections are also highlighted separately. Here we see that the contrast generally deepens as the stellar brightness increases. For the faintest stars in our sample ($m_H\geq7$ mag), the contrasts of disk detection images are similar to the contrasts of non-detections. For slightly brighter stars with $6 \leq m_H < 7$ mag, contrasts of detections tend to be among the deepest ${\sim}50$\% of the subset. We disregard the contrasts for the detection images in the $5 \leq m_H < 6$ mag subset because they are contaminated and of limited usefulness, as discussed at the end of this section. For the brightest stars ($m_H < 5$ mag), however, all detection images contained contrasts in the best quartile for this stellar magnitude range. Examining the detection contrasts across the three useful subsets, we find a trend in which detection contrasts deepen relative to non-detection contrasts as host star brightness increases. Thus, more disks may be detectable around our bright target stars but would require data with deeper contrasts to detect.

As a final caveat, we note that the \Qr contrast will be artificially worsened in data sets where there is significant \Ur brightness, given our method of measuring the contrast from the \Ur channel. As we discuss further in \ref{appendix-a}, this \Ur signal is likely from systematic ``leakage'' of \Qr signal into that channel when recovering the Stokes parameters. Thus, some of our quoted contrasts may be unduly pessimistic, especially for the brightest disks (where the \Ur signals appear brightest). This is the case for the two previously noted detections in the $5 \leq m_H < 6$ bin, HD~100546 and HR~4796. We include their contrasts here for the sake of completeness but consider them to be rough upper limits on the true contrasts achieved.

\subsection{Disk Morphology Modeling} \label{sect:morph_mcmc}

We used models to estimate morphological parameters for five disks that GPIES resolved in scattered light for the first time (HD~111161, HD~117214, HD~143675, HD~145560, and HD~156623; discussed in Section \ref{sect:detection_overview}). For each individual disk, we followed the framework described in \citet{esposito2018} and ran a Markov Chain Monte Carlo (MCMC) sampler comparing the Stokes \Qr, \Q, and \U images with analogous models created by the radiative transfer and ray-tracing code MCFOST. We do not include total intensity, which is only available for two of the disks and requires additional forward modeling of the PSF-subtraction process.

The dust volume density of our models follows the form 
\begin{equation}
\rho(r,z) \propto \frac{\exp\Big({-}\Big(\frac{|z|}{H(r)}\Big)^{\gamma} \Big)}{\big[\left(r/r_c\right)^{-2\alpha_{in}}+\left(r/r_c\right)^{-2\alpha_{out}}\big]^{1/2}}\ , \\
\label{eq:disk_density} \\
\end{equation}

\noindent where $r$ is the radial coordinate in the equatorial plane, $z$ is the height above the disk midplane, $\gamma$ is fixed at 1, and $r_c$ is a critical radius that divides the ring into inner and outer regions with separate density power-law indices of $\alpha_{\mathrm{in}}$ and $\alpha_{\mathrm{out}}$, respectively \citep{augereau1999}. The disk scale height varies radially as $H(r) = H_0 (r/r_H)^\beta$ with $r_H$ being a reference radius at which the scale height equals $H_0$. In our models, we set $\beta=1$ so the scale height is a constant fraction of the radius throughout the disk. We set that fraction at $H_0/r_H = 0.055$ for most disks but use $H_0/r_H = 0.027$ for the particularly sharply defined HD~117214 ring. We chose these fractional heights to be consistent with values of 0.03--0.10 that have been estimated for other disks including HR 4796, Fomalhaut, AU Mic, $\beta$ Pic, and HD 35841 \citep{augereau1999, kalas2005_fomb, krist2005, millar-blanchaer2015, esposito2018}.

As \citet{augereau1999} show, the maximum of the dust density does not occur at $r_c$ but at a peak radius that we label $R_0$:
\begin{equation}
R_0 = \bigg(\frac{\Gamma_{\mathrm{in}}}{{-}\Gamma_{\mathrm{out}}}\bigg)^{(2\Gamma_{\mathrm{in}} - 2\Gamma_{\mathrm{out}})^{-1}} r_c\ , \\
\label{eq:R0} \\
\end{equation}

where $\Gamma_{\mathrm{in}}=\alpha_{\mathrm{in}}+\beta$ and $\Gamma_{\mathrm{out}}=\alpha_{\mathrm{out}}+\beta$. It is $R_0$ that we use for our radius analyses in Section \ref{sect:sample_results} and beyond when a disk's dust distribution has been modeled with an expression akin to Equation (\ref{eq:disk_density}).

Our focus was the basic disk morphology, so we assumed the disks to be circular, azimuthally symmetric rings centered on the star and then varied six model parameters in each MCMC. Inclination ($i$) and PA set the disk's orientation in the sky plane, with PA defined as the angle measured eastward (i.e. counterclockwise in our images) from north to the projected major axis such that $90\degr + \mathrm{PA}$ is the PA of the projected semiminor axis on the disk's presumed front side (with ``front'' chosen here to be the brightest side in \Qr). The critical radius $r_c$ is the transition radius of the smooth two-component (``broken'') power law in dust volume density (Equation \ref{eq:disk_density}). Parameters $r_{\mathrm{in}}$ and $r_{\mathrm{out}}$ set the inner and outer radii, respectively, where that dust volume density profile rapidly tapers toward zero (following a Gaussian function with $\sigma=2$ au). Finally, the total dust mass ($M_d$) is varied to scale the dust volume density up or down and thus change the global surface brightness of the optically thin disks. We do not expect this dust mass to accurately reflect the true mass, as it is heavily influenced by grain properties that are empirically determined rather than physically motivated, which we describe next.

A proper morphological fit requires reasonably accurate polarized scattering phase functions due to the degeneracies between parameters. To compute these phase functions, we selected dust grain properties that roughly reproduced the empirical phase functions by eye. This was good enough to constrain the basic morphological disk parameters listed earlier, but we stress that the values we selected for these grain properties are not strictly physically motivated, and we do not expect them to reflect the true values for these disks. Nonetheless, we report them here and in Table \ref{tab:mcmc_params} so others may reproduce our models. All models used Mie scattering theory, pure astrosilicate grains \citep{draine1984_si}, a maximum grain size of 1.0 mm, and a grain size distribution power-law index of -3.5. Although Mie scattering typically fails to reproduce real phase functions with high fidelity (e.g., \citealt{milli2017b, esposito2018, ren2019}), it is computationally tractable and we found its accuracy sufficient for our purposes. No gas was simulated in the disks.

The four grain properties that we tuned on a disk by disk basis were the minimum grain size ($a_\mathrm{min}$), porosity, and radial inner and outer dust density power-law indices ($\alpha_\mathrm{in}$, $\alpha_\mathrm{out}$). Porosity is handled as inclusions of void in the grain volume through effective medium theory (Bruggeman rule; \citealt{bruggeman1935}). For HD~111161 and HD~143675, we manually tuned all four grain properties to get a reasonable by-eye match to the data and then fixed them during the MCMC. We could not adequately reproduce the phase functions for HD 117214 and HD 156623 by manually tuning $a_\mathrm{min}$ and porosity, so we turned them into free MCMC parameters to improve the phase function agreement. The resulting dust parameter values are listed in Table \ref{tab:mcmc_params}. We found that the optimal $a_\mathrm{min}$ and porosity values for HD~117214 also matched the HD~145560 disk well, so we adopted them for the latter as well. While the values that we adopt for $a_\mathrm{min}$ and porosity adequately reproduce our observed phase functions, we caution that they may not be unique solutions, as the two properties are often degenerate when considering only a single wavelength and polarization state (\citealt{hughes2018_review} and references therein).

We ran parallel-tempered MCMC samplings with the Python package \texttt{emcee} (v2.2.1) that, depending on the disk, employed 2--4 walker temperatures, 100--120 walkers per temperature, and 750--1300 iterations per walker from which the final posterior distributions were drawn after discarding 1000--1700 ``burn-in'' iterations \citep{foreman-mackey2013}. We found this relatively short sampling sufficient to converge on well-constrained inclinations and PAs. The radius parameters often remained poorly constrained, which we attributed to limitations in the model's parameterization and the data S/N. The results for each disk are quoted in Sections \ref{sect:new_disks} and \ref{sect:morph_other_three} as the 16th, 50th, and 84th percentiles of each parameter's marginalized posterior distribution.

% MCMC model parameter table.
\begin{deluxetable}{lcccc}
%\tablewidth{0pt}
\tablecaption{MCMC Model Dust Parameters}
\tablehead{
% Header Line 1
\colhead{Target} & \colhead{$a_\mathrm{min}\ (\micron)$} & \colhead{Porosity} & \colhead{$\alpha_\mathrm{in}$} & \colhead{$\alpha_\mathrm{out}$}
%\vspace{-0.3cm} 
% Header Line 2
% & \colhead{($\micron$)} &  &  &
}
\startdata
HD 111161 & 2.00 & 0.90 & 2.5 & -3.0 \\
HD 117214 & 1.60* & 0.86* & 4.5 & -4.5 \\
HD 143675 & 2.00 & 0.01 & 2.5 & -3.0 \\
HD 145560 & 1.60 & 0.86 & 3.5 & -3.0 \\
HD 156623 & 1.19* & 0.84* & 1.5 & -3.5 \\
\enddata
%\vspace{-0.8cm} 
\tablecomments{Values with * are the median of the posterior distribution found from the MCMC. All other values were fixed.\label{tab:mcmc_params}}
\end{deluxetable}

%\vskip4\baselineskip
\section{Survey Results}

\subsection{Debris Disk Detection Overview}
\label{sect:detection_overview}

\begin{figure*}[ht]
\centering
\includegraphics[width=\textwidth]{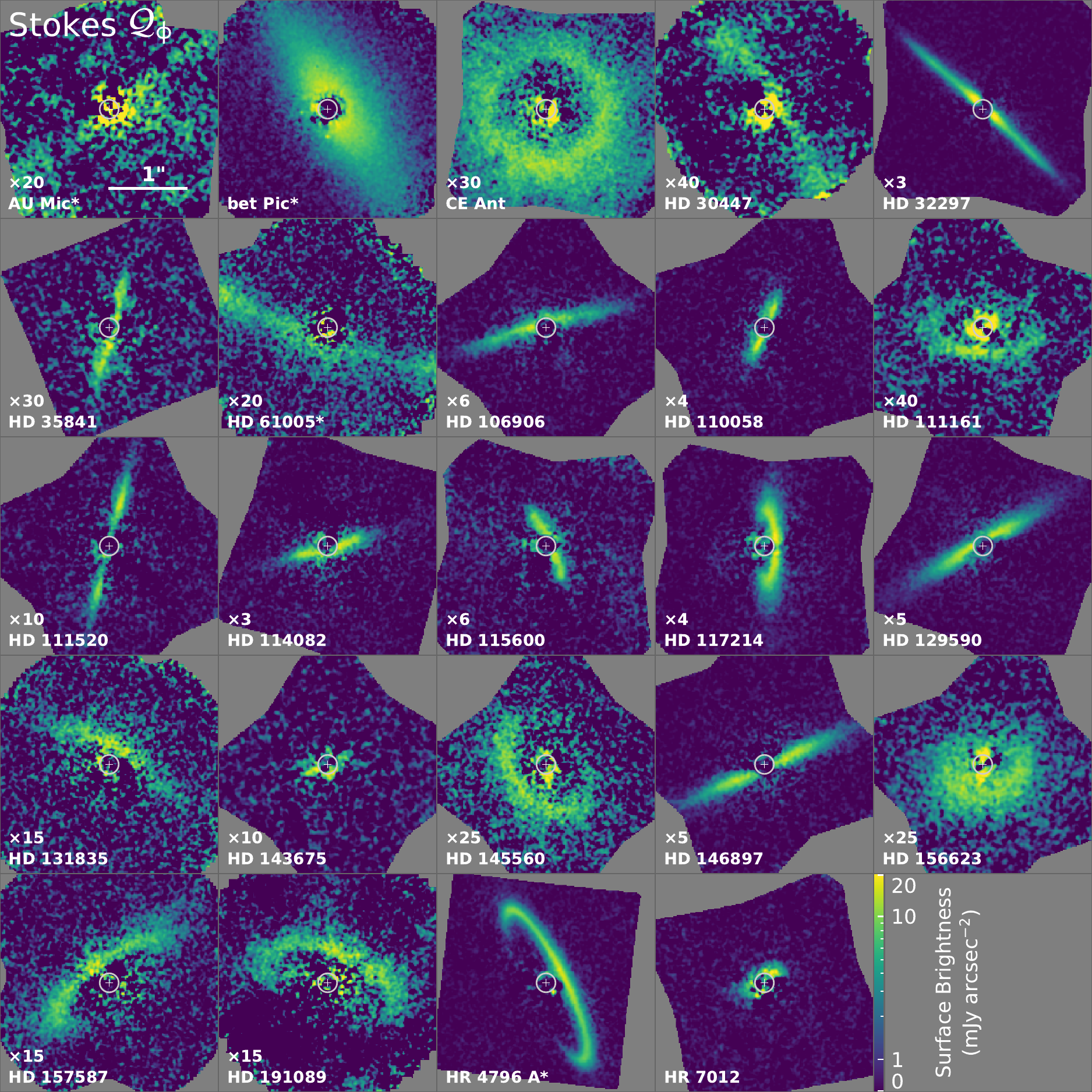}
\caption{GPIES debris disk polarized intensity detections as seen in $H$-band Stokes \Qr. North is up, east is left, and all panels are on the same angular size scale. Disks observed during GPI commissioning have asterisks after their names. All panels display the disk surface brightness using the same color map that is logarithmic between 1 and 20 mJy arcsec$^{-2}$ but linear between 0 and 1 mJy arcsec$^{-2}$; however, all but the three brightest disks have been scaled linearly before plotting by a factor noted above the target name. The white circles mark the GPI $H$-band FPM edge and the crosses mark the star location.}
\label{fig:det_gallery_Qr}
\end{figure*}

\begin{figure*}[h!]
\centering
\includegraphics[width=\textwidth]{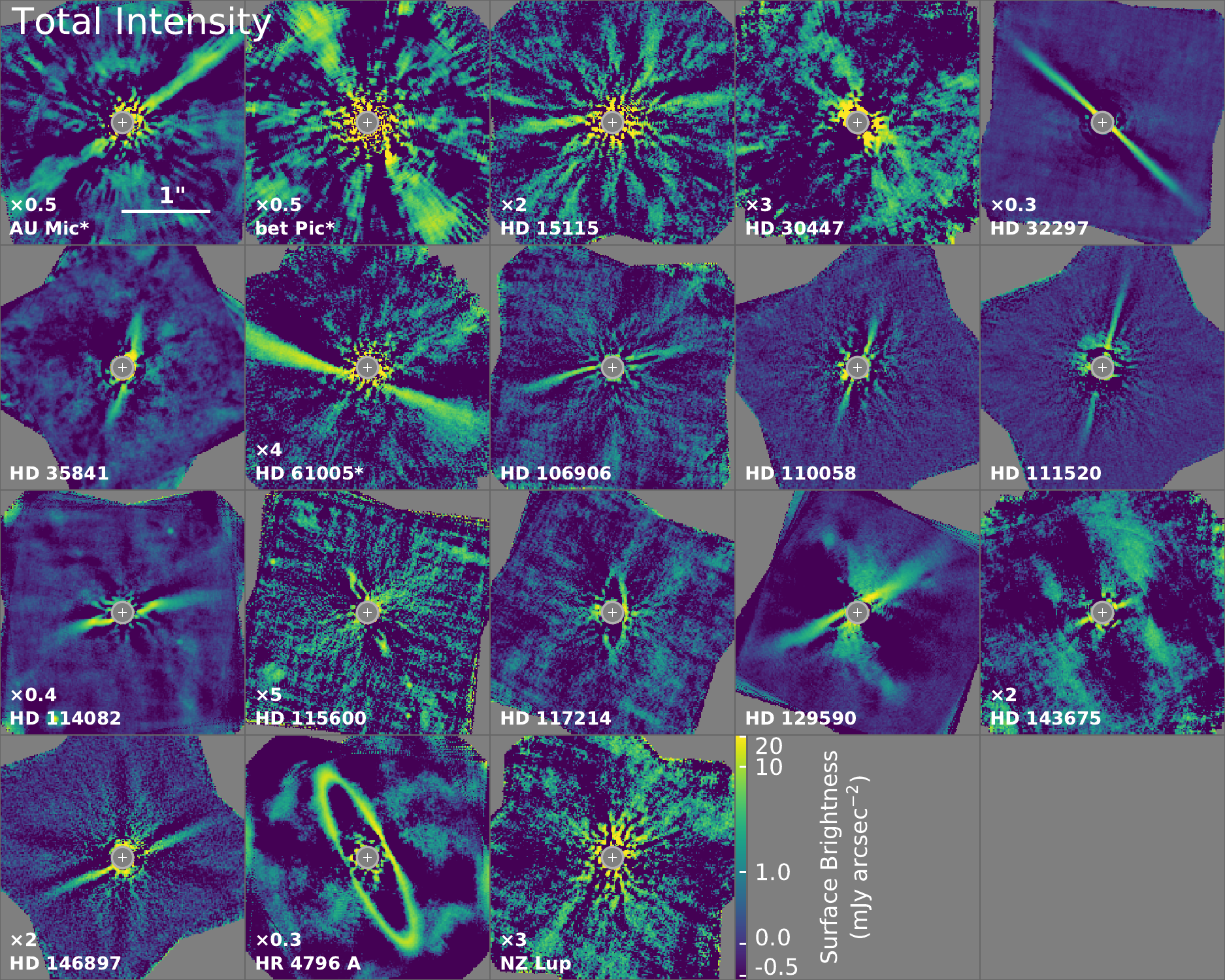}
\caption{GPIES debris disk detections in total intensity after stellar PSF subtraction. North is up, east is left, and all panels are on the same angular size scale. Disks observed during GPI commissioning have asterisks after their names. All panels display the disk surface brightness using the same color map that is logarithmic between 1 and 20 mJy arcsec$^{-2}$ but linear between -0.5 and 1.0 mJy arcsec$^{-2}$; however, some disk brightnesses have been scaled linearly before plotting by a factor noted above the target name (which may differ from previous scale factors applied in Fig \ref{fig:det_gallery_Qr}. The white circle marks the GPI $H$-band FPM edge. Except for $\beta$~Pic~b (a saturated point source overlapping the disk at $r=0\farcs4$), all point sources are confirmed background stars or suspected to be so (pending proper motion confirmation).}
\label{fig:det_gallery_totI}
\end{figure*}

We have resolved circumstellar disks in scattered light around 29 of the \Nobs target stars that we observed in the disk program. Twenty-six of these detections fall into the ``debris disk'' class. Table \ref{tab:datasets} indicates if the detections were made in polarized intensity (P) and/or total intensity (I), where the latter detection could derive from spec-mode and/or pol-mode data. Images of these disks in $H$-band Stokes \Qr polarized intensity and total intensity (Stokes \I) are presented in Figures \ref{fig:det_gallery_Qr} and \ref{fig:det_gallery_totI}, respectively. Of these \Ndebris disks, 2 were detected only in total intensity (\object{HD 15115} and \object{NZ Lup}), and \Npolonly were detected only in polarized intensity (\object{CE Ant}, \object{HD 111161}, \object{HD 131835}, \object{HD 145560}, \object{HD 156623}, \object{HD 157587}, \object{HD 191089}, and \object{HR 7012}). The fewer total intensity detections are largely due to disk inclination, which we discuss in Section \ref{sect:inc_tau_effects}. We include the Stokes \Ur images in \ref{appendix-a} and discuss therein sources of the signals we see in that channel for our brightest disks.

As introduced in Section \ref{sect:non-debris_targets}, separate from our debris disk sample are three detected disks that are more accurately classified as protoplanetary or transitional disks due to their high IR-excess magnitudes and gas contents: AK~Sco, HD~100546, and HD~141569. We discuss their results separately in Section \ref{sect:proto}. Given that they represent an earlier evolutionary phase and are outliers with respect to many of our debris disk sample statistics, we do not include them in the following analyses unless stated otherwise.

The strongest commonality between our detected targets is their high \Lir. This is clear from the histogram in Figure \ref{fig:targ_histograms} and further documented in Table \ref{tab:tau_det_rate}. We had a 100\% \Qr detection rate among 15 debris disk targets observed with \Lir$\geq \num{2.0e-3}$. That rate decreased with \Lir until it reached 0\% for the 56 targets with \Lir $<\num{3.0e-4}$. This latter group constituted just over half of our observed sample. We did have one total intensity detection in that range, NZ Lup at \Lir $=\num{1.3e-4}$, albeit with low S/N. Overall, our survey apparently reached a sensitivity floor at \Lir $=10^{-4}$, below which we had no detections of any kind out of 27 targets.

The trend with \Lir is the same for disks that had been detected in scattered light with other instruments before being selected as GPIES targets and for disks that had no resolved images in scattered light. Unsurprisingly, our \Qr detection rate was higher overall for previously resolved disks than for unresolved disks\footnote{\citet{engler2018} published a visible-light detection of HR 7012 after we had selected the target for GPIES but before we actually observed it, so we consider it ``unresolved'' in this context.}: 53\% (16/30) vs. 11\% (8/71). Some GPIES non-detections of previously resolved disks were simply due to the disk being too large for the GPI field of view (e.g., Fomalhaut, HD 107146, and HD 181327). Other reasons for non-detections are discussed in Section \ref{sect:non-detections}. Overall, combining polarized and total intensities, our debris disk detection rate was 26\% (26/101). This is not necessarily the true occurrence rate for scattered-light debris disks, as our observed sample is biased (e.g., toward disks previously resolved in scattered light), and we have not corrected for completeness. That said, our detection rate is similar to general debris disk occurrence rates measured at ${\sim}17\%$--36\% across spectral types K to A \citep{thureau2014, montesinos2016, hughes2018_review, sibthorpe2018}.

\begin{deluxetable}{lcc}
%\tablecolumns{6}
%\tablewidth{0pt}
\tablecaption{\label{tab:tau_det_rate}Debris disk detection rate by \Lir and scattered-light imaging history}
\tablehead{
 & \colhead{Prev. Resolved} & \colhead{Prev. Unresolved} \vspace{-0.3cm} \\
\colhead{\Lir} & \colhead{\% Det. ($N_\mathrm{obs}$)} & \colhead{\% Det. ($N_\mathrm{obs}$)}
% \colhead{\Lir} & \colhead{Det. Rate} & \colhead{$N_\mathrm{obs}$}
\vspace{-0.1cm}
}
\startdata
$\geq \num{5e-3}$ & 100\% (3) & 100\% (1) \\
\num{1e-3} -- \num{5e-3} & 75\% (12) & 71\% (7) \\
\num{5e-4} -- \num{1e-3} & 100\% (2) & 29\% (7) \\ 
\num{1e-4} -- \num{5e-4} & 40\% (10) & 0\% (31) \\
$<\num{1e-4}$ & 0\% (2) & 0\% (25) \\
\enddata
%\vspace{-0.8cm} 
\tablecomments{Percent of observed GPIES targets with a detection, broken down by \Lir (rows). Targets are further separated into those that had been resolved in scattered light before their selection as a GPIES target and those that had not (columns). The total number of targets observed for a given group is in parentheses. TWA 25 is excluded because we have no \Lir measurement for it. Also excluded are the protoplanetary/transitional disks, all three of which are detections in the highest \Lir bin and previously resolved in scattered light.}
\end{deluxetable}

The S/N of our detections varies by disk according to factors such as intrinsic brightness of the disk, integration time, and observing conditions. High-S/N data allow for comprehensive analyses of each disk, many of which are already published. These disks are \object{AU Mic} \citep{wang2015_aumic}, \object{$\beta$ Pic} \citep{millar-blanchaer2015}, \object{HD 32297} \citep{duchene2020}, \object{HD 35841} \citep{esposito2018}, \object{HD 61005} \citep{esposito2016}, HD~100546 \citep{follette2017, rameau2017}, \object{HD 106906} \citep{kalas2015}, \object{HD 111520} \citep{draper2016}, HD~131835 \citep{hung2015b}, HD~141569 \citep{bruzzone2019}, HD~157587 \citep{millar-blanchaer2016c}, HD~191089 \citep{ren2019}, and \object{HR 4796} \citep{perrin2015_4796}. The scattered-light discoveries of HD~111161, \object{HD 143675}, and HD 145560 are presented together in \citet{hom2020}. Individual analyses of the remaining high-S/N disks \object{HD 110058}, \object{HD 129590}, and \object{HD 146897} are in progress now.

On the other hand, low S/N limits the analysis possible from some data. In polarized intensity, this pertains to AU~Mic, \object{HD 30447}, and HD~143675. The particularly low-S/N cases in total intensity, where PSF subtraction with ADI techniques often suppresses the disk signal, are $\beta$~Pic, HD~15115, HD~30447, HD~141569, and NZ~Lup. We present these data here for completeness but extracting significant results will require deeper observations or alternative PSF subtraction.

Seven GPIES detections represented the first scattered-light images of those disks at the time we detected them: HD~106906, HD~111161, \object{HD 117214}, HD~131835, HD~143675, HD~145560, and HD~156623. All are Sco--Cen members. Those were the first resolved images at any wavelength for HD~106906, HD~111161, and HD~143675; the others were previously resolved with ALMA (HD~117214, HD~145560, and HD~156623; \citealt{lieman-sifry2016}) or in the mid-IR with Michelle (HD~131835; \citealt{hung2015a}). In this work, we examine the previously unpublished subset of HD~117214 and HD~156623. For details of the HD~111161, HD~143675, and HD~145560 scattered-light discoveries, see the recent work \citet{hom2020}. For details of HD~106906 and HD~131835, see \citet{kalas2015} and \citet{hung2015b}, respectively.

% Detection gallery Qr

% Detection gallery Total Intensity

\clearpage

\subsection{Two New Scattered-light Debris Disks} \label{sect:new_disks}

\subsubsection{HD 117214}

\begin{figure}[h!]
\centering
\includegraphics[width=\columnwidth]{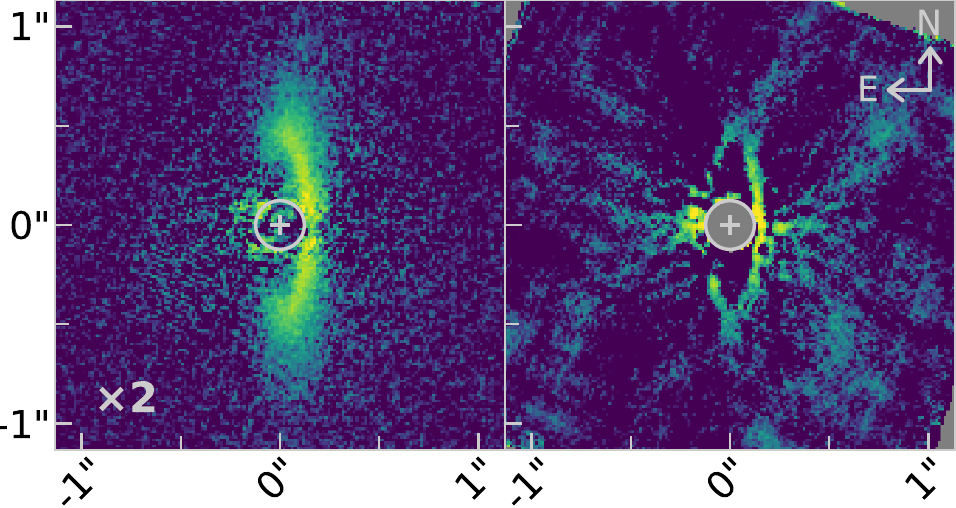}
\caption{HD~117214 debris disk seen in $H$-band Stokes \Qr (left) and spectroscopic-mode total intensity (right). The \Qr image's flux was scaled up by a factor of 2 before being plotted on the same logarithmic brightness scale as the total intensity image in ADU s$^{-1}$.}
\label{fig:hd117214}
\end{figure}

The HD~117214 (SpT = F6V) debris disk is resolved in both \Qr\ and total intensity as a relatively bright, narrow ring (Figure \ref{fig:hd117214}). Given the strong brightness asymmetry between the west and east edges in \Qr, we assume the west to be the front edge. The asymmetry is weaker in total intensity, possibly having been suppressed by PSF-subtraction biases, but the west edge still appears brightest. The ring does not show a noticeable stellocentric offset along the major axis. Faint nebulosity appears exterior to the ansae, preferentially toward the front edge of the ring, suggesting dust outside the main ring similar to HD~32297 and HR~4796~A \citep{schneider2014}. We note that while this paper was undergoing peer review, \citet{engler2020} published a simultaneous scattered-light detection of HD~117214 with VLT/SPHERE. Comparison with our results shows their imaged disk features to be qualitatively similar and their model-derived morphological parameters to be consistent (within reasonable expectations for different wavelengths, models, and approaches).

Applying the MCMC modeling methods described in Section \ref{sect:morph_mcmc}, we find $i=71\fdg0 \substack{+1.1 \\ -0.4}$ and major axis $\mathrm{PA}=179\fdg8 \pm 0\fdg2$, adopting the $\pm34\%$ confidence intervals for the uncertainties. The inner radius is $20.8 \substack{+12.4 \\ -0.6}$ au but could have a substantially larger value of 42.3 au within the 99.7\% confidence level. The outer radius is established as a 99.7\% confidence lower limit of $r_\mathrm{out} > 149.8$ au. This limit is the effective edge of the GPI field of view at $1\farcs4$ projected separation, so we likely do not see the full extent of the disk. Our models may also systematically underestimate $r_\mathrm{in}$ and overestimate $r_\mathrm{out}$ because we made the surface density power-law indices relatively steep at 4.5 (inner) and -4.5 (outer) to match the ring's sharp edges. This has the additional effect of creating low surface brightness far from the critical radius that is on the order of the data background noise and thus contributes weakly to the model's likelihood, thereby placing little constraint on the boundaries of that surface brightness. The remaining MCMC parameter values were $r_c=58.7 \substack{+0.6 \\ -1.1}$ au and log($M_d/M_\odot$) $=-5.64 \substack{+0.09 \\ -0.02}$ (although we reiterate that $M_d$ is dependent on empirically driven grain properties and is not designed to reflect the true dust mass).

HD~117214 has the highest peak \Qr surface brightness of the seven disks that GPIES resolved for the first time. Its peak \Qr surface brightness is over eight times greater than that of HD~145560, and given that both host stars have an F5V spectral type, Sco--Cen moving group membership, and ring-shaped disks, we investigate the physical source of the brightness differences between their two disks in greater detail in Section \ref{sect:117214_v_145560}. The HD~117214 disk is also noteworthy for the detection of both its front and back edges in total intensity, bearing some resemblance to the HR~4796~A ring but on smaller angular and physical scales.

% HD 117214 images.

% HD 156623 images.
\begin{figure}[h!]
\centering
\includegraphics[width=\columnwidth]{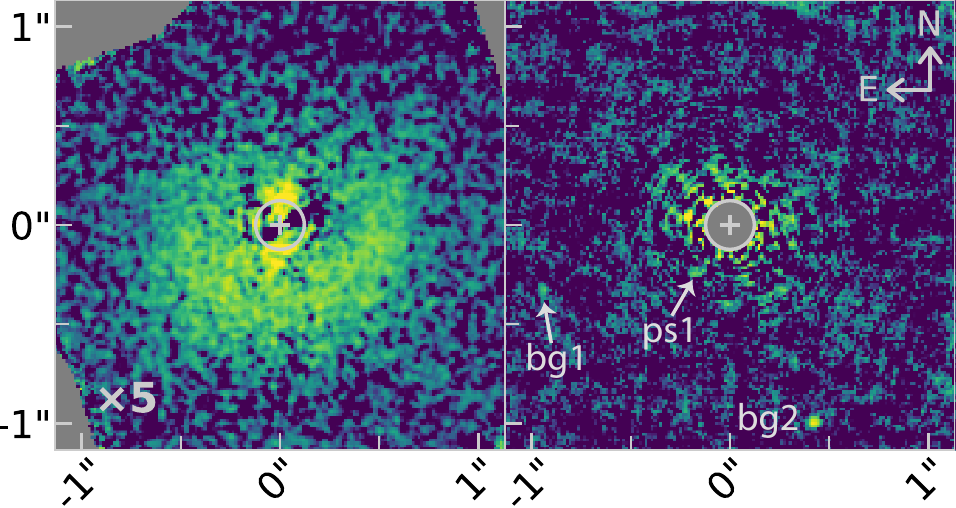}
\caption{HD~156623 debris disk seen in $H$-band Stokes \Qr (left) and the non-detection in spectroscopic-mode total intensity (right). The \Qr image's flux was scaled up by a factor of 5 before being plotted on the same logarithmic brightness scale as the total intensity image in ADU s$^{-1}$. Two point sources in the total intensity image are confirmed background objects, with a third out of view to the west at $r\approx1\farcs8$, and the innermost point source (ps1) is a suspected background object that has yet to be confirmed.}
\label{fig:hd156623}
\end{figure}

\subsubsection{HD 156623}

We resolved the HD~156623 (SpT = A0V) debris disk for the first time in scattered light, seen as a low-inclination disk detected only in \Qr and with a brighter south edge that we assume to be the forward-scattering edge (Figure \ref{fig:hd156623}). The GPI image reveals a radially broad ring without a clear inner hole, which are both departures from the other debris disks in our sample. The morphology could be considered closer to that of the HD~100546 disk, considered to be in the protoplanetary phase, yet HD~156623's disk surface brightness is two orders of magnitude lower. Given that HD~156623 is only 1.2 mag fainter in $H$, its disk likely has a much lower dust density than that of HD~100546. There is also evidence that HD 156623 is one of roughly a dozen ``hybrid'' debris disks with substantial CO gas masses that may still contain some primordial gas \citep{kospal2013} and/or have secondary CO gas shielded from photodissociation by neutral C \citep{kral2019}. ALMA detected an extremely strong resolved $^{12}$CO (2--1) signature indicative of a CO gas mass $\geq\num{3.9e-4}~M_{\oplus}$ and marginally resolved the disk with moderate S/N in 1.3 mm continuum imaging \citep{lieman-sifry2016, moor2017}.

We measure values of $i=34\fdg9 \substack{+3.6 \\ -9.5}$ and $\mathrm{PA}=100\fdg9 \substack{+1.9 \\ -2.2}$ from the modeling MCMC, both of which are in agreement with estimates from ALMA CO line emission \citep{lieman-sifry2016}. The model dust mass is log($M_d/M_\odot$)$=-6.30 \substack{+0.21 \\ -0.05}$ (again, $M_d$ is dependent on empirically based grain properties and is not designed to reflect the true dust mass). The disk's critical radius is tightly constrained by the models to be $r_\mathrm{c}=64.4 \pm 1.8$ au, and the observed surface brightness does decrease interior to that radius, suggesting that this disk is ring-like. On the other hand, the models only constrained the inner radius to be $<26.7$ au (99.7\% confidence limit) with a best-fit value near the MCMC's prior boundary of 12 au (set by the edge of the GPI FPM), confirming that we do not observe an inner hole. \citet{lieman-sifry2016} also did not resolve an inner hole in their ALMA continuum data and placed an upper limit of 19 au on the inner radius with a best-fit value of 9 au (after updating to the Gaia DR2 distance). This lack of observational or model-based evidence for a dust-poor inner hole makes HD~156623 unique among the GPIES-detected sample, and we discuss its possible connection to the disk's high gas content in Section \ref{sect:discuss_gas}. Similar to HD~117214, we can only place a lower limit of $>139.3$ au (99.7\% confidence) on the outer radius, which is statistically consistent with the ALMA dust-continuum estimate of $142\pm28$ au (as is our best-fit value of 175 au). Combining the limits for $r_\mathrm{in}$ and $r_\mathrm{out}$, we find a lower limit of ${\sim}113$ au for the ring's radial width.

HD~156623's galactic latitude of $-5\degr$ places it along a crowded line of sight through the galaxy, so it is not surprising that we count four other total intensity point sources in our field of view. We confirmed the three outermost sources to be unbound background stars by their relative motions between two GPIES epochs 736 days apart. The fourth source is also suspected to be a background star but had a stellar separation of only $0\farcs31$ ($\mathrm{PA}\approx143\fdg1$) in our 2017 April 21 observation and was not detected in our 2019 April 27 observation, when it was expected to be at only $0\farcs27$ separation (and thus less detectable) if it is a distant background star. Follow-up observations are needed to firmly establish this source's relationship to HD~156623.

\subsection{Morphological Modeling of Three More Recent GPIES Discoveries}
\label{sect:morph_other_three}

In addition to HD~117214 and HD~156623, we applied our MCMC modeling to the recently discovered scattered-light disks around HD~111161, HD~143675, and HD~145560 to retrieve their basic morphological properties. We report the resulting values here, and selected parameters are summarized in Table \ref{tab:measures} (see Section \ref{sect:sample_results} for explanation). \citet{hom2020} present additional analyses of the observations and morphologies of these three disks.

\begin{itemize}
    \item \emph{HD 111161}: We measure values of $i=62\fdg1 \substack{+0.3 \\ -0.2}$ and major axis $\mathrm{PA}=83\fdg2 \substack{+0.5 \\ -0.6}$, again assuming $\pm34\%$ confidence intervals for the uncertainties. The ring's inner edge is tightly constrained to $r_\mathrm{in} = 71.4 \substack{+0.5 \\ -1.0}$ au. A critical radius of $r_c=68.8 \substack{+1.6 \\ -1.5}$ au, just interior to $r_\mathrm{in}$, means the model preferred a single negative power-law index for the entire ring's dust density profile. The outer radius is more loosely constrained at $r_\mathrm{out}=217.9 \substack{+15.5 \\ -15.3}$ au, with a lower probability solution at ${\sim}127$ au that more closely traces the outer edge of the bright disk emission. The inclination and PA are not strongly correlated with either radius parameter. The MCMC also returned log($M_d/M_\odot$) $=-6.29 \pm 0.03$.
    
    \item \emph{HD 143675}: Our modeling returns parameters of $i=87\fdg2 \substack{+0.6 \\ -0.7}$ and $\mathrm{PA}=113\fdg2 \substack{+0.5 \\ -0.4}$, placing the disk's front edge on the north side of the star. The ring's inner radius is $r_\mathrm{in}=44.0 \substack{+3.5 \\ -7.6}$ au, with a low-probability tail down to 15 au that likely results from degeneracies created by the disk's edge-on appearance. The outer radius is more tightly constrained at $r_\mathrm{out}=52.1 \substack{+1.4 \\ -1.0}$ au. This may underestimate the outer radius because the MCMC only considers the \Qr data, which shows an approximately 20\% smaller outer extent than the total intensity data. Even so, this disk is one of the most compact in our sample in terms of both physical and angular radial extent. The critical radius is essentially unconstrained with a $3\sigma$ lower limit of $r_c>11.1$ au, likely because the small angular extent and edge-on orientation provide little leverage to define the precise shape of the dust density radial profile. Given the tighter constraints on $r_\mathrm{in}$ and $r_\mathrm{out}$, we assume for analyses in Section \ref{sect:sample_results} that the peak of the dust density profile is between those two radii. Finally, the MCMC returned log($M_d/M_\odot$)$=-7.17 \pm 0.03$.
    
    \item \emph{HD 145560}: For this moderately inclined disk, we estimate $i=43\fdg9 \substack{+1.5 \\ -1.4}$ and $\mathrm{PA}=41\fdg5 \substack{+1.0 \\ -1.2}$. This is the third lowest inclination of our detected sample (to only CE~Ant and HD~156623). Our inclination is within $1\sigma$ of the ALMA measurement of $50\degr \substack{+6 \\ -7}$ \citep{lieman-sifry2016}, with the difference possibly stemming from the disparity in angular resolution. Our PA disagrees with the ALMA value of $20\degr \substack{+7 \\ -6}$ by ${\sim}3\sigma$ but we propose that our measurement is the more accurate result because the GPI data clearly resolve the ring ansae while the ALMA data do not. Our models return radii of $r_\mathrm{in}=68.6 \substack{+2.9 \\ -1.3}$ au, $r_c=81.1 \pm 1.2$ au, and $r_\mathrm{out}>196.2$ au (a 99.7\% confidence lower limit). The disk's large $r_\mathrm{out}$ is partially driven by a faint nebulosity exterior to the ring's front edge (i.e., SE of the star), where the forward-scattering dust remains detectable even at large radii (Figure \ref{fig:det_gallery_Qr}); however, it is also generally poorly constrained because the model's low surface brightness at large radii reaches the background noise level, making little difference in likelihood between a model with $r_\mathrm{out}=150$ au and one with $r_\mathrm{out}=300$ au. The dust mass value is log($M_d/M_\odot$) $=-5.52 \substack{+0.04 \\ -0.02}$.
\end{itemize}

\subsection{Individual Results for Other GPIES Debris Disks}

Our observations produced several other noteworthy results for individual disks that we did not model in this work:

\begin{itemize}
    \item \emph{CE Ant (TWA 7)} is detected only in the \Qr data, confirming the nearly face-on ring structure previously discovered with \textit{HST} NICMOS \citep{choquet2016} and mapped in $J$ and $H$-band polarized light with VLT/SPHERE \citep{olofsson2018}. Our data independently confirm tentative features evident in the VLT/SPHERE images: (1) peak azimuthal brightness and radial width south of the star at $170\degr{<}\mathrm{PA}{<}190\degr$, (2) minima in azimuthal brightness and radial width NE of the star at $0\degr{<}\mathrm{PA}{<}60\degr$, and (3) radially extended nebulosity in the SW quadrant that \citet{olofsson2018} compare to a spiral arm.
    
    \item \emph{HD 30447} shows a double-lobed morphology in the total intensity data that is unusual but possibly consistent with the bifurcated structure in a previous \textit{HST} NICMOS image from \citet{soummer2014}. Such a shape can be an effect of ADI PSF subtraction, which preferentially subtracts disk light near the star, thus creating a ``pinched'' appearance along the disk's projected minor axis \citep{milli2012}. If that is the case here, then the true shape of the disk is more likely to be the simple ring seen in the \Qr data. From the disk's NW--SE brightness asymmetry in \Qr, we assume the NW edge to be the front edge, a distinction that is difficult to make from the NICMOS data. Additional high-resolution imaging is warranted to investigate this disk's morphology and also to recover a polarization fraction measurement (with a wide range of scattering angles being accessible for rings like this with $i\approx80\degr$).
    
    \item \emph{HD 111520} is an edge-on disk with a significant brightness deficit close to the star. Its \Qr brightness peaks near $0\farcs55$ but it decreases by at least a factor of 2 interior to that separation. This supports a similar initial finding from a shorter GPI observation in \citet{draper2016}. This polarized brightness trend is different from that of other edge-on disks in our sample, which either remain nearly flat (HD~146897) or increase in brightness as separation decreases (HD~32297). HD~111520's total intensity, on the other hand, is more typical for forward-scattering dust and increases as separation decreases from $1\farcs0$ to $0\farcs30$ (interior to which residual speckle noise dominates the measurement). This ignores corrections for over-/self-subtraction by the KLIP algorithm, so the true disk brightness is likely even higher at small separations than measured. Thus, this disk's polarization fraction must decline at small scattering angles more than that of other disks; explaining this feature may require invoking atypical grain properties.
    
    \item \emph{\object{HD 114082}} is detected for the first time in polarized intensity, confirming the belt-like morphology that was resolved in total intensity with SPHERE/IRDIS \citep{wahhaj2016}. Our total intensity data show an asymmetry in which the dust-scattered light external to the belt ansae is 50\%--80\% brighter on the west side than the east. Possibly as a result, the scattered-light emission extends $1\farcs2$ to the west but only $1\farcs0$ to the east. The three point sources detected in total intensity are confirmed background objects based on their proper motions between our 2018 January 29 epoch and a 2016 February 14 epoch from archival SPHERE/IRDIS data (PI: J.~Milli).
    
    \item \emph{\object{HD 115600}} is also detected for the first time in polarized intensity, with previous total intensity detections produced by GPI \citep{currie2015a} and SPHERE \citep{gibbs2019}. Under the assumption of forward-scattering dust, our polarized intensity data identify the front side (i.e. near side) of the disk to be the NW side. This contradicts both previous total intensity studies which identified the SE side of the disk to be the front side. However, the S/N of their data inside of the ansae is poor relative to our detection, and our polarized intensity data is able to probe smaller inner working angles, where the forward scattering is strongest. Thus, we consider the NW side to be the disk's true front side.
    
    \item \emph{HR 7012} has an unusually small radial extent in polarized intensity compared to the rest of our sample. At a distance of 28.5 pc, this disk's $0\farcs42$ outer radius translates to 12 au. As a disk around an A7V star, we would expect its outer radius to be at least as large as the ${\sim}52$ au outer radii of HD 143675 (A5IV/V) and HD 115600 (F2/3V), but it is a factor of 4 smaller. In fact, the median \emph{inner} radius of our detected debris disks is 57 au, meaning that the entire HR 7012 ring would fit within the inner holes of most other GPIES-detected rings. Our outer radius estimate is consistent with the SPHERE/ZIMPOL polarized visible-light measurement of ${\sim}12$ au by \citet{engler2018}. Those authors noted that this star has a comoving K5Ve low-mass companion at $>2000$ au separation (CD-64~1208; \citealt{torres2006}). Dynamical interactions between that companion and the disk might explain its truncation, especially if the companion has an elliptical orbit or had a closer orbit in the past.
    
\end{itemize}

\subsection{Protoplanetary and Transitional Disk Detections} \label{sect:proto}

\begin{figure*}
    \centering
    \includegraphics{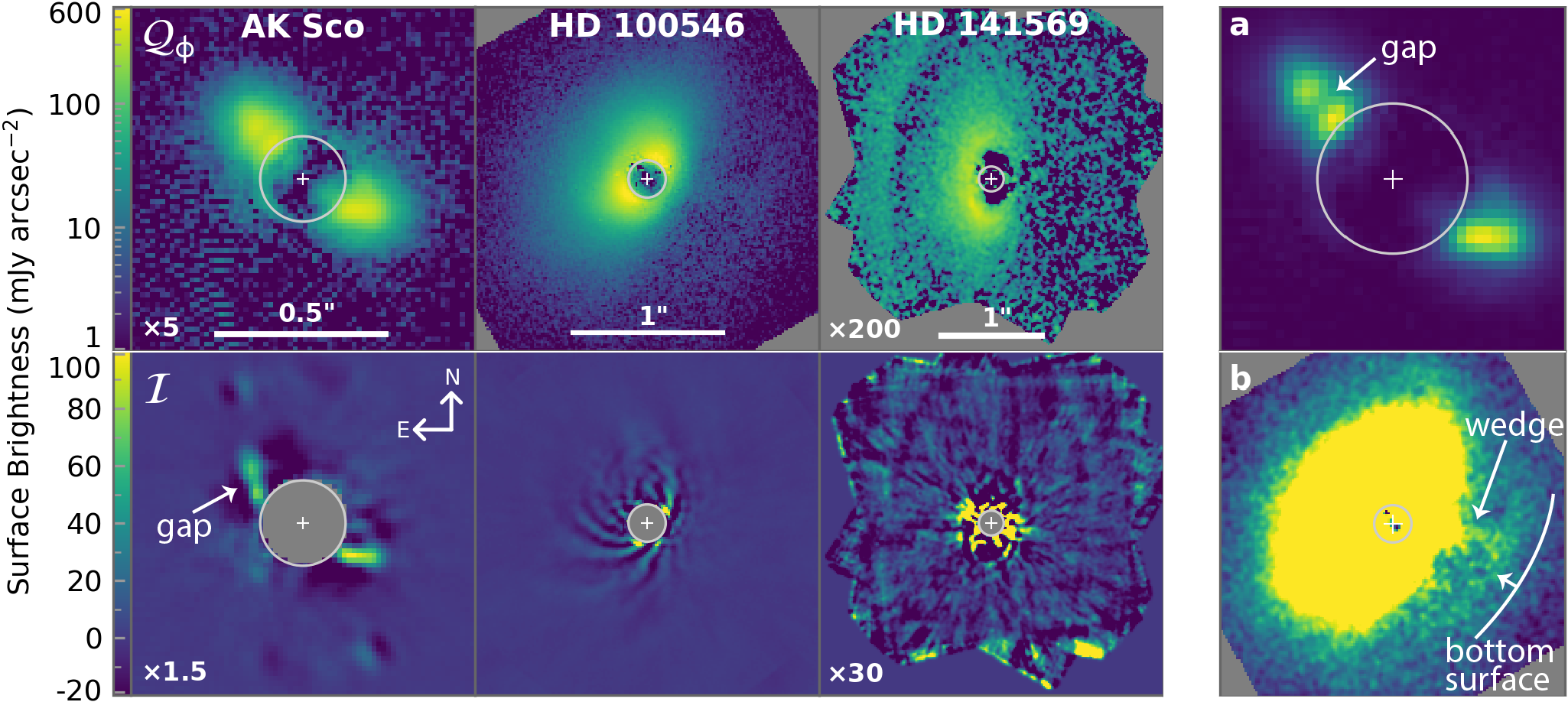}
    \caption{(Left) Three protoplanetary or transitional disks detected by GPIES. The top row shows the \Qr polarized surface brightness on a logarithmic scale, and the bottom row shows the total intensity surface brightness on a separate linear scale. The surface brightnesses of AK Sco and HD 141569 were multiplied by the factors in the lower-left corners of their panels before plotting. Both HD 141569 images were smoothed with a Gaussian ($\sigma=1$ pixel) to suppress background noise. An arrow in the AK Sco total intensity image points to the observed gap, and the \Qr artifacts SE of the star are microphonic noise induced by instrument vibrations. (Right) (a) A zoomed-in view of the AK Sco \Qr disk on an arbitrary linear brightness scale to highlight the gap. (b) The HD 100546 \Qr disk on an exaggerated logarithmic brightness scale to highlight the scattering by the bottom surface and the wedge of light between the top and bottom surfaces.} \label{fig:proto}
\end{figure*}

In addition to debris disks, GPIES also detected three objects that are perhaps better classified as protoplanetary or transitional disks: AK Sco, HD 100546, and HD 141569 (Figure \ref{fig:proto}). All are detected in both polarized and total intensity, although the HD 141569 total intensity detection is marginal. All three stars are classified as Herbig Ae/Be stars (\citealt{vioque2018} and references therein). These disks are also distinguished by large IR excesses (\Lir$>\num{1e-2}$) and high gas masses \citep{zuckerman1995, panic2010, czekala2015a}. The only other GPIES target in this excess range is HD 146897, which has a bright dust disk and also a tentative CO gas detection \citep{lieman-sifry2016} that suggests it may be a debris disk in a relatively early evolutionary state.

% Protoplanetary/Transitional disk gallery.

Briefly, we find the following results for each disk:

\begin{itemize}
    \item \textit{AK Sco} is a highly inclined disk detected out to a projected separation of ${\sim}0\farcs35$ (45 au) around an equal-mass spectroscopic binary. Based on the brightness asymmetry observed in total intensity, we see only the forward-scattering front edge of a ring. The SW side of the ring appears brighter than the NE side, and there is a marginally resolved gap in the NE side at $r\approx0\farcs17$. This gap appears in both \Qr and total intensity, and its \Qr surface brightness is ${\sim}20$--30\% fainter than the surrounding regions. The gap also coincides with a similar feature unremarked on but present in the VLT/SPHERE $H$ and $YJ$ total intensity images from \citet{janson2016}. Two total intensity point sources SSE of the star are likely background stars but require additional astrometric follow-up. They are out of the frame in Figure \ref{fig:proto} but have separations and PAs of approximately ($0\farcs98$, $154\fdg0$) and ($1\farcs03$, $162\fdg5$) for the brighter and fainter sources, respectively.
    
    \item \textit{HD 100546} has a low-inclination disk that appears smooth in \Qr but shows spiral structures in total intensity. The smooth \Qr surface brightness  confirms the result from \citet{millar-blanchaer2017} in an independent reduction of the same data. That study showed the uneven surface brightness previously reported by \citet{currie2015b} to be the result of data reduction artifacts created by an inaccuracy in the DRP that has since been corrected.
    
    We marginally detect a faint azimuthal band of emission west of the star, spanning a PA range of ${\sim}192\degr$--$285\degr$ and with a sharp inner boundary at $r\approx0\farcs64$ and diffuse outer boundary near $r\approx0\farcs9$. We propose that this is light scattered by the bottom surface of the disk's near side, similar to that seen in polarimetric imaging of the IM Lup protoplanetary disk \citep{avenhaus2018}. The location, shape, and relative brightness of the feature qualitatively match those seen in 1.6 $\micron$ polarized intensity models of this disk by \citet{tazaki2019}. There is also a faint ``wedge'' of light WSW of the star (centered on $\mathrm{PA}\approx263\degr$) that is consistent with the wedge seen in visible polarimetry by \citet{garufi2016}; presumably, this wedge spans the shadowed disk midplane between the near side's top and bottom surfaces. Deeper imaging and further modeling are warranted. Analyses of the $H$ total intensity and other GPI $Y$ polarized intensity data have previously been presented in \citet{follette2017} and \citet{rameau2017} regarding planet candidates in this system.
    
    \item \textit{HD 141569} appears as a bright inner ring and parts of a fainter, larger radius ring. These correspond to the two innermost rings as seen by, e.g., \citet{perrot2016}, while the disk's outermost ring \citep{konishi2016, mazoyer2016} is outside of our field of view. The total intensity detection is marginal, but we resolve the sharp edges of both rings. A more detailed analysis of the GPIES data and modeling of the disk are presented in \citet{bruzzone2019}.
\end{itemize}

\subsection{Non-detections}
\label{sect:non-detections}

We did not detect scattered-light disks (of any class) around \Nnondet of the \Nobs observed stars. Additionally, some GPIES disks were detected in one data set but not another, so we ended up with 83 viable data sets that did not yield polarized intensity disk detections: 70 pol snapshots and 13 deep pol observations (Table \ref{tab:time_pa}). We had 175 data sets yield no total intensity detection: 80 pol snapshot, 21 deep pol, and 74 spec-mode observations. We first examine the properties of the data sets themselves for observational explanations of non-detections. Then we consider our sensitivity to disks given their physical characteristics, such as angular size, surface brightness, polarizability of the grains, etc.

Comparing the data sets with detections to those without, we find differences that may explain some non-detections. The three main properties that we explore are total integration time, total parallactic angle rotation ($\Delta$PA), and final \Qr contrast, shown in Table \ref{tab:time_pa}. The latter two are correlated with integration time but also depend on other factors (target declination and observation timing in the first case, seeing conditions and AO performance in the second case). We examine snapshot and deep pol-mode data sets separately in this context because of their inherent differences in integration time. On average, detections and non-detections had the same integration times to within a few minutes, so an uneven distribution of integration time does not appear to be the primary cause of non-detections, when examined in aggregate.

One substantial discrepancy appears in $\Delta$PA. The deep pol-mode data sets averaged $29\fdg0$ of $\Delta$PA for total intensity non-detections compared to $54\fdg2$ for detections. The $\Delta$PA primarily affects total intensity sensitivity because the ADI PSF-subtraction algorithms that we employed best preserve disk brightness when it appears at a diverse set of PAs. A $\Delta$PA less than the azimuthal width of the disk at a given projected separation can lead to the stellar PSF-subtraction process significantly attenuating the disk intensity, and in some cases (such as azimuthally broad and symmetric disks) removing the disk signal completely (e.g., \citealt{marois2006, lafreniere2007, milli2012, esposito2014}). This attenuation may explain some total intensity non-detections, particularly those of disks with low inclinations and/or small angular sizes (where the angular azimuthal disk width is large).

HD 114082 serves as an example. The nearly edge-on ring was strongly detected in \Qr (Figure \ref{fig:det_gallery_Qr}) but showed no significant total intensity signal in the same deep pol observation. It was later clearly detected in total intensity in a spec-mode data set (Figure \ref{fig:det_gallery_totI}) that was only 31\% longer in integration time but had 110\% more parallactic rotation than the deep pol data set ($25\fdg8$ vs. $12\fdg3$). The raw (per frame) contrast in the spec data was notably better than that of the deep pol data, however, so while the increased $\Delta$PA likely played a role in the detection, disentangling and quantifying the multiple effects at play are beyond the scope of this work.

In terms of polarized intensity detections, the depth of the final \Qr contrast appears to be a more important factor for brighter stars (see Section \ref{sect:contrasts} for details), but IR-excess magnitude plays the dominant role regardless. As discussed previously, GPIES detections strongly favor the targets with the highest \Lir. This is true across the full range of observed stellar magnitudes. The \Qr contrast of a data set, though, becomes more strongly correlated with detection as stellar brightness increases, as we showed in Figure \ref{fig:contrast_rstokes}. This implies that some of our targets with \Qr non-detections would have been detected if we had reached deeper contrasts in their observations, particularly for stars with $H < 7$ mag. Exactly how much deeper is unclear---it stands to reason that the depth of contrast required increases as \Lir decreases within a given stellar magnitude bin, but we have not explored this question due to the relatively small number of detections per bin.

\begin{deluxetable}{lccc}
\tablecaption{\label{tab:time_pa}Average Data Set Properties}
\tablehead{
% Header Line 1
\colhead{Mode} & \colhead{Status (\# data sets)} & \colhead{$t_{int}$ (minute)} & \colhead{$\Delta$PA ($\degr$)}
%\vspace{-0.1 cm}
}
\decimals
\startdata
 \multirow{2}{*}{Pol Snap: \Qr} & Detections (13) & 13.5 & 8.7 \\
 & Non-Detections (70) & 13.5 & 9.1 \\
 \midrule
 \multirow{2}{*}{Pol Deep: \Qr} & Detections (23) & 42.3 & 49.9 \\
 & Non-Detections (13) & 40.3 & 21.1 \\
 \midrule[1pt]
 \multirow{2}{*}{Pol Snap: \I} & Detections (3) & 10.7 & 5.5 \\
 & Non-Detections (80) & 13.6 & 9.2 \\
 \midrule
 \multirow{2}{*}{Pol Deep: \I} & Detections (15) & 41.4 & 54.2 \\
 & Non-Detections (21) & 41.7 & 29.0 \\
 \midrule
 \multirow{2}{*}{Spec: \I} & Detections (17) & 46.3 & 46.7 \\
 & Non-Detections (74) & 41.5 & 43.4 \\
\enddata
%\vspace{-0.8cm} 
\tablecomments{Average integration time and parallactic rotation of all GPIES data sets, broken down by observing mode and disk detections vs. non-detections in either \Qr polarized intensity or \I total intensity.}
\end{deluxetable}

% HD 181327 w/ STIS overlay.
\begin{figure}[h!]
\centering
\includegraphics[width=\columnwidth]{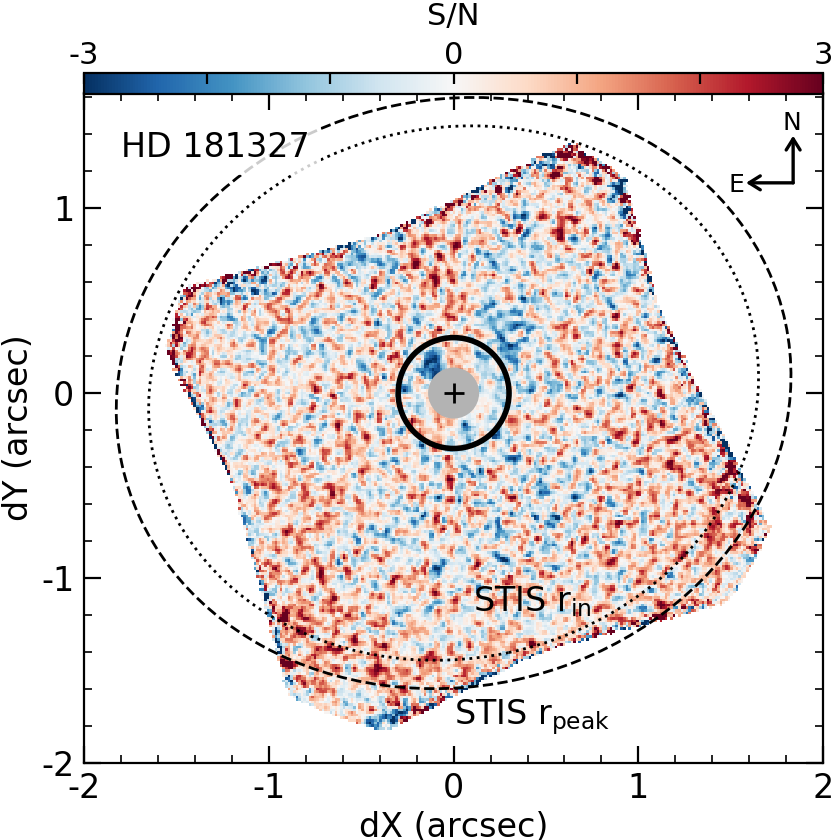}
\caption{GPI \Qr S/N map of HD 181327 on a linear scale with ellipses marking the known dust ring's inner edge (dotted) and peak radius (dashed) from STIS imaging \citep{schneider2014}. We found some positive intensity with 2$\sigma$--3$\sigma$ significance in the SE and SW corners of the GPI field of view but we do not consider this a detection of the ring, and no significant dust was detected interior to the ring's location. The solid black line marks the STIS inner working angle, the gray filled circle denotes the GPI focal plane mask size, and the black cross is the star position.}
\label{fig:hd181327}
\end{figure}

Related to the effects of contrast, we can attribute non-detections of some \textit{HST}-resolved disks to their intrinsically low scattered-light surface brightnesses. These include 49~Cet \citep{choquet2017, pawellek2019}, HD~377, TWA~25 \citep{choquet2016}, and V419~Hya (HD~92945; \citealt{golimowski2011, schneider2014}). For example, the TWA~25 disk has a peak surface brightness of ${\sim}40\ \mu\mathrm{Jy}\ \mathrm{arcsec}^{-2}$ at $r=1\arcsec$ and an overall mean of $28\pm5\ \mu\mathrm{Jy}\ \mathrm{arcsec}^{-2}$, based on the model of NICMOS 1.6 $\mu$m (F160W) observations presented by \citet{choquet2016}. Those NICMOS data, and consequent models, are subject to underestimating the disk brightness due to oversubtraction side effects from the stellar PSF subtraction; nevertheless, they remain our best estimators of the disk's true brightness. The GPIES \Qr image reached a $3\sigma$ sensitivity level of 31 $\mu\mathrm{Jy}\ \mathrm{arcsec}^{-2}$ at $1\arcsec$. Thus, the disk's light would have to be nearly 100\% polarized for us to achieve a significant detection based on the NICMOS-derived brightness. We know from other GPI data that $H$-band polarization fractions are nearly all $<50$\%, so the polarized surface brightness of that disk likely lies below the sensitivity level of our GPI observation. A similar explanation applies to 49~Cet, HD~377, and V419~Hya, highlighting the sensitivity differences between ground-based polarimetry and space-based total intensity observations. In the case of 49~Cet, we also did not reach the contrast needed to detect the ring in total intensity at $r<1\farcs4$, similar to the SPHERE results that did not detect the disk inward of $1\farcs4$ with approximately the same integration time in the $H$ band \citep{choquet2017} or more than double our greatest integration time and field rotation in the $Y$ band \citep{pawellek2019}.

Similarly, we did not detect a pair of disks that have been resolved at millimeter wavelengths within the GPI field of view. HD~138813 was resolved at 3$\sigma$--5$\sigma$ in 1.3 mm ALMA dust continuum and $^{12}$CO $J$=2--1 transition observations by \citet{lieman-sifry2016}. They estimated an inner radius of $67 \substack{+20\\-19}$ au and inclination of $28\degr$. Despite the presence of those large grains, abundant gas, an \Lir$=\num{1.3e-3}$, and median \Qr contrasts, we did not detect the disk. From the same study, HD~142315 had no gas detection but appeared to be a highly inclined ring in dust continuum and had a total 1.3 mm flux equal to that of HD~111161, which we detected with GPI in \Qr. We did not detect this disk despite reaching roughly median \Qr contrasts in the same amount of integration time with which we detected HD~111161. These results suggest that HD~138813 and HD~142315 are fainter in polarized scattered light than other disks that are bright at millimeter wavelengths, possibly due to differences in the grain size distribution, scattering phase functions, and/or polarizability. We consider this question further in Section \ref{sect:117214_v_145560}.

We detected the HD~15115 and NZ~Lup disks with GPI in total intensity but had non-detections in polarized intensity. After our HD~15115 observations were conducted, \citet{engler2019} published a SPHERE $J$-band polarized intensity detection at projected separations ${>}1\arcsec$ (from data with three times the total integration time of our pol-mode data set). Assuming their measured $J$-band total (i.e. integrated) polarization fraction of $0.29 \pm 0.09$ to be the average fraction in the $H$ band, we can use the contrast of our $H$-band total intensity detection to roughly estimate the disk's $H$-band polarized contrast at $0\farcs8$. Doing so, we predict the per-pixel contrast would be roughly at the $1\sigma$ level (${\sim}10^{-7}$) in our \Qr image, likely explaining the lack of a significant detection. No polarized intensity detection has yet been published for NZ Lup's disk, although \citet{boccaletti2019} presented additional total intensity images, so we can only speculate that the disk's polarized signal is fainter than our achieved sensitivity.

Some non-detections are also due to the disk's dust being located outside the GPI field of view. We know this to be the case for the bright, previously resolved scattered-light rings around Fomalhaut, HD~107146, and HD~181327, which have minimum projected separations of roughly $7\farcs5$, $3\farcs5$, and $1\farcs5$, respectively \citep{ardila2004, kalas2005_fomb, schneider2006}. It was also likely true for large disks resolved only at thermal wavelengths, like HD~95086, which has a depression in millimeter flux interior to $1\farcs7$ \citep{zapata2018}. Upon their selection as targets, we had assumed we would not detect those known rings. Instead, we observed those targets searching for interior dust that was hidden from previous instruments but that GPI might detect with its small inner working angle and polarimetric sensitivity. We consider these observations non-detections because no such inner dust was detected (and not because the known rings went undetected). Additional dust may still be present in the observed regions of these systems, just with relatively low surface brightnesses.

In the case of HD~181327, we found some positive intensity with 2$\sigma$--3$\sigma$ significance in the SE and SW corners of our \Qr image (Figure \ref{fig:hd181327}), corresponding to projected separations of $r=1\farcs5$--$1\farcs7$. Given the relatively low significance and our limited spatial coverage, we do not count this as a detection in our statistics. Although STIS also detected scattered light from $1\farcs5$ (the edge of the bright ring) down to $r=1\farcs0$ \citep{schneider2014, ren2018_nmf}, we did not detect substantial polarized light in this region. This part of the disk has a low surface brightness ($<0.8$ mJy arcsec$^{-2}$) in STIS's broad optical bandpass, and the polarization fraction is almost certainly less than 100\%, so the polarized intensity signal from the inner disk is presumably below the sensitivity limit of our 19 minute observation, as established by similar GPIES data.

% AVOID ORPHANED SECTION HEADING
\vspace{40pt}

\subsection{Sample-wide Results on Debris Disk Properties} \label{sect:sample_results}

Our resolved images allow us to directly measure the scattered-light surface brightness produced by small dust grains in a disk and thereby infer the spatial distribution of those grains. This information is clearly valuable on the individual level; see $\beta$ Pic's warp \citep{burrows1995} and AU Mic's clumps \citep{boccaletti2015a} as just two examples. With GPIES, however, we have the opportunity to study \Ndebris\ debris disks as a group and look for trends on the population level.

In Table \ref{tab:measures}, we have consolidated measurements of disk morphological properties for GPIES detections. These properties are $i$, PA, minimum and maximum projected separations where we detect scattered light in the GPIES data ($r_\mathrm{min}$ \& $r_\mathrm{max}$), inner disk radii from scattered light ($R_\mathrm{in}$) and thermal emission ($R_\mathrm{in, mm}$), radii where the dust density peaks based on scattered light ($R_\mathrm{0}$) and thermal emission ($R_\mathrm{0, mm}$), and the blackbody effective radius estimated from SED fits ($R_\mathrm{bb}$, described in Section \ref{sect:targ_selection}). Their quoted values come from a variety of sources, listed in Table \ref{tab:measures}. The uncertainties mark the $\pm34$\% confidence intervals or $1\sigma$ errors where other studies assumed Gaussian posterior distributions. For measurements from GPI data, uncertainty on the instrument's true north angle has been combined in quadrature with the PA measurement uncertainty (see Section \ref{sect:pol_reduction}). Uncertainty on GPI's pixel scale (0.05\% fractional error) is similarly included in uncertainties of disk radii; however, it is insignificant compared to the radius measurement errors (typically $\gtrsim1$\%).

Due to the extended 3D nature of these disks and projection effects on their apparent structure, the most stringent constraints on their morphologies come from modeling the dust's spatial distribution and propagating it to a scattered-light simulation via some estimate of the dust's scattering (and polarization) phase function. This is a time- and computation-intensive process that typically requires a parameter grid search or MCMC for each disk. Thus, we only directly perform this process in this work for the five recently resolved disks discussed in Sections \ref{sect:new_disks} and \ref{sect:morph_other_three}. For disks with previously published GPIES results or papers in preparation, we preferentially adopt those GPIES values. For disks with published results from other resolved scattered-light imaging (of similar resolution and S/N) but no individual GPIES study, we adopt the non-GPI published values after ensuring that they qualitatively agree with our data. In the case of thermal emission radii, we must rely on other studies because that quantity cannot be measured from GPI data. This multisourced approach may introduce additional scatter into the ensemble of measurements. We plan for a future study to consistently model all disks in our sample for a more in-depth analysis of their properties.

The $R_\mathrm{bb}$ values were computed using new single-component fits to existing photometry, except for that of HD~106906. For this target, the fit included \textit{IRAS} photometry at 60 and 100~$\micron$ \citep{hindsley1994} that appears anomalously high compared to measurements from \textit{Spitzer} MIPS at 70~$\micron$ \citep{chen2014} and \textit{Herschel} PACS at 100 and 160~$\micron$. The resulting $R_\mathrm{bb}$ of 152 au would be the second largest out of the \Ntarg star GPIES target list and 61 au greater than the next largest, so we suspect it is biased by the \textit{IRAS} data, which may have had a contaminating source in its large beam. In its place, we adopted the outer-component $R_\mathrm{bb}$ value of 44.4 au from \citet{chen2014} that is based on a two-component fit to only \textit{Spitzer} photometry, which we believe to be more accurate.

\subsubsection{Ring Morphologies}

In terms of basic morphology, we find all of the GPIES debris disks except HD 156623 to be consistent with dust rings that have inner holes. A ring shape is immediately evident in polarized intensity for disks with $i\lesssim75\degr$ because our unobscured view of the disk's inner regions shows that they are substantially depleted in dust (e.g., HD 117214 and HD 145560). Total intensity images also reveal inner holes of higher inclination disks that are not obvious in polarized intensity (e.g., HD 35841 and HD 129590), although the holes' radial extents and depths are often exaggerated by PSF-subtraction effects. HD 156623 is one exception because we find significant polarized intensity surface brightness all the way inward to the FPM edge on all sides of the star. For the nearly edge-on disks, a ring shape is inferred either from models of the dust distributions (e.g., HD 143675) or from breaks in the slopes of their surface brightness radial profiles interpreted as primary locations of dust production via collisions \citep{strubbe2006}. In the GPIES sample, we directly observe minimum projected separations as small as 2 au for our nearest detected disk and 27 au even for our farthest detection, set by an effective inner working angle of ${\sim}0\farcs15$--${\sim}0\farcs20$ in pol-mode (some data have stronger speckle and/or instrumental noise around the FPM than others). Additional dust may exist interior to those separations in asteroid belt analogs; however, their detection will require future high-contrast imagers with even smaller inner working angles or possibly interferometry.

% Detection Properties table.
\startlongtable
\begin{deluxetable*}{lDDccDDDDr}
\setlength{\tabcolsep}{4pt}
\tablecaption{\label{tab:measures}Resolved Disk Properties}
\tablehead{
% Header Line 1
\colhead{Name} & \twocolhead{$i\pm\sigma$} &  \twocolhead{$\mathrm{PA}\pm\sigma$} & \colhead{$r_\mathrm{min}$} & \colhead{$r_\mathrm{max}$} & \twocolhead{$R_\mathrm{in}$} & \twocolhead{$R_\mathrm{in,mm}$} &
\twocolhead{$R_\mathrm{0}$} & \twocolhead{$R_\mathrm{0,mm}$} &
\colhead{$R_\mathrm{bb}$} \vspace{-0.3cm} \\
% Header Line 2
 & \twocolhead{(deg)} & \twocolhead{(deg)} & \colhead{($\arcsec$)} & \colhead{($\arcsec$)} & \twocolhead{(au)} & \twocolhead{(au)} & \twocolhead{(au)} & \twocolhead{(au)} & \colhead{(au)}
}
\decimals
\startdata
AK Sco$^{\dagger}$ & 82.0 $\substack{+5.0 \\ -5.0}$\ \ (1) & 53.\ \ (1) & FPM & 0.3 & 6.3$\substack{+5.7 \\ -2.8}$\ \ (1) & $\mbox{--}$ & 128.0$\substack{+228.0 \\ -93.0}$\ \ (1) & $\mbox{--}$ & 11 \\
AU Mic & 89.4 $\substack{+0.1 \\ -0.1}$\ \ (2) & 128.6 $\substack{+0.2 \\ -0.2}$\ \ (2) & 0.2 & 1.3 & 30.2$\substack{+12.0 \\ -22.5}$\ \ (3) & 23.7$\substack{+1.0 \\ -13.6}$\ \ (4) & 30.2$\substack{+12.0 \\ -22.5}$\ \ (3) & 41.9$\substack{+0.8 \\ -31.8}$\ \ (4) & 6 \\
$\beta$ Pic & 85.3 $\substack{+0.3 \\ -0.2}$\ \ (5) & 30.4 $\substack{+0.1 \\ -0.1}$\ \ (5) & FPM & 1.6 & 23.6$\substack{+0.9 \\ -0.6}$\ \ (5) & 63.2$\substack{+2.9 \\ -3.1}$\ \ (6) & 23.6$\substack{+0.9 \\ -0.6}$\ \ (5) & 106.3$\substack{+1.4 \\ -1.3}$\ \ (6) & 14 \\
CE Ant & 13.1 $\substack{+3.0 \\ -3.0}$\ \ (7) & 91.0 $\substack{+9.0 \\ -9.0}$\ \ (7) & 0.6 & 1.1 & 21.8$\substack{+1.9 \\ -1.9}$\ \ (7) & $\leq60.0$\ \ (8) & 29.8$\substack{+1.6 \\ -1.3}$\ \ (7) & $\leq100.0$\ \ (8) & 3 \\
HD 15115$^{\mathrm{a}}$ & 86.5 $\substack{+0.5 \\ -0.5}$\ \ (9) & 278.9 $\substack{+0.1 \\ -0.1}$\ \ (9) & 0.3 & 1.3 & 64.0$\substack{+16.0 \\ -3.0}$\ \ (9) & 43.9$\substack{+5.8 \\ -5.8}$\ \ (10) & 98.0$\substack{+2.5 \\ -2.5}$\ \ (9) & 65.7$\substack{+4.5 \\ -4.5}$\ \ (10) & 34 \\
HD 30447 & 83.0 $\substack{+6.0 \\ -6.0}$ & 212.3 $\substack{+5.0 \\ -5.0}$ & 0.3 & 1.2 & $\leq103.0$ & $\mbox{--}$ & 83.0$\substack{+20.0 \\ -20.0}$ & $\mbox{--}$ & 30 \\
HD 32297 & 88.4 $\substack{+0.3 \\ -0.3}$\ \ (11) & 47.9 $\substack{+0.2 \\ -0.2}$\ \ (11) & FPM & 1.6 & 77.5$\substack{+4.0 \\ -4.0}$\ \ (11) & 78.5$\substack{+8.1 \\ -8.1}$\ \ (12) & 98.4$\substack{+0.3 \\ -0.3}$\ \ (11) & 122.0$\substack{+3.0 \\ -3.0}$\ \ (12) & 25 \\
HD 35841 & 84.9 $\substack{+0.2 \\ -0.2}$\ \ (13) & 165.8 $\substack{+0.2 \\ -0.2}$\ \ (13) & FPM & 0.7 & 60.3$\substack{+1.1 \\ -2.2}$\ \ (13) & $\mbox{--}$ & 60.3$\substack{+1.1 \\ -2.2}$\ \ (13) & $\mbox{--}$ & 26 \\
HD 61005 & 84.3 $\substack{+0.3 \\ -0.3}$\ \ (14) & 70.7 $\substack{+0.8 \\ -0.8}$\ \ (14) & FPM & 1.5 & 53.0$\substack{+11.0 \\ -11.0}$\ \ (14) & 41.9$\substack{+0.9 \\ -0.9}$\ \ (12) & 48.0$\substack{+26.0 \\ -16.0}$\ \ (14) & 67.0$\substack{+0.5 \\ -0.5}$\ \ (12) & 12 \\
HD 100546$^{\dagger}$ & 41.9 $\substack{+0.0 \\ -0.0}$\ \ (15) & 325.1 $\substack{+0.0 \\ -0.0}$\ \ (15) & FPM & 1.2 & 11.0$\substack{+1.0 \\ -1.0}$\ \ (16) & 23.0$\substack{+0.0 \\ -0.0}$\ \ (17) & 17.0$\substack{+1.0 \\ -1.0}$\ \ (16) & 33.0$\substack{+0.0 \\ -0.0}$\ \ (17) & 9 \\
HD 106906 & 84.6 $\substack{+0.4 \\ -0.4}$\ \ (18) & 284.2 $\substack{+0.2 \\ -0.2}$\ \ (18) & FPM & 1.1 & 66.6$\substack{+3.7 \\ -2.8}$\ \ (19) & $\mbox{--}$ & 72.3$\substack{+3.7 \\ -2.8}$\ \ (19) & $\mbox{--}$ & 44 \\
HD 110058 & $\geq 84.$ & 155.0 $\substack{+1.0 \\ -1.0}$\ \ (20) & FPM & 0.6 & $\leq39.0$\ \ (20) & $\leq36.0$\ \ (21) & 39.0$\substack{+6.0 \\ -6.0}$\ \ (20) & $\leq36.0$\ \ (21) & 16 \\
HD 111161 & 62.1 $\substack{+0.3 \\ -0.2}$ & 83.2 $\substack{+0.5 \\ -0.6}$ & 0.5 & 0.9 & 71.4$\substack{+0.5 \\ -1.0}$ & $\mbox{--}$ & 72.4$\substack{+1.7 \\ -1.6}$ & $\mbox{--}$ & 44 \\
HD 111520 & $\geq 88.$\ \ (22) & 165.0 $\substack{+1.0 \\ -1.0}$\ \ (22) & FPM & 1.1 & $\leq71.0$\ \ (22) & 45.0$\substack{+15.0 \\ -15.0}$\ \ (21) & 81.0$\substack{+10.0 \\ -10.0}$\ \ (22) & 45.0$\substack{+15.0 \\ -15.0}$\ \ (21) & 9 \\
HD 114082 & 83.3 $\substack{+0.4 \\ -3.8}$\ \ (23) & 105.7 $\substack{+1.5 \\ -0.5}$\ \ (23) & FPM & 0.7 & 28.7$\substack{+2.9 \\ -3.7}$\ \ (23) & $\mbox{--}$ & 30.7$\substack{+4.4 \\ -3.7}$\ \ (23) & $\mbox{--}$ & 14 \\
HD 115600 & 80.0 $\substack{+1.0 \\ -1.0}$\ \ (24) & 27.5 $\substack{+1.1 \\ -1.1}$\ \ (24) & 0.2 & 0.5 & 39.0$\substack{+4.0 \\ -4.0}$\ \ (24) & $\mbox{--}$ & 46.0$\substack{+2.0 \\ -2.0}$\ \ (24) & $\mbox{--}$ & 17 \\
HD 117214 & 71.0 $\substack{+1.1 \\ -0.4}$ & 179.8 $\substack{+0.2 \\ -0.2}$ & FPM & 0.9 & 20.8$\substack{+12.4 \\ -0.6}$ & $\mbox{--}$ & 60.2$\substack{+0.6 \\ -1.1}$ & $\mbox{--}$ & 19 \\
HD 129590 & 75.7 $\substack{+1.2 \\ -1.2}$\ \ (25) & 121.7 $\substack{+0.0 \\ -0.0}$\ \ (25) & FPM & 1.5 & 47.4$\substack{+5.5 \\ -5.5}$\ \ (25) & $\leq41.0$\ \ (21) & 66.9$\substack{+4.0 \\ -4.0}$\ \ (25) & $\leq41.0$\ \ (21) & 17 \\
HD 131835 & 75.1 $\substack{+0.9 \\ -0.9}$\ \ (26) & 61.4 $\substack{+0.4 \\ -0.4}$\ \ (26) & 0.4 & 1.0 & 75.0$\substack{+2.0 \\ -4.0}$\ \ (26) & 26.0$\substack{+12.0 \\ -12.0}$\ \ (21) & 107.7$\substack{+1.4 \\ -1.3}$\ \ (26) & 26.0$\substack{+12.0 \\ -12.0}$\ \ (21) & 25 \\
HD 141569$^{\dagger, \mathrm{b}}$ & 60.0 $\substack{+10.0 \\ -10.0}$\ \ (28) & 5.0 $\substack{+10.0 \\ -10.0}$\ \ (28) & 0.2 & 0.8 & 20.0$\substack{+10.0 \\ -10.0}$\ \ (28) & 16.0$\substack{+18.0 \\ -15.0}$\ \ (29) & 51.0$\substack{+8.0 \\ -12.0}$\ \ (28) & 45.0$\substack{+7.0 \\ -6.0}$\ \ (29) & 34 \\
HD 143675 & 87.2 $\substack{+0.6 \\ -0.7}$ & 113.2 $\substack{+0.5 \\ -0.4}$ & FPM & 0.4 & 44.0$\substack{+3.5 \\ -7.6}$ & $\mbox{--}$ & 48.1$\substack{+5.4 \\ -11.7}$ & $\mbox{--}$ & 7 \\
HD 145560 & 43.9 $\substack{+1.5 \\ -1.4}$ & 41.5 $\substack{+1.0 \\ -1.2}$ & 0.3 & 0.9 & 68.6$\substack{+2.9 \\ -1.3}$ & 50.0$\substack{+10.0 \\ -8.0}$\ \ (21) & 85.3$\substack{+1.3 \\ -1.2}$ & 50.0$\substack{+10.0 \\ -8.0}$\ \ (21) & 10 \\
HD 146897 & 84.0 $\substack{+3.0 \\ -3.0}$\ \ (30) & 113.9 $\substack{+2.2 \\ -2.2}$\ \ (31) & FPM & 1.3 & 67.0$\substack{+19.0 \\ -18.0}$\ \ (32) & 64.0$\substack{+12.0 \\ -14.0}$\ \ (21) & 85.0$\substack{+17.0 \\ -17.0}$\ \ (32) & 64.0$\substack{+12.0 \\ -14.0}$\ \ (21) & 13 \\
HD 156623 & 34.9 $\substack{+3.6 \\ -9.5}$ & 100.9 $\substack{+1.9 \\ -2.2}$ & FPM & 0.8 & $\leq26.7$ & $\leq20.0$\ \ (21) & 80.2$\substack{+2.3 \\ -2.2}$ & $\leq20.0$\ \ (21) & 18 \\
HD 157587 & 70.0 $\substack{+2.2 \\ -2.2}$\ \ (33) & 127.0 $\substack{+0.3 \\ -0.3}$\ \ (33) & 0.2 & 0.7 & 79.0$\substack{+1.0 \\ -1.0}$\ \ (33) & $\mbox{--}$ & 79.0$\substack{+1.0 \\ -1.0}$\ \ (33) & $\mbox{--}$ & 25 \\
HD 191089 & 59.0 $\substack{+4.0 \\ -2.0}$\ \ (34) & 70.0 $\substack{+4.0 \\ -3.0}$\ \ (34) & 0.7 & 1.1 & 26.0$\substack{+4.0 \\ -4.0}$\ \ (34) & $\mbox{--}$ & 43.9$\substack{+0.3 \\ -0.3}$\ \ (34) & $\mbox{--}$ & 13 \\
HR 4796 A & 76.5 $\substack{+0.1 \\ -0.1}$\ \ (35) & 26.1 $\substack{+0.1 \\ -0.1}$\ \ (35) & 0.2 & 1.3 & 74.4$\substack{+0.6 \\ -0.6}$\ \ (35) & 75.4$\substack{+1.0 \\ -1.0}$\ \ (36) & 78.5$\substack{+0.7 \\ -0.7}$\ \ (35) & 78.5$\substack{+0.2 \\ -0.2}$\ \ (36) & 27 \\
HR 7012 & 76.2 $\substack{+1.7 \\ -1.7}$\ \ (37) & 112.3 $\substack{+1.5 \\ -1.5}$\ \ (37) & FPM & 0.3 & 8.0$\substack{+2.0 \\ -1.9}$\ \ (37) & $\mbox{--}$ & 10.8$\substack{+1.6 \\ -1.6}$\ \ (37) & $\mbox{--}$ & 3 \\
NZ Lup$^{\mathrm{c}}$ & 87.0 $\substack{+1.0 \\ -1.0}$\ \ (38) & 146.5 $\substack{+0.1 \\ -0.1}$\ \ (38) & 0.3 & 1.4 & 73.0$\substack{+2.0 \\ -2.0}$\ \ (38) & $\mbox{--}$ & 103.0$\substack{+17.0 \\ -17.0}$\ \ (38) & $\mbox{--}$ & 11 \\

\enddata
%\vspace{-0.8cm} 
\tablecomments{Disk properties were measured from a combination of GPIES and non-GPIES data; where multiple sets of comparable measurements existed, we chose those made from GPIES data. Values with no parenthetical reference were measured in this work from GPIES data. $\dagger$ denotes protoplanetary/transitional disks that are excluded from most analyses. Column descriptions from left to right: target name, inclination with 1$\sigma$ uncertainties, PA with 1$\sigma$ uncertainties (see Section \ref{tab:mcmc_params} for our PA convention), minimum projected separation in GPIES scattered light, maximum projected separation in GPIES scattered light, scattered-light inner disk radius, thermal emission inner disk radius, scattered-light peak dust density radius, thermal emission peak dust density radius, and SED-inferred blackbody dust radius. Disks with $r_\mathrm{min}=$ ``FPM'' are detected down to the FPM edge at $0\farcs123$. Special cases:\\
$^{\mathrm{a}}$ HD 15115 inner radii are for a presumed inner belt and maximum density radii are for an outer belt.\\
$^{\mathrm{b}}$ HD 141569 radii are for the inner ring only (as seen in the GPI images) and the millimeter uncertainties are ${\sim}95$\% confidence intervals.\\
$^{\mathrm{c}}$ NZ Lup's $R_{\mathrm{in}}$ is for the inner belt only and $R_0$ is the mean of both belts (with the uncertainty spanning the range of their individual $R_0$ values) in the two-belt ``gap'' model from \citet{boccaletti2019}.}

\tablerefs{(1) \citet{janson2016}, (2) \citet{krist2005}, (3) nominal value is from \citet{boccaletti2018} and uncertainties encompass range of values from \citet{augereau2006}, \citet{schuppler2015}, \citet{sezestre2017}, (4) mean of two models (with positive surface density slopes) by \citet{daley2019} and uncertainties encompass a possible inner ring at ${\sim}10$--14 au, (5) \citet{millar-blanchaer2015}, (6) \citet{matra2019b}, (7) \citet{olofsson2018}, (8) \citet{matra2019a}, (9) \citet{engler2019}, (10) \citet{macgregor2019}, (11) \citet{duchene2020}, (12) \citet{macgregor2018b}, (13) \citet{esposito2018}, (14) \citet{esposito2016}, (15) \citet{pineda2014}, (16) \citet{garufi2016}, (17) \citet{walsh2014}, (18) \citet{kalas2015}, (19) \citet{lagrange2016}, (20) \citet{kasper2015}, (21) \citet{lieman-sifry2016}, (22) \citet{draper2016}, (23) \citet{wahhaj2016}, (24) \citet{gibbs2019}, (25) mean of two models by \citet{matthews2017}, (26) \citet{hung2015b}, (27) \citet{feldt2017}, (28) \citet{bruzzone2019}, (29) \citet{white2018b}, (30) \citet{thalmann2013}, (31) Wolff et al. in prep, (32) \citet{engler2017}, (33) \citet{millar-blanchaer2016c}, (34) \citet{ren2019}, (35) \citet{perrin2015_4796}, (36) \citet{kennedy2018}, (37) \citet{engler2018}, (38) \citet{boccaletti2019}.}

\end{deluxetable*}

\subsubsection{Disk Radii} \label{sect:radius_results}

The GPIES-detected disks span a range of radial extents. When comparing them we choose to focus on the radius $R_0$ corresponding to the peak of the dust surface density profile as determined from scattered-light imaging. As the radius where dust is most highly concentrated, this is a more physically relevant parameter than the inner or outer radii, which are dependent on the sensitivity of the observations. For most disks, $R_0$ is derived from a model employing either a single or broken power-law distribution to describe the dust density profile that is then fit to the scattered-light image, such as the modeling described by Equation \ref{eq:disk_density} in Section \ref{sect:morph_mcmc}. Several special cases are worth noting. For NZ Lup, we aim to represent the entire disk by taking $R_0$ as the mean of both belts (with individual $R_0$ values of 86 and 121 au) in the two-belt ``gap'' model by \citet{boccaletti2019}. For HD~143675, our MCMC only determined a lower limit for the dust density critical radius $r_c$, so we assumed $R_0$ to be the average of the better constrained $r_{\mathrm{in}}$ and $r_{\mathrm{out}}$. In a handful of cases (e.g., HD~35841; \citealt{esposito2018}) where $r_c \leq r_{\mathrm{in}}$, effectively enforcing a single density power-law index for the entire disk, we set $R_0=r_{\mathrm{in}}$. For the remaining disks, $R_0$ was estimated from measurements of the peak surface brightness: these are HD~15115 (for only the outer belt from \citealt{engler2019}), HD~30447 (this work), HD~110058 \citep{kasper2015}, and HD~111520 \citep{draper2016}. It should be kept in mind that projection effects and inaccurately estimated scattering phase functions are potential sources of uncertainty in all scattered-light radius measurements and may not be fully accounted for in the quoted errors.

As a secondary measurement for comparison, we used an inner radius of the dust distribution that we label $R_{\mathrm{in}}$ (distinct from, but sometimes equivalent to, the model-specific parameter $r_{\mathrm{in}}$). Given that planets (if present) are typically expected to orbit interior to the large-radius dust rings (i.e. Kuiper Belt analogs) that GPI detects, knowing the inner radius of the dust is particularly interesting from a planetary system standpoint. The radius of the outer edge is interesting in other respects (e.g., total dust mass, grain orbit eccentricities, influence of exterior planets) but the large radii involved (${>}1000$ au for some disks; e.g., \citealt{schneider2014}) and faint, smooth surface brightness distributions of the disks' outer regions (essentially requiring \textit{HST} imaging) make the outer radius a poor parameter choice for our GPI-based analysis. For each disk, we adopt a value for $R_{\mathrm{in}}$ from one of three sources. Our first preference is to adopt the inner radius where dust density rapidly falls off toward zero in a disk model, such as $r_{\mathrm{in}}$ from MCFOST models (Section \ref{sect:morph_mcmc}). If no such value is available but the dust density has been modeled with a broken power-law function similar to Equation (\ref{eq:disk_density}), then we estimate $R_{\mathrm{in}}$ to be the (inner) radius where the dust surface density is half of the maximum (which occurs at $R_0$). As a last option for disks that have not been modeled, we take the deprojected radius at which the observed scattered-light surface brightness reaches the level of the local background noise, i.e., $\mathrm{S/N}\approx1$. In cases where multiple belts are present or suspected, we use the inner radius of the innermost belt that is resolved in scattered light (see footnotes of Table \ref{tab:measures}).

% Scattered-light inner radii vs stellar luminosity.
\begin{figure}[h!]
\centering
\includegraphics[width=\columnwidth]{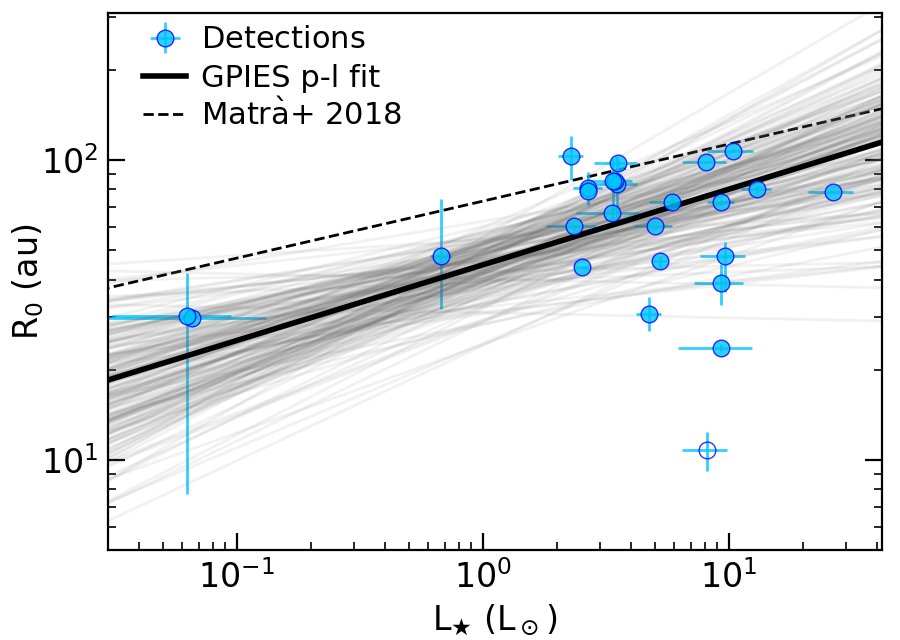}
\caption{Scattered-light radii for the peak dust surface density of GPIES debris disks versus stellar luminosity (excluding the three protoplanetary/transitional disks). Our best-fit power-law function is in solid black and 200 samples drawn from the corresponding normal distributions are shown in gray. We excluded HR~7012 (unfilled circle at $R_0=11$ au) from the fit as an outlier (see text for details). The radius--luminosity power law for planetesimal belt central radii from \citet{matra2018} is shown for comparison (dashed line); it is offset upwards because they define their radii differently and we consider different samples.}
\label{fig:radii_luminosity}
\end{figure}

We find that $R_0$ ranges from 11 au to 98 au across our debris disk detections with a mean of 64 au (Table \ref{tab:measures}) that is slightly larger than the Kuiper Belt (${\sim}30$--50 au; \citealt{levison2008}). Those $R_0$ measurements are plotted against $L_\star$ in Figure \ref{fig:radii_luminosity}. We tested for a trend in $R_0$ with $L_\star$, although this approach is limited by substantial scatter in disk radii among stars with $L_\star = 2$--14 $L_\odot$ and only four measurements outside of that range. We first fit a power-law function to our data, assuming no uncertainties on $L_\star$ and excluding HR~7012's disk ($R_0\approx11$ au) as a phenomenological outlier because it may have been gravitationally disturbed by interaction with its nearby K-type companion, as discussed earlier. This fit returns a power-law index of $0.25\pm0.09$, which is statistically distinguishable from a zero slope at the $2.8\sigma$ level. To assess whether a positive power-law actually describes the data better than a flat line, we compared the Bayesian information criterion (BIC) for the best-fit power-law function to that of the best-fit flat line (not shown). The BIC accounts for differences in the number of degrees of freedom between models when considering goodness of fit. The resulting $\Delta \mathrm{BIC}$ = 3.6 (power-law $-$ flat line) is an argument slightly in favor of the flat-line model, which returns larger residuals but has only one free parameter.

Consequently, we find marginal evidence of a scattered-light trend similar to that reported by \citet{matra2018} for planetesimal belt radii derived from thermal imaging. They found a power-law index of $0.19\pm0.04$ for ring radius as a function of $L_\star$ (dashed line in Figure \ref{fig:radii_luminosity}, using their posterior 50th percentile values), which is consistent with our scattered-light index within $1\sigma$. We caution that this is not a perfectly fair comparison because \citet{matra2018} examine a different disk sample that only partially overlaps ours and they also define their radii differently---either as the average between $r_{\mathrm{in}}$ and $r_{\mathrm{out}}$ (for models with power-law radial surface density distributions and sharp cutoffs) or the best-fit centroid of a Gaussian surface density distribution. On the other hand, the statistical possibility that there is no correlation between $L_\star$ and scattered-light $R_0$ would be consistent with the \citet{pawellek2014} finding of no significant correlation between $L_\star$ and \textit{Herschel} PACS-resolved disk radii (based on a power-law index of $0.04 \pm 0.04$). We note that they used yet another sample and radius definition --- the central radius of a narrow ring model convolved with the PACS PSF and fit to a 100 $\micron$ image using a grid search, also making this an imperfect comparison. Future attempts to make concrete statements about a scattered-light trend would benefit from folding in additional measurements beyond GPIES detections, especially for disks with host $L_\star<2~L_\odot$.

Looking at individual disks more closely, three appear to have $R_0$ that are notably smaller than those of other disks with similar host $L_\star$. One such disk is HR~7012, which we already noted may have been truncated by interaction with a nearby K-type companion. The two other disks, $\beta$~Pic (24~au) and HD~114082 (31~au), are nearly edge on, so it is possible their radii are underestimated due to the unfavorable geometric projection. If we assume, however, that the disk radii and stellar properties we are using are correct, then the range of $R_0$ from 24 to 98 au for disks with host star $2~L_\odot \lesssim L_\star \lesssim 10~L_\odot$ suggests there is intrinsic scatter of at least a factor of 4 in the radial location of small grains around luminous stars. This scatter increases to a factor of almost 9 if we include HR~7012.

% Scattered-light peak radius v SED-inferred radius
\begin{figure}[h!]
\centering
\includegraphics[width=\columnwidth]{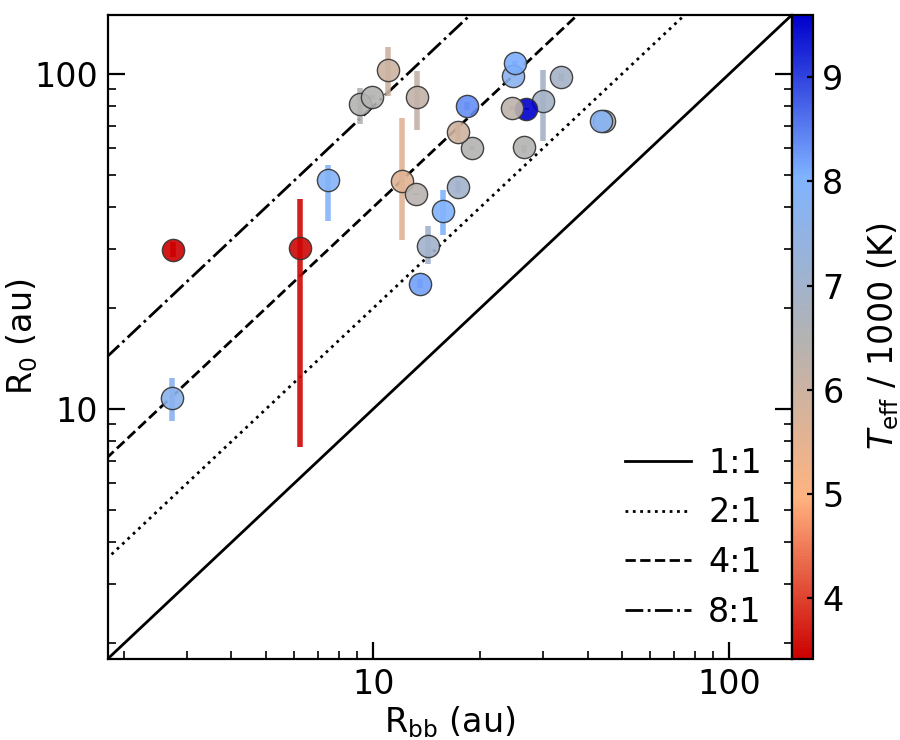}
\caption{Peak radii of the scattered-light dust surface density for the GPIES debris disk detections compared with their blackbody dust radii inferred from SED fits. Lines mark reference ratios. Points are colored according to the host star's effective temperature. The scattered-light radius is larger than the blackbody radius for all disks, with the average ratio being 4.33:1.}
\label{fig:radii}
\end{figure}

We also have a calculation of each disk's blackbody dust radius from an SED fit ($R_{\mathrm{bb}}$), which primarily traces the disk's thermal emission. For these fits, we assumed single-component models composed of blackbody grains. In Figure \ref{fig:radii}, we compare $R_0$ to $R_{\mathrm{bb}}$ and find that $R_0$ is an average of 4.33 times larger. All GPIES disks have $R_0$:$R_{\mathrm{bb}} > 1.5$, and the largest ratio is ${\sim}11$:1. This is consistent with similar comparisons made by \citet{rodriguez2012} and \citet{cotten2016} and shows that using blackbody grain models to fit SEDs consistently underestimates the radial location of small dust in debris disks. The scattered-light relationship is analogous to that seen in thermal emission by \citet{booth2013} and \citet{morales2016}, who found resolved radii from far-IR \textit{Herschel} imaging to be 1--2.5 times larger than modified blackbody radii (also noted by \citealt{rodriguez2012}). \citet{pawellek2014} and \citet{pawellek2015} used \textit{Herschel} data to provide even more evidence that this ratio of thermally resolved to blackbody radius is consistently greater than unity for debris disks, with the latter study demonstrating that this trend holds for the scattered-light radius of a few disks as well. Those two studies also showed that these radius ratios decrease with $L_\star$, which is hinted at in Figure \ref{fig:radii} by a weak positive gradient in $T_{\mathrm{eff}}$ (as proxy for $L_\star$) as one moves toward lower $R_0$:$R_{\mathrm{bb}}$ ratios; however, we are again limited by our detected sample's relatively narrow range of $L_\star$ so we do not investigate this further. Overall, our findings imply that the integrated photometry of the SED primarily traces dust interior to the cold outer belts we observe with GPI and/or that the particles in those outer belts are emitting less efficiently at IR and (sub)millimeter wavelengths, hence are warmer, than considered in the blackbody models (e.g., \citealt{zuckerman2001} and references therein).

\begin{figure}[h!]
\centering
\includegraphics[width=\columnwidth]{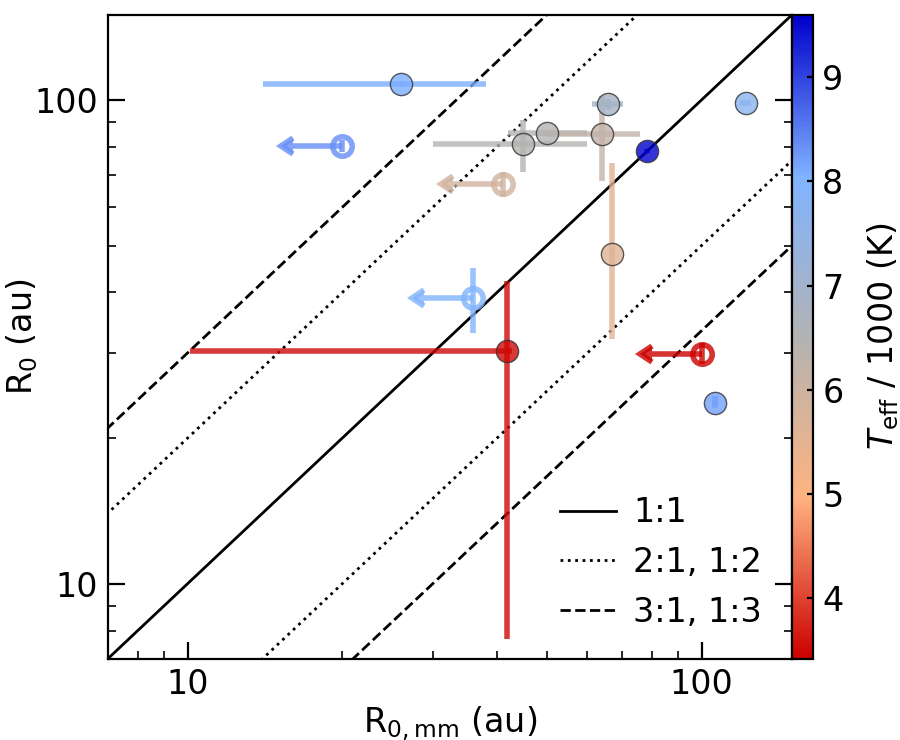}
\caption{Peak radii of the scattered-light dust surface density for the GPIES debris disk detections compared with peak surface density radii from thermal millimeter observations. Lines mark reference ratios. Points are colored according to the host star's effective temperature and unfilled circles denote upper limits on one or both axes. The two radii are similar for many disks but the average ratio of 1.39:1 suggests a slightly larger scattered-light radius.}
\label{fig:peak_radii_th}
\end{figure}

\begin{figure}[h!]
\centering
\includegraphics[width=\columnwidth]{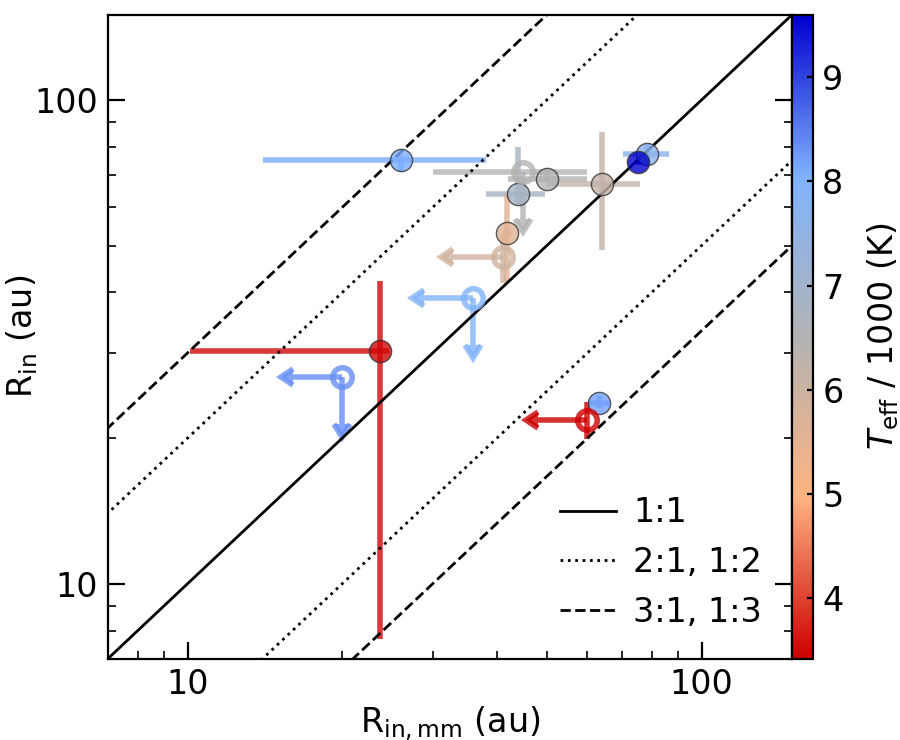}
\caption{Same as Figure \ref{fig:peak_radii_th} but for the inner radius of the disk. The average ratio of scattered-light to thermal radius is 1.29:1 in this case.}
\label{fig:radii_th}
\end{figure}

We can also compare $R_0$ to the analogous surface density peak radius derived from resolved thermal emission ($R_{\mathrm{0,mm}}$, defined the same way as for scattered light) in published literature, shown in Figure \ref{fig:peak_radii_th}. In this case, we have fewer data points because only about half of our detected sample have resolved thermal images with measured radii. We find the scattered-light radius to be 1.39 times larger on average than the thermal radius, with only the HD~32297 and $\beta$~Pic disks having $R_0$:$R_{\mathrm{0,mm}}$ significantly less than unity. Additionally, comparing the inner radii of $R_{\mathrm{in}}$ and $R_{\mathrm{in, mm}}$ in Figure \ref{fig:radii_th} gives a similar result with an average scattered-light:thermal ratio of 1.29:1 and only $\beta$~Pic with a ratio less than unity. These results indicate that the orbits of small grains are either the same as or slightly wider than those of large grains that are presumed to be in a planetesimal belt. Thus, it is likely that the small grains are either born from the large grains or both are born cospatially from even larger progenitor bodies. Once created, the small grains would be preferentially pushed onto elliptical orbits by stellar radiation pressure (and ejected entirely if smaller than the blowout size), thus ending up at larger average stellocentric radii than the large grains \citep{strubbe2006}.

% Scattered-light peak radius v thermal peak radius

% Scattered-light inner radius v thermal inner radius

Regarding the two outlier disks, HD~32297's ratio of $R_0$:$R_{\mathrm{0,mm}}=0.8$ runs counter to the broader trend, but its ratio of $R_{\mathrm{in}}$:$R_{\mathrm{in,mm}}=1.0$ is similar to that of HR~4796 and HD~146897. Given the previously mentioned difficulties in accurately measuring the scattered-light radius of an edge-on disk like this one, we defer interpretation of this disk's scattered light and thermal emission differences to a point in time when its morphology is more definitively established.

The $\beta$ Pic disk is a more significant outlier, with peak (0.2) and inner radius (0.4) ratios firmly less than unity. Again, this could be a result of its edge-on orientation leading to underestimated scattered-light radii. Assuming the measured ratios are correct, however, then perhaps the primary source of the scattered light is spatially distinct from the source of the millimeter emission; \citet{wahhaj2003} and \citep{okamoto2004} presented evidence of multiple planetesimal belts within the disk. On the other hand, this is already a complex system, hosting the  $11\pm2$ $M_\mathrm{J}$ \citep{snellen2018} planet $\beta$~Pic~b \citep{lagrange2010} on an inclined, eccentric orbit with a semimajor axis of 9 au \citep{bonnefoy2014, wang2016, lagrange2019} that has been shown to influence the warped inner disk \citep{mouillet1997, augereau2001, dawson2011, nesvold2015}, plus recently announced evidence for a second planet ($\beta$~Pic~c) with mass ${\sim}9$ $M_\mathrm{J}$ and semimajor axis of ${\sim}2.7$ au with eccentricity of 0.24 \citep{lagrange2019b}. Collisional models by \citet{nesvold2015} showed that secular perturbations from $\beta$~Pic~b will excite collisions between planetesimals in the disk and clear large particles from the region interior to ${\sim}59$ au but also produce a significant amount of micron-sized dust in the same region. Consequently, submillimeter brightness is depressed inside of 59 au while scattered-light brightness is not. If this scenario is correct, then the values that we find for the radius ratios would be the natural result of the planet stirring the $\beta$ Pic disk. The fact that we find these ratios to be so much less than unity for only this one disk invites the question of whether the other disks do not have massive planets orbiting inside their inner holes or their holes are created by a different mechanism. It may be the case that planets in those systems are simply less eccentric than the $e=0.08$ assumed for $\beta$~Pic~b in those models; \citet{nesvold2015} found no reduction in submillimeter surface brightness at $r<59$ au when the planet's orbit was circular. Alternatively, the relatively narrow rings with radius ratios $ > 1$ and dust-poor inner regions may result from dust--gas interaction and not be directly related to planets (e.g., \citealt{lyra2013}). More observed examples of planets orbiting inside debris disks (or a continued lack thereof) that have measured radii at scattered-light and millimeter wavelengths would help answer this question. The HR~2562 and HD~206893 systems containing brown dwarfs \citep{konopacky2016, milli2017a} are promising candidates but need better characterization of their disks (both were GPIES disk non-detections).

\subsubsection{Effects of Inclination and IR Excess on Detection}
\label{sect:inc_tau_effects}

An open question in high-contrast imaging is why some debris disks, known to exist from thermal IR and millimeter data, are not detected in scattered light while others are. There are many factors that lead to differences in scattered-light signals, making this a difficult question to answer. With the GPIES sample, we can directly examine the effects of two of these factors on disk detectability: line-of-sight inclination and IR excess.

\begin{figure}[h!]
\centering
\includegraphics[width=\columnwidth]{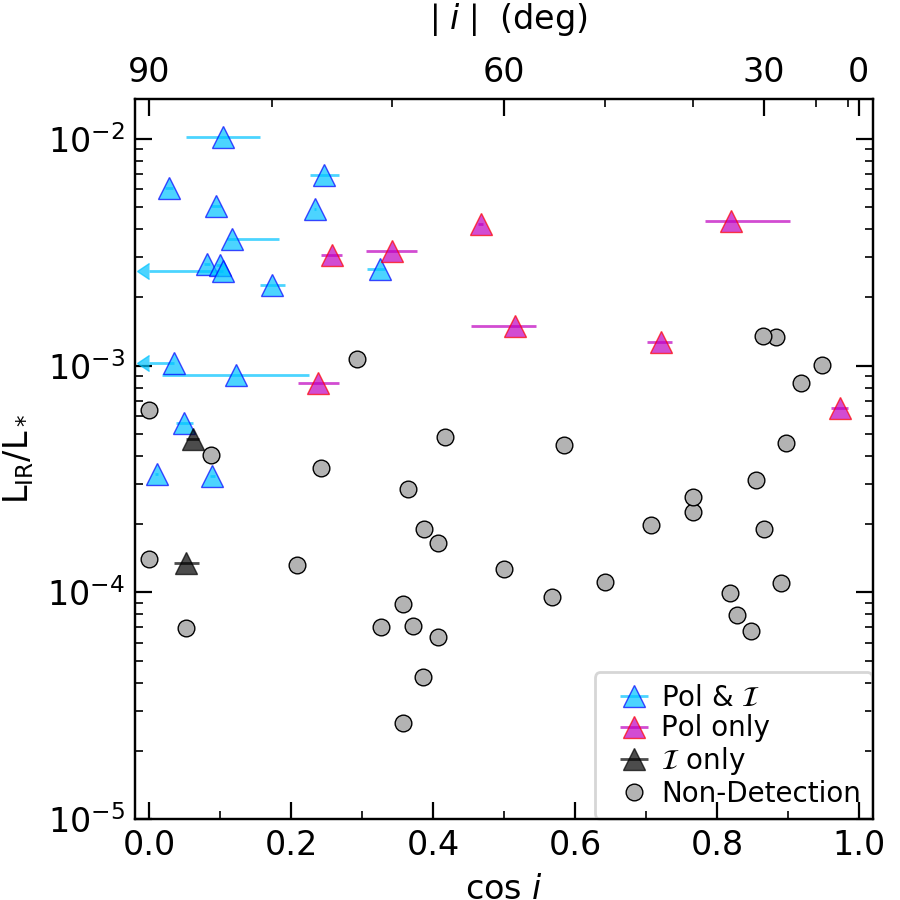}
\caption{IR excess versus resolved disk inclination for all GPIES debris disks (triangles) and those non-detections (circles) for which inclinations were available in the literature. Detections are divided into polarized intensity only, total intensity only, and both. Randomly oriented disks should be uniformly distributed in $\cos{i}$ but GPIES detections, particularly in total intensity, are biased toward higher inclination.}
\label{fig:inc_vs_tau}
\end{figure}

In Figure \ref{fig:inc_vs_tau} we show GPIES detection status as a function of inclination and IR-excess magnitude. All \Ndebris of the GPIES debris disk detections have inclination measurements and appear on the plot; however, only the 36 non-detections (out of \Nnondet total) with inclinations from the literature are plotted. The correlation between detections and \Lir that was highlighted in Section \ref{sect:detection_overview} is seen again here. There are, however, additional detectability trends with disk inclination for us to explore.

% Inclination vs Lir/L*

% Inclination histogram.
\begin{figure}[h!]
\centering
\includegraphics[width=\columnwidth]{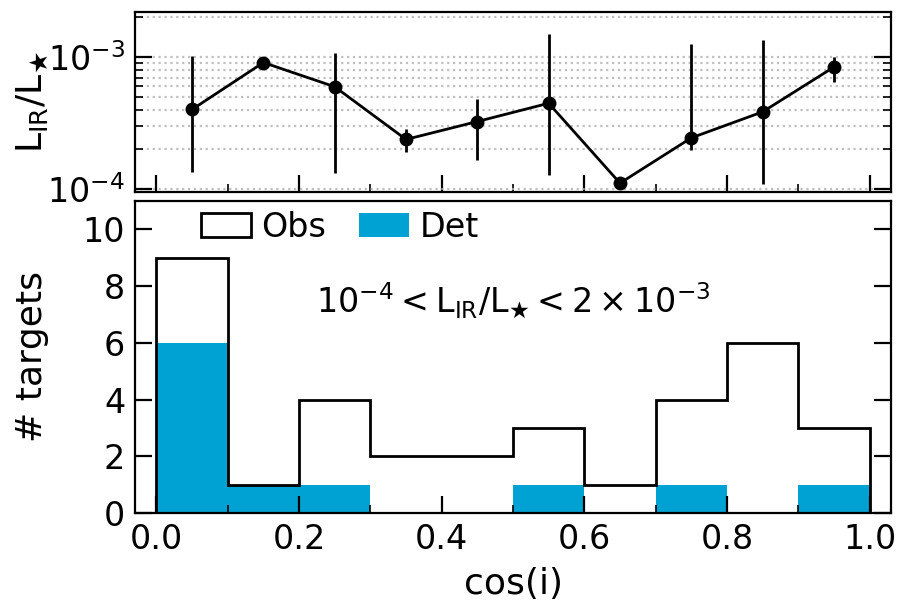}
\caption{(Bottom) Number of debris disk targets observed (black outline) and scattered-light disks detected (blue filled) as a function of $\cos{i}$ for targets with \Lir between $10^{-4}$ and $\num{2e-3}$. (Top) The median \Lir of observed targets in each $\cos{i}$ bin, with vertical lines marking the minimum and maximum values in the bin.}
\label{fig:inc_hist}
\end{figure}

The distribution of detections in Figure \ref{fig:inc_vs_tau} shows a clear trend of higher inclinations being more detectable than low inclinations. This is amplified in Figure \ref{fig:inc_hist} by a histogram of the nominal $\cos{i}$ for the observed and detected targets. To make the sample more uniform in \Lir and reduce its impact on the detection rates, we narrowed our examination to targets with \Lir between $10^{-4}$ and $\num{2e-3}$. In this subsample, the majority of GPIES detections have $i\geq72\degr$ ($\cos i\leq0.3$) despite the majority of observed disks having $i<72\degr$. To quantify the difference between the observed and detected distributions, we can use the Kolmogorov--Smirnov (K-S) test to compare their cumulative distribution functions (CDFs); the higher the resulting K-S statistic\footnote{Defined as the maximum absolute difference at any point between the two CDFs.}, the more dissimilar the distributions. In this case, the K-S statistic of 0.38 rejects the null hypothesis that the two distributions are the same at a 91\% confidence level. The median target \Lir varies between $\cos{i}$ bins but does not show a consistent trend that indicates the high $i$ targets are intrinsically dustier than the low $i$ targets (Figure \ref{fig:inc_hist}, top). Thus, having partly controlled for \Lir, inclination still appears to be a factor in detection. Additionally, we do not think our sample was significantly biased by observing targets that were previously resolved in scattered light and that favored high $i$. For one thing, most, if not all, of those targets would have been observed in GPIES anyway based on their \Lir and detectability metric values. For another, if high inclination makes disks more detectable in scattered light, then it follows naturally that the previously resolved disks would favor those inclinations.

The preference for high inclinations becomes even more extreme when considering total intensity detections. In this same subsample, all seven total intensity detections have $i\gtrsim83\degr$. To the same point, the four disks with lower inclinations were detected only in polarized intensity. Overall, these results imply that higher inclination increases scattered-light detectability and does so more strongly in total intensity than polarized intensity. We consider the physical explanations and implications of this inclination effect in Section \ref{sect:discuss_detection_factors}.

\section{Discussion}

\subsection{Scattered-light Disk Detection and Comparison to Thermal Wavelengths} \label{sect:discuss_detection_factors}

The results of our survey provide new insight into the factors that affect scattered-light disk detection for both polarized and total intensity. In turn, those factors inform us about the physical properties of debris disks.

We found IR excess to be the most important target property for our scattered-light disk detections, assuming the disk resides within the GPI field of view. This sheds light on debris disk particle size distributions. A star's IR excess is primarily created by the thermal emission from particles with sizes in the tens of microns up to several millimeters. However, these particles contribute little to a disk's scattered-light emission, which is instead driven by dust ${\sim}0.1$--10 $\micron$ in size. The finding that IR excess, a mainly thermal property, also predicts nonthermal scattered-light brightness favors a strong coupling between the two regimes of particle size. This supports the premise of collisional cascades in debris disks as the source of dust; large particles are ground down into smaller particles, thus connecting the two populations (e.g., \citealt{dohnanyi1969, williams1994, kenyon2004a}). The IR-excess trend also means that, as typically assumed, debris disks are optically thin at both near-IR and millimeter wavelengths because increasing the disk mass, i.e., \Lir, increases the number of scattering/emitting particles and the brightness increases accordingly (assuming disk radius is held constant). This represents a step toward filling out the mass budget for these planetary systems, although doing so requires the nontrivial addition of realistic assumptions about the dust opacity.

We found high-angular-resolution millimeter-wavelength detections to be good predictors of scattered-light detections, and better than unresolved or marginally resolved far-IR detections. To test this, we examined the 14 observed Sco--Cen debris disk targets that overlapped between GPIES and the \citet{lieman-sifry2016} ALMA survey at 1.3 mm. Eleven disks were detected in both GPI \Qr and ALMA continuum (with ${\geq}3\sigma$ peak significance for the latter): HD numbers 110058, 111161, 111520, 114082, 115600, 117214, 129590, 131835, 145560, 146897, and 156623. One more disk, HD 106906, was detected clearly by GPIES but only marginally at $2.8\sigma$ by ALMA. Another two were detected by ALMA but not by GPIES (HD 138813 and HD 142315). Overall, this demonstrates that disks with enough large grains to produce a 1.3 mm ALMA detection in ${\sim}10$ minutes of integration generally have enough small grains to produce a GPI detection in ${\sim}10$--30 minutes. This is based primarily on observations of high-mass, high-luminosity F- and A-type stars and may not necessarily hold for lower luminosity stars that have dynamically colder disks and thus fewer small grains produced through a collisional cascade \citep{krijt2014, pawellek2015, thebault2016}. We also note that the median \Lir for this subsample is $\num{2.9e-3}$ and the ALMA imaging focused on separations ${\lesssim}2\arcsec$, making their targets particularly well suited for detection with GPI.

Unresolved and marginally resolved far-IR detections from \textit{Herschel} were moderately predictive of scattered-light detections but not as effective as the resolved millimeter imaging. For a fair comparison, we start from the Sco--Cen subsample just described and take only the 11 targets that have \textit{Herschel} PACS observations in Table \ref{tab:obs_targs}: HD numbers 106906, 110058, 111520, 114082, 115600, 117214, 131835, 138813 (a GPIES non-detection), 142315 (another GPIES non-detection), 145560, and 146897. Three of these disks were detected at ${\geq}3\sigma$ at 70~$\micron$, two of which GPIES also detected. Similarly, GPIES detected five of the seven disks detected at PACS 100~$\micron$ and four of the six detected at 160~$\micron$. Once we expand the subsample to include all 48 targets with \Lir$\geq10^{-4}$ that were observed by GPIES and detected with \textit{Herschel}, however, the percentages of GPIES detections drop to 27\%, 33\%, and 35\% among \textit{Herschel} detections at 70, 100, and 160~$\micron$, respectively. We expect this is because \textit{Herschel} PACS spans the wavelength range in which debris disk thermal emission typically peaks, so it remains more sensitive to disks with lower \Lir than scattered-light imagers working at shorter wavelengths and ALMA at longer wavelengths. Indeed, PACS has resolved disks with \Lir below $10^{-6}$ (e.g., \citealt{eiroa2013}). The broader sample may also contain more large-radius disks with inner holes that are unresolved by \textit{Herschel}'s ${\sim}5\arcsec$ beam FWHM (at 70~$\micron$) but extend beyond GPI's ${\sim}1\farcs8$ maximum observable radius.

Disk inclination and angular size also affect scattered-light detectability and likely account for some GPIES non-detections of disks detected at thermal wavelengths. Of the \textit{Herschel}-detected disks with \Lir $\geq10^{-4}$ that we observed, the median inclination of GPIES detected disks is $84\degr$ as opposed to $48\degr$ for GPIES non-detections. The effects of inclination are conflated with those of \Lir and other properties. That said, the difference does suggest that low inclinations hinder scattered-light detections of some disks with substantial IR excesses and far-IR detections. 

A key factor is that line-of-sight column density is directly related to inclination for optically thin debris disks. The disks are typically much wider radially than they are vertically, so viewing a disk at higher inclination means seeing more scattering particles within a given solid angle, thus increasing the observed surface brightness. Viewing the same dust distribution edge on versus face on could change the observed surface brightness by an order of magnitude. For example, a parametric model fit to the edge-on $\beta$~Pic disk is fainter by $\sim$2 mag arcsec$^{-2}$ (a factor of $\sim$6) when it is inclined from $i=90\degr$ to $i=30\degr$ \citep{kalas1996}. On a related morphological note, a low inclination exacerbates the issue of disk signal falling outside the GPI field of view by increasing the minimum projected separation of the dust ring from the star.

We know less about the roles played in detectability by the dust's scattering properties and the disk structure. In addition to the line-of-sight column density effect, the correlation between inclination and detectability may be partially due to disk phase functions that are primarily forward scattering. Such phase functions have been measured for several disks and are consistent with multiple sources of solar system dust \citep{hughes2018_review}. What is less clear is how much phase functions vary between disks and the effect on the surface brightness. The same is true for the polarizability of the dust, which is another property blended with the total intensity phase function. Other studies have shown a handful of disk polarization fractions to be ${<}10\%$ at small scattering angles and 30\%--50\% at $60\degr$--$90\degr$ scattering angles \citep{graham2007, maness2009, milli2015, esposito2016, esposito2018}. Substantial variations in phase function and polarizability beyond those observed at this point might explain some other scattered-light non-detections. They also could point to differences in dust size, composition, and structure between disks, something we explore briefly for two GPIES disks in Section \ref{sect:117214_v_145560}. Future measurements for more disks over a wide range of scattering angles are needed to make substantial progress.

In terms of disk structure, models often assume a single or broken power law for the dust density as a function of radius but there is almost certainly more complex structure in these disks on some size scale. As imaging angular resolution increases with technological advances, we expect to find that many disks that appear to be smooth and continuous now are actually full of gaps and clumps. This kind of progression has already been seen with protoplanetary disks; see HL Tau \citep{alma2015_hltau} and the DSHARP survey \citep{andrews2018_dsharp1} as prime examples via ALMA's high resolution and image fidelity. For debris disks, there are the recently identified moving clumps around AU~Mic \citep{boccaletti2015a, boccaletti2018, wisniewski2019} and new imaging suggesting that presumed single dust belts are actually double belts around HD~15115 \citep{engler2019, macgregor2019} and NZ~Lup \citep{boccaletti2019}. Future instruments, perhaps on 30 m class or large-diameter space telescopes, may further reveal how small-scale structure affects the scattered-light and thermal signatures of debris disks.

% CDF's for metric, Lir, etc.
\begin{figure*}[ht!]
%\centering
\includegraphics[width=\textwidth]{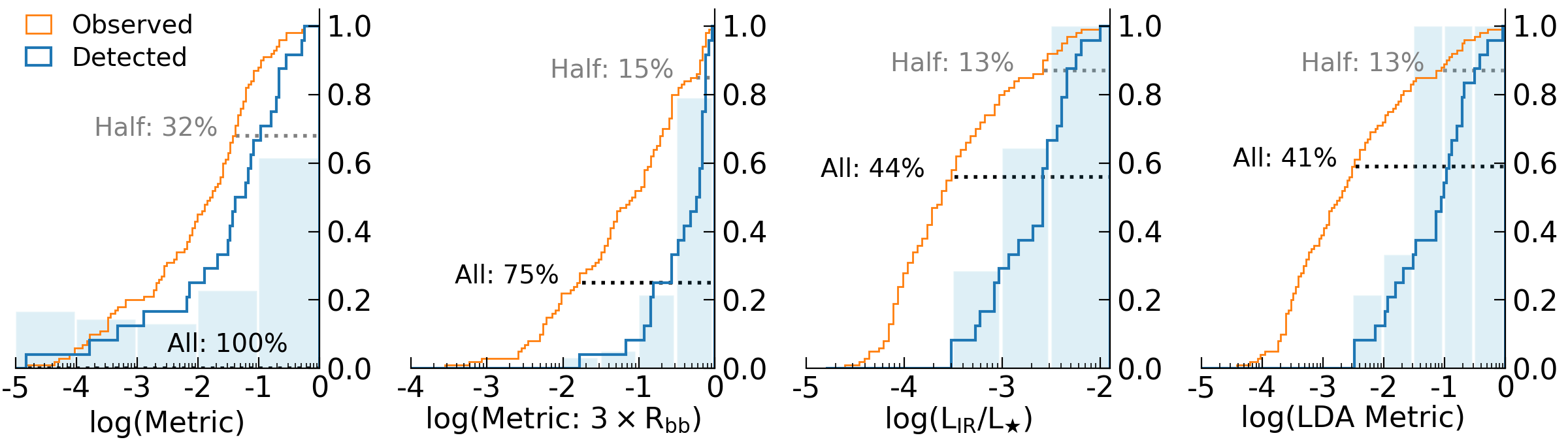}
\caption{Normalized cumulative distribution functions for the observed GPIES sample (orange) and the \Qr -detected debris disks (blue line) as a function of four target quantities (from left to right): the original detectability metric used in the survey, a revised metric that employs a disk radius of three times the blackbody radius, \Lir, and a metric based on linear discriminant analysis (LDA). Horizontal dotted lines mark the percentage of targets needing observation before detecting half (12, gray) and all (24, black) of the \Qr detected disks had the targets been prioritized by the given quantity. Shaded histograms show the detection fraction per x-axis bin. The narrower the detection CDF compared to the observed CDF, the stronger the correlation between disk detectability and that quantity. In this case, prioritizing targets purely by \Lir would have been more efficient than prioritizing by our detectability metric, and the linear discriminant analysis produces a slightly more efficient (but similar) result.}
\label{fig:cdf_ks}
\end{figure*}

Our results also encourage further polarimetric observations and continued advancement of total intensity PSF-subtraction techniques. GPI's polarimetry mode proved highly effective for scattered-light detection and allowed us to detect at least eight disks that otherwise went undetected, including four that we detected in scattered light for the first time. In particular, polarimetry is more sensitive than total intensity ADI to low inclinations and small separations. It also provides complementary information to total intensity on disk morphology (Figures \ref{fig:det_gallery_Qr} and \ref{fig:det_gallery_totI}) and grain properties. In parallel, application of new methods to subtract the total intensity stellar PSF without heavily biasing the disk brightness will greatly benefit future disk science. Recently developed methods like non-negative matrix factorization \citep{ren2018_nmf} and mask-and-interpolate \citep{perrin2015_4796} may do so, thereby providing more accurate measurements of disk total intensity and polarization fraction that better constrain dust properties, especially for spatially extended and low-inclination disks that are predominantly inaccessible now.

\subsection{Predictive Power of Detectability Metrics} \label{sect:discuss_metric}

Now that GPIES is completed and we have gained new insight from its results, we can reassess the effectiveness of our original detectability metric at predicting polarized intensity detections (Section \ref{sect:metric}). While we did not end up observing targets purely in order of metric value, we did use it as an informal guide when prioritizing observations. Based on its apparent correlation with detections, we will use target \Lir as an initial comparison point. The target's \Lir was used to compute its metric value but many other factors were also incorporated, so we expect the two quantities to perform differently. One way to assess the predictive value of the quantity being tested is to compare the CDF for observed targets to the CDF for detected disks as a function of that quantity (Figure \ref{fig:cdf_ks}). In this case, the more predictive the quantity, the narrower the detected CDF (i.e., most skewed toward high quantity values) compared to the observed CDF. We consider the two disks detected only in total intensity (HD~15115 and NZ~Lup) as observed non-detections because we are interested in polarized intensity detectability here. We also completely exclude one target that lacks an \Lir measurement (TWA~25) and the three transition disks we detected.

A visual comparison of the CDFs indicates that \Lir is a more predictive quantity than the detectability metric described in Section \ref{sect:metric}. To quantify the difference between the observed and detected CDFs, we can again use the K-S test. The K-S statistic for our detectability metric is 0.30, while the statistic for  \Lir is 0.61. We also note from the CDFs that we would have had to observe every target in the ``Observed'' sample to detect all \Qr-detected disks if we observed targets purely in decreasing order of the original metric, as several detected disks were among the 15 lowest metric values. In contrast, all of our detections lie in the top 56\% of the observed sample's \Lir values. Put another way, if we had observed targets purely in descending order by \Lir, we would have had to observe 44 fewer stars to detect all GPIES disks than if we had observed in order of metric value. Ultimately, this means that our metric is a poorer predictor of scattered-light disk detectability than \Lir is on its own. In fact, the original metric performed only slightly better than if we had chosen targets completely randomly from our list, a strategy that averaged a K-S statistic of 0.15 when simulated.

With new knowledge about the disks detected in our survey, we revised our detectability metric to see if we could improve its predictive power and, separately, we statistically determined the target properties most correlated with detections. For a revised metric, we tested various changes and their effects on the CDFs. For example, considering solely the peak polarized intensity of the disk model ($\sqrt{\mathcal{Q}^2 + \mathcal{U}^2}$), rather than its contrast to the star's total intensity PSF, showed no improvement. Using a grain size distribution of $dN \propto ds^{-3.5}$ between $s_\mathrm{blow}$ and 1 mm produced a minor improvement over the original where all grains had $s=s_\mathrm{blow}$. Additionally, using the maximum grain size instead of $s=s_\mathrm{blow}$ to estimate the total dust mass, on the assumption that most of the mass is contained in the largest grains, produced negligible improvement.

One change that did have a noticeable effect was systematically increasing the radii of the dust rings in our disk models. Specifically, instead of using the SED-derived blackbody radius $R_\mathrm{bb}$ directly, we used $3\times R_\mathrm{bb}$ as the central radius of the narrow dust ring. We based this change on our finding that scattered-light disk radii are on average ${\sim}3$ times larger than the $R_\mathrm{bb}$ radius (Figure \ref{fig:radii}). The result was a substantial improvement in the predictive power of this revised detectability metric, evidenced by a K-S value of 0.49 and the CDFs shown Figure \ref{fig:cdf_ks}. With this method, we would have detected 12 disks within the top 15\% of our observed sample and all 24 \Qr disks within the top 75\%. Thus, it appears that our original detectability metric inappropriately penalized some targets because their disks were erroneously considered to have angular sizes too small for detection with GPI. The revised metric adjusts the angular sizes to better match the reality of scattered-light detections and performs better as a result.

We also took a separate approach to determining the target properties that most strongly predict disk detectability, this time using linear discriminant analysis (LDA), a ``supervised'' relative of principal component analysis. LDA is an eigvenvector technique that determines the linear combination of parameters that best separate two or more classes of objects. Here, our two classes were disks that we detected in polarized intensity and those that we did not. To select our input parameters, we started from a list of approximately 10 target parameters, including stellar mass, distance, and age, among others. However, LDA relies on independent input parameters and, as a result, many of the parameters were removed. After a non-exhaustive exploration of possible input parameters, we settled on four: the host star's effective temperature ($T_{\mathrm{eff}}$), log(\Lir), and $I$-band magnitude ($I$), and the blackbody radius of the disk in angular size units of arcseconds ($R_\mathrm{bb}$). Although some correlations may still remain between these final parameters, we found through trial and error that further reducing the number of parameters resulted in less predictive power.

Before carrying out the LDA, we first standardized all of the data using the mean and standard deviation of each parameter. Once we performed the LDA, we found that a single eigenvector explained nearly 100\% of the variance between classes, and thus this eigenvector contains the linear weights to apply to each standardized input parameter to create a new metric. The weights of the four standardized input parameters ($T_{\mathrm{eff}}$, log(\Lir), $I$, $R_\mathrm{bb}$) were found to be (-0.019, 0.998, 0.053, 0.040). Unsurprisingly, \Lir has the greatest influence (i.e. largest magnitude), followed by $I$, $R_\mathrm{bb}$, and $T_{\mathrm{eff}}$. The weight for $T_{\mathrm{eff}}$ is so close to zero that its negative sign may be solely due to noise on the coefficient. The CDFs of the LDA metric applied to our detected disks and the full observed sample are shown in Figure~\ref{fig:cdf_ks}. This new LDA metric outperforms the other three metrics by requiring the fewest targets to be observed to achieve all \Qr detections, although it only slightly outperforms plain \Lir and shares the same K-S statistic of 0.61. This similarity to \Lir is to be expected given the heavy weight assigned to \Lir by the LDA. Despite these encouraging results for GPI, they likely have limited applications to other instruments because the exact coefficients applied to the input parameters --- in particular $I$ and $R_\mathrm{bb}$ --- will almost certainly depend on instrumental characteristics set by design and hardware. Nonetheless, this analysis demonstrates that a simple linear combination of parameters may be sufficient to develop detectability metrics for future surveys.

\subsection{Gas in Scattered-light Debris Disks}
\label{sect:discuss_gas}

Detections of substantial molecular gas reservoirs have called into question the traditional picture of circumstellar disks being extremely gas poor by the ${\sim}10$ Myr old debris phase (e.g., \citealt{moor2017, rebollido2018}, and references therein).
Eight GPIES debris disk detections contain significant amounts of gas: $\beta$~Pic, CE~Ant, HD~32297, HD~110058, HD~131385, HD~146897, and HD~156623 have CO detections, while HR~7012 has [OI] emission (see \citealt{riviere-marichalar2012, lieman-sifry2016, macgregor2018b, matra2019a, matra2019b}, among others). We also observed seven other gas-bearing debris disks that resulted in GPIES non-detections: 49~Cet, $\eta$~Tel~A ([\ion{C}{2}] emission), Fomalhaut, HD~95086, HD~138813, HD~181327, and HR~1082 (see \citealt{zuckerman1995, smith2009, kospal2013, riviere-marichalar2014, lieman-sifry2016, marino2016, matra2017b, booth2019}, among others). Together, these constitute nearly all gas-bearing debris disks known to date. For most of them, the gas is presumed to be secondary in nature, produced recently from collisions and outgassing. Even the highest gas masses among debris disks may be of secondary origin, as it has been shown that neutral C can shield CO gas from photodissociation (in addition to CO self-shielding) and cause CO to accumulate \citep{kral2019}.

% Tau v spectral type with gas/no-gas.
\begin{figure}
\centering
\includegraphics[width=\columnwidth]{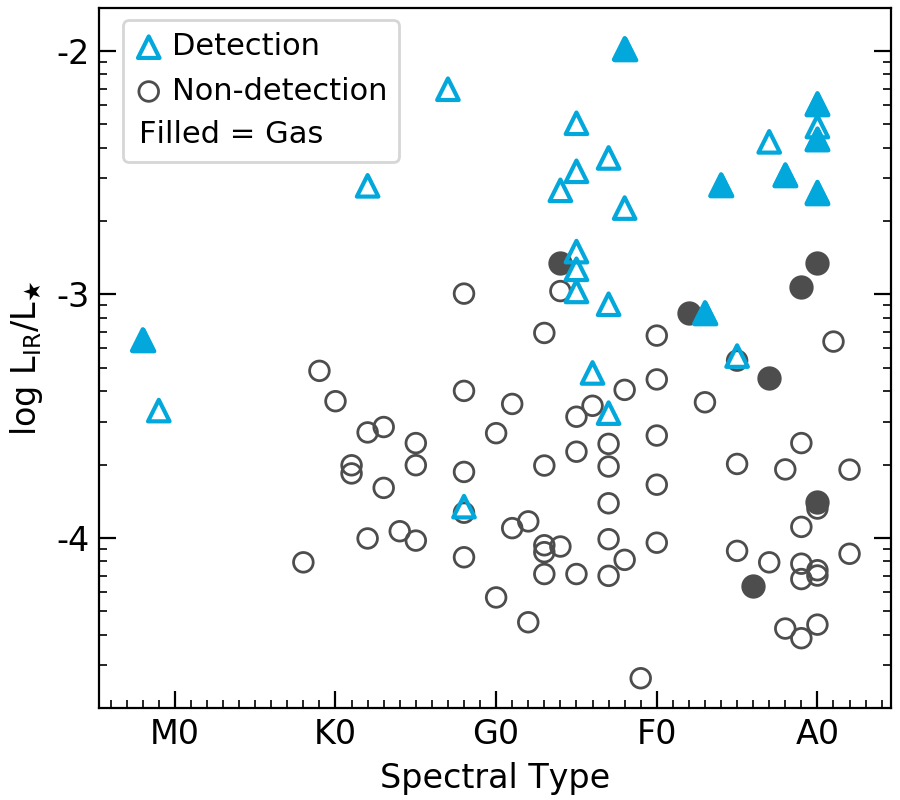}
\caption{Observed GPIES debris disk targets plotted by IR-excess magnitude and host spectral type. GPIES detections are blue triangles and non-detections are gray circles. Filled symbols denote disks with significant gas detections.}
\label{fig:gas_tau}
\end{figure}

The previously identified trend of scattered-light detection correlating with \Lir also holds for gas-bearing disks, as shown in Figure \ref{fig:gas_tau}. We can see this in more detail by examining the four gas-bearing GPIES non-detections that were plausibly detectable with GPI based on their known radii: 49~Cet, HD~138813, HR~1082 (HD~21997), and $\eta$~Tel~A. The first two have \Lir$>10^{-3}$, giving them higher IR-excess magnitudes than several GPIES detections. HR~1082's \Lir of $\num{4.6e-4}$ is also similar to a number of GPIES detections, although its $30\degr$ inclination may have contributed to our non-detection. $\eta$~Tel~A has the lowest \Lir of the group at $\num{1.4e-4}$ (similar to GPIES detection NZ~Lup) but appears to be highly inclined in mid-IR imaging \citep{smith2009}. Despite these four disks having qualities favorable to scattered-light imaging, only 49~Cet has been detected in scattered light so far, and it was found to have a relatively low surface brightness (see Section \ref{sect:non-detections}). The faintness of these disks shows that abundant gas is not necessarily correlated with high scattered-light brightness. Additionally, relatively bright scattered-light disks like HD~114082, HD~117214, HD~129590, and HR~4796 (all with \Lir$>10^{-3}$) have been searched for gas with ALMA and resulted in CO non-detections \citep{lieman-sifry2016, kennedy2018}.

These results are relevant to theories about the source of gas in debris disks. A proposed mechanism for producing secondary (i.e. nonprimordial) gas in such disks is the collisional vaporization of dust via the collisional cascade of solid bodies (e.g., \citealt{czechowski2007, kral2017b}). According to \citet{kral2017b}, the gas production rate in such a scenario is proportional to the dust production rate. Thus, with all else being equal, disks with high gas abundances should also have large amounts of micron-size dust grains and be relatively bright in scattered light compared to gas-poor disks. We do not find such a clear correlation between gas abundance and scattered-light brightness, however, suggesting that additional mechanisms are needed to explain the observations. For example, shielding of CO from photodissociation by neutral C (or through CO self-shielding) could boost gas abundance without similarly increasing the dust abundance \citep{kral2019}. On the other hand, the amount of CO contained in solid bodies might be significantly lower in some disks than others, such that similar collision rates will produce less gas. We look forward to future studies to settle these questions.

Regarding only our detections and the measured properties considered in this study, we find no obvious traits that set the gas-bearing debris disks apart from their gas-poor counterparts. On the whole, their morphologies, radii, and inclinations are consistent with the broader detected sample.

On an individual basis, though, HD~156623's scattered-light ring appears unusually wide---in terms of radial breadth relative to inner radius---compared to the rest of the GPIES disks, and it does not have a pronounced inner hole (although modeling shows it to still be consistent with a ring shape). This differs from the two other GPIES disks with similarly low inclinations, CE~Ant and HD~145560, both of which have narrower rings and larger inner holes. These disks offer good comparisons, as higher inclination disks are poorly suited for accurately measuring radial widths without detailed modeling. HD~181327 is another example of a narrow face-on ring, based on \textit{HST} data.

The CO mass of HD~156623 is unexceptional overall compared to our other observed gas-bearing disks, but it is estimated to be ${\sim}5$--500 times higher than that of CE~Ant and ${\sim}100$--500 times higher than HD~181327 \citep{marino2016, moor2017, matra2019a}, while HD~145560 only has a CO non-detection \citep{lieman-sifry2016}. Additionally, the best qualitative morphological matches to HD 156623 in our sample are HD 100546 and the inner component of HD 141569, both transitional disks with much higher gas-to-dust ratios. While this might suggest a connection, more rigorous analysis is needed to determine whether HD 156623's radial width is a result of or coincidental to its gas content. Investigating such connections in a broader context will also require more examples of scattered-light debris disks with accurate radius and CO measurements.

\subsection{Evidence of Differing Grain Populations in HD~117214 and HD~145560}
\label{sect:117214_v_145560}

One of the more promising avenues for characterizing debris disks is through a multiwavelength analysis. This is particularly feasible now with the maturation of high-contrast imagers like GPI and new high-resolution, high-sensivity millimeter imaging from ALMA. By comparing the appearance of a disk at near-IR and millimeter wavelengths, we are comparing its small and large grain populations. With the following example, we illustrate how comparing disk scattered-light fluxes to millimeter fluxes can inform us about grain properties and possible mechanisms influencing them.

We were motivated to investigate the HD~117214 and HD~145560 pair of disks by an apparent inconsistency between their millimeter ALMA fluxes and their scattered-light GPI fluxes. \citet{lieman-sifry2016} measured a 1.23~mm ALMA flux of $270 \pm 50~\mu$Jy for HD~117214 and $1850 \pm 120~\mu$Jy for HD~145560. This was curious, because the HD~117214 disk appeared markedly brighter than HD~145560 in our $H$-band \Qr intensity images from GPI (Figure \ref{fig:det_gallery_Qr}). Adding to our confusion was the similarity of the host stars: both are Sco--Cen members with an F5V spectral type and age of 12--18~Myr.

This drove us to ask two questions: (1) is the ratio of polarized scattered-light flux to millimeter flux truly different between the HD~117214 and HD~145560 disks? (2) And if that is the case, what does it tell us about the properties of the two disks?

To answer the first question, we need to accurately measure the $H$-band \Qr fluxes from our GPIES data, which we label $F_{\mathcal{Q}\phi}$. We start from each disk's ``median likelihood model'' output by the MCMCs we described in Section \ref{sect:morph_mcmc}, which is a single model constructed using the 50th percentile values of each individual parameter's marginalized posterior distribution (reported in Section \ref{sect:new_disks}). In the cases of these two disks, the median models are nearly indistinguishable from the maximum likelihood models. We then integrate the model's total flux and take 95\% of that value as our measurement of $F_{\mathcal{Q}\phi}$. We use only 95\% of the total model flux because the contour that encompasses this flux includes the parts of the model that unambiguously reproduce the GPI-observed surface brightness while excluding an extended halo of low surface brightness that is unconstrained by the data (being at or below the background noise level). With this approach, we restrict ourselves to the model flux that we can confidently associate with the disk. If this underestimates the true disk fluxes, it should do so similarly for both disks (within the broad tolerances of this test).

Dividing the $F_{\mathcal{Q}\phi}$ of HD~117214 by that of HD~145560, we find the ratio of their $H$-band \Qr fluxes to be 2.42. Such a direct comparison is not fair, however, because the disks have different radii, inclinations, and host star $H$-band luminosities, all of which affect the amount of light scattered. On top of this, the host stars have different distances, which has an inverse-square effect on the observed disk fluxes. Fortunately, we have measurements of all of these quantities and can correct for the differences between targets to compare their intrinsic disk properties. To do so, we will define the ``true'' $H$-band \Qr flux ratio of the disks as

\begin{equation} \label{eq:flux_ratio}
% \begin{aligned}
    f_{\mathcal{Q}\phi} = \frac{F_{\mathcal{Q}\phi,A}}{F_{\mathcal{Q}\phi,B}} \bigg(\frac{d_A}{d_B}\bigg)^2 \bigg(\frac{R_{0,A}}{R_{0,B}}\bigg)^2 \bigg(\frac{F_{H,A}}{F_{H,B}}\bigg)^{-1} C_i
% \end{aligned}
\end{equation}

where A denotes HD~117214 values and B denotes HD~145560 values, $F_H$ is stellar $H$-band flux, and $C_i$ is the estimated fractional change in $F_{\mathcal{Q}\phi}$ when increasing a disk model's inclination from the $i$ of HD~117214 to that of HD~145560.

For these two stars, $d_A / d_B = 0.893$ and $R_{0,A} / R_{0,B} = 0.706$. We simplify the radius correction by only considering the single radius $R_0$ even though the disks have finite radial widths of tens of au. A more nuanced approach requires a better constraint on the disks' outer radii, which we have acknowledged is difficult with our simple model. We only care about the ratio of the stellar fluxes here, so we convert their apparent $H$-band magnitudes (Table \ref{tab:datasets}) to absolute $H$ magnitudes (i.e. $M_H = m_H - 5 \log(d/10~\mathrm{pc})$) and then convert those absolute magnitude into an $H$-band flux ratio (at 10 pc) as $F_{H,A}/F_{H,B} = 10^{-(M_{H,A} - M_{H,B})/2.5} = 1.76$.

To get a value for $C_i$, we need to estimate how much of an effect inclination has on $F_{\mathcal{Q}\phi}$. The main source of any such effect is the polarized scattering phase function, as our approach of using the integrated flux effectively negates effects of dust column density. We estimate the strength of the effect by first computing a new MCFOST model that has the same parameter values as the median likelihood model for HD~145560 except the inclination is increased to $i=71\fdg0$ to match that measured for HD~117214. We measure $F_{\mathcal{Q}\phi}$ for this new model and call it $F'_{\mathcal{Q}\phi,B}$. Taking the ratio of this flux to the actual HD~145560 flux, we get $F'_{\mathcal{Q}\phi,B}/F_{\mathcal{Q}\phi,B}=1.25$, which tells us that the HD~145560 disk would be ${\sim}25$\% brighter in polarized scattered-light if it were instead viewed at the higher inclination of HD~117214. Repeating an analogous process for HD~117214, where we compute a new model using the HD~145560 inclination of $i=43\fdg9$, we find $F_{\mathcal{Q}\phi,A}/F'_{\mathcal{Q}\phi,A}=1.18$ (again dividing the flux at the higher inclination by the flux at the lower inclination). Thus, we estimate that the HD~117214 disk is ${\sim}18$\% brighter when viewed at its actual inclination than if it were seen at the lower inclination of HD~145560. To be conservative, we set $C_i=1.21$ as the average of the estimates from the two disks.

With all of our conversion factors in hand, we substitute them into Equation (\ref{eq:flux_ratio}) to arrive at $f_{\mathcal{Q}\phi} = 0.273$. This means that, if the HD~117214 and HD~145560 disks had the same inclinations, physical radii, distances from the observer, and stellar illumination, the HD~145560 disk would actually have an $H$-band \Qr flux 3.66 times greater than that of HD~117214. This is opposite the apparent brightness ratio from our images and demonstrates the impact of the considered parameters on a disk's scattered-light brightness.

Shifting our attention to the thermal emission, we see from \citet{lieman-sifry2016} that the ratio of integrated 1.23-mm fluxes is $F_{\mathrm{mm},A}/F_{\mathrm{mm},B} = 0.146$. Once again, though, this is not a fair comparison without adjusting for differences in disk and star properties. Doing so is simpler at millimeter wavelengths because, without a scattering phase function at work, there is no inclination effect on the flux, assuming that the disks are optically thin. Thus, we can define the ``true'' mm flux ratio as

\begin{equation} \label{eq:flux_ratio_mm}
    f_{\mathrm{mm}} = \frac{F_{\mathrm{mm},A}}{F_{\mathrm{mm},B}} \bigg(\frac{d_A}{d_B}\bigg)^2 \bigg(\frac{R_{0,\mathrm{mm},A}}{R_{0,\mathrm{mm},B}}\bigg)^2 \bigg(\frac{L_{\star,A}}{L_{\star,B}}\bigg)^{-1}
\end{equation}

where again A denotes HD~117214 values and B denotes HD~145560, $R_{\mathrm{0,mm}}$ is the peak surface density radius from thermal emission this time, and $L_{\star}$ is the bolometric luminosity of the star. We are interested in the bolometric luminosity here because the thermally emitting grains are absorbing energy from a broad range of wavelengths.

We adopt $R_{0,\mathrm{mm},B} = 56$ au for HD~145560 from \citet{lieman-sifry2016}, which, given their uncertainties of $[+11, -9]$ au and the average ratio of $R_0:R_{\mathrm{0,mm}}=1.39$ from Section \ref{sect:radius_results}, is consistent with our scattered-light value of $R_0 = 85.3 \substack{+1.3 \\ -1.2}$ au. No thermal radius estimate is available for HD~117214 because it was not sufficiently resolved \citep{lieman-sifry2016}, so we scale our scattered-light radius of $R_0 = 60.2 \substack{+0.6 \\ -1.1}$ down by the average ratio of 1.39 to get our adopted value of $R_{0,\mathrm{mm},A} = 43.3$ au. Thus, we end up with $R_{0,\mathrm{mm},A} / R_{0,\mathrm{mm},B} = 0.77$. We also have $L_{\star,A} / L_{\star,B} = 1.44$ based on the values in Table \ref{tab:obs_targs}, although that ranges from 1.14 to 1.77 within the $1\sigma$ uncertainties (largely due to the $\pm0.90 L_\odot$ uncertainty for HD~117214). As in the scattered-light case, $d_A / d_B = 0.893$. Substituting these values into Equation \ref{eq:flux_ratio_mm}, we find $f_{\mathrm{mm}} = 0.0479$. This means that the observed 1.23 mm flux of the HD~145560 disk would be 20.9 times higher than that of the HD~117214 disk if they were at the same radius around the same star. In this case, the difference in bolometric luminosity has a large effect; however, even taking $L_{\star,A} / L_{\star,B}$ to be its lowest possible value of 1.14 (based on $1\sigma$ uncertainties) only decreases the 1.23 mm flux ratio of the stars to 16.1.

Now that we have consistent ratios comparing the disk fluxes to each other in both polarized scattered light ($f_{\mathcal{Q}\phi}$) and thermal emission ($f_{\mathrm{mm}}$), we can finally answer our first question: is the ratio of polarized scattered light to millimeter emission truly different between these two disks? Indeed, the value $f_{\mathcal{Q}\phi} / f_{\mathrm{mm}} = 5.70$ tells us that the HD~145560 disk's ratio of polarized scattered-light flux to millimeter flux is 5.7 times higher than the same ratio for the HD~117214 disk. One could view this as the HD~145560 disk producing more millimeter emission than expected given its amount of observed polarized scattered light. On the other hand, it could be seen as a dearth of polarized scattered light given the amount of millimeter emission. The reverse could be said of HD~117214 (e.g., too much scattered light given its millimeter flux). We made several assumptions and simplifications in order to put the two disks on equal ground in terms of inclination, radius, stellar illumination, and observed distance. That said, a factor of 5.7 difference in $F_{\mathrm{mm}}/F_{\mathcal{Q}\phi}$ after controlling for those aspects suggests physical differences in the disks' constituent particles.

This brings us to our second question: what does this tell us about the properties of the two disks? We have already accounted for their main morphological differences in radius and inclination, so we turn next to the disks' grain properties.

One possible explanation is a steeper grain size distribution in HD~117214 compared to the HD~145560 disk, giving the former a higher relative fraction of micron-size grains versus millimeter-size grains. Grain size distributions are often assumed to follow a power law of the form $N(a) \propto a^{-q}$. Recent studies of millimeter emission alone (not considering the micron-size grains) have measured ranges of $q=3.24$--3.64 for
A--F stars \citep{macgregor2016b} and 3.11--3.26 for A--G
stars \citep{marshall2017}, so significant differences between disks are plausible. Theoretically, $q$ is linked to the physical structure and velocity dispersion of particles in the disk (e.g., \citealt{pan2005, pan2012}. In a dynamically hot disk, for example, higher velocities will produce higher energy collisions more frequently and thus create more small particles compared to a dynamically cold disk. Such could be the case in HD~117214's disk, where perhaps particles are stirred gravitationally by planets (e.g., \citealt{thebault1998, moro-martin2007, mustill2009, daley2019}) or self-stirred by large planetesimals (e.g, \citealt{kenyon2004a, kenyon2004b, pan2012, krivov2018}). Alternately, some mechanism may be dampening collisions in the HD~145560 disk or accelerating the removal of its small dust particles.

Apart from the wider grain size distribution, differences within just the micron-size dust populations could be impactful. A disk's scattered-light brightness depends partly on the scattering phase function and polarizability of its constituent dust grains, which in turn depend on grain size, structure (e.g., fluffy aggregates vs. compact spheres), and material composition. We consider it unlikely that HD~117214 and HD~145560 grains have severely different intrinsic phase functions: a compilation by \citet{hughes2018_review} of existing measurements of debris disks and solar system dust populations shows most phase functions to be similar. Even if both of the disks have extraordinary phase functions, it is questionable whether this alone could produce a factor of ${>}5$ difference in $F_{\mathcal{Q}\phi}$ based on the largest known outliers. The same can be said for dust polarizability (i.e. polarization fraction) among debris disks, though fewer measurements have been published (e.g., \citealt{graham2007, maness2009, milli2015, draper2016, esposito2016, esposito2018, duchene2020}). That said, moderately different phase functions could be working in concert with other effects. For example, the near-IR albedo of the dust could be higher for HD~117214; water-ice and astrosilicate albedos can differ by a factor of 2 at a wavelength of 1.6 $\micron$ \citep{roberge2010}. If the optical albedo is higher, too, then this would have the added effect of decreasing thermal emission by reducing grains' absorption.

While we only examine a single pair of disks here, the GPIES debris disk sample offers an opportunity for additional comparisons with millimeter data that will hopefully confirm or reject some of the many possible scenarios regarding grain properties.

\subsection{Substellar Companions in GPIES Disk Systems}

Empirical and theoretical connections between planets and debris disks make it natural to search for such links in the GPIES results. We focus on directly imaged, wide-separation (i.e. semimajor axes $>$ a few astronomical units) planets and brown dwarfs because our disks generally have radii much larger than the semimajor axes to which radial velocity and transit methods are sensitive. Additionally, for simplicity, we will use the term ``substellar companion'' here to include both giant planets and brown dwarfs; the distinctions between the two classes of object are not particularly relevant to this discussion. Wide-separation substellar companions, if massive enough and/or close enough to the disk, could dynamically stir planetesimals and intensify the collisional cascade, producing a positive correlation; when a debris disk is prominent in scattered light, the greater the chances of detecting a substellar companion, too. Or, substellar companions could very efficiently remove disk material in the primordial phase, resulting in a negative correlation; if a disk is depleted in the debris disk phase, the greater the chances of detecting a substellar companion. Another type of connection to study is whether the presence or absence of substellar companions has an effect on the radial, azimuthal, and vertical architectures of debris disks.

Eight out of the 101 stars in our observed debris disk sample have confirmed directly imaged substellar companions. Six of these systems produced GPIES disk non-detections: 51~Eri \citep{macintosh2015}, Fomalhaut \citep{kalas2005_fomb}, HD~95086 \citep{derosa2015b}, HD~206893 \citep{milli2017a}, HR~2562 \citep{konopacky2016}, and HR~8799 \citep{marois2008, marois2010_8799}. Disks were detected in two systems: $\beta$~Pic \citep{lagrange2009a} and HD~106906 \citep{bailey2014}. Comparing the two groups at face value, the occurrence rate of wide-separation substellar companions is 2/26 (7.7\%) for systems with a GPIES-detected scattered-light debris disk and 6/75 (8.0\%) for systems without one. In other words, with respect to systems containing significant IR excesses, those with debris disks detected by GPIES are not more likely to contain wide-separation substellar companions than those with no GPIES detection. We can adjust our numbers by considering all scattered-light detections (not just GPIES), as the Fomalhaut, HD~107146, and HD~181327 disks are known but outside the GPI field of view. In this case, the occurrence rate of wide-separation substellar companions is 3/29 (10.3\%) for systems with a scattered-light disk detection and 5/72 (6.9\%) for systems without one. In either case, there is no appreciable difference between the detected- and undetected-disk rates, considering the small number of companions. Thus, our sample indicates that debris disks are equally likely to be detected in scattered light regardless of whether there is a wide-separation substellar companion. It is possible that the companions in these systems do not come close enough to the disks to interact significantly; we do not test this because some of the companions' orbits and disk locations are not yet well constrained. That said, other factors, such as significant differences in initial conditions, may be more important in determining which stars at $\gtrsim10$ Myr have prominent scattered-light debris disks.

These percentages are roughly consistent with the 3.86\%--9.76\% giant-planet occurrence rate established by \citet{meshkat2017} for stars with \emph{Spitzer}-detected debris disks. The fact that we measure similar substellar companion occurrence rates for systems with and without scattered-light disk detections does not contradict that study's other finding of a lower occurrence rate for stars with no known debris disk. This is because they considered the presence of any debris disk and not just those resolved in scattered light; i.e., our selection of targets by IR excess means that nearly all of the GPIES-observed targets (regardless of GPIES detection) are included in the \citet{meshkat2017} category of ``debris disk hosts'' based on their IR color criteria\footnote{\citet{meshkat2017} defined their ``debris disk'' stars as having \textit{WISE} $\mathrm{W1}-\mathrm{W4}\geq0.3$ mag for $J-K_s<0.8$ mag, or $\mathrm{W1}-\mathrm{W4}\geq0.6$ mag for $J-K_s>0.8$ mag.}. 

The two substellar-companion-bearing disks in our sample display different morphologies, one of which is similar to another GPIES disk. $\beta$~Pic~b creates an inclined warp in the system's inner disk compared to its outer disk. A similar morphology may be seen in HD~110058, which has no known substellar companion but shows a counterclockwise warp to both sides of the edge-on disk, perhaps indicating misaligned inner and outer components \citep{kasper2015}. In the GPIES data, the warp is more prominent and the emission is more radially extended in total intensity than \Qr (Figures \ref{fig:det_gallery_Qr} and \ref{fig:det_gallery_totI}), which suggests we detect primarily the inner component in \Qr. With no clear view of this edge-on disk's inner edge, however, we cannot constrain the semimajor axis of a potential warp-inducing substellar companion other than to say it is likely interior to the disk's apparent ansae in \Qr, i.e., ${\lesssim}40$ au. HD~106906~b, on the other hand, is exterior to the dust ring imaged by GPI (738 au projected separation from the star) and may be responsible for the ring's eccentricity, as well as substantial radial and vertical asymmetries \citep{kalas2015}. HD~106906 is also unusual in our sample for being a nearly equal-mass spectroscopic binary \citep{lagrange2016}. In terms of other properties like \Lir, $R_\mathrm{in}$, age, and surface brightness (considering distance), though, the $\beta$ Pic and HD 106906 disks do not stand out from the rest of the GPIES sample.

Other GPIES-detected disks also have architectures that could be linked to dynamical interactions with a substellar companion. All of the debris disks can be described as rings with inner holes or dust depletions, and substellar companions are theoretical causes of such features (e.g., \citealt{roques1994, quillen2006, dong2016}). Given our range of $R_\mathrm{in}$, these companions could have semimajor axes of several astronomical units up to ${\sim}70$ au depending on the disk. Apart from dust-poor holes, the most common features suggestive of companion--disk interaction are stellocentric offsets implying eccentric rings \citep{wyatt1999, kalas2004}. This is the case for HD~61005 \citep{esposito2016}, HD~157587 \citep{millar-blanchaer2016c}, and HR~4796 (first measured by \citealt{schneider2009}, with GPIES confirmations in \citealt{perrin2015_4796} and P. Arriaga et al. 2020, accepted.), all with offsets ${>1}$ au. Future analyses may reveal offsets in other GPIES rings, too, but the high inclinations of many of them complicate this by concealing the ansae and prohibiting measurements along the projected minor axis. A case in point is HD~32297, which has been interpreted as an elliptical disk with its major axis nearly parallel to our line of sight that leads to almost no projected offset \citep{lee2016, lin2019}.

Such difficulties with interpreting edge-on disk morphologies emphasize the value of polarimetry data for constraining the presence of planets in ``low''-inclination debris disks ($i\lesssim75\degr$). Morphological properties like eccentricity and radius are often obscured or degenerate with the dust's scattering properties when disks are nearly edge on. This leads to ambiguity when considering evidence for substellar companions. The morphology can be determined with greater certainty in low-inclination rings where the ansae and both front and back edges are clearly defined. The GPIES sample shows that these inclinations are best detected in polarized intensity from a ground-based instrument.

Illustrating the importance of knowing the morphology are model explanations for some of the more extreme debris disk morphologies (e.g., ``moths,'' ``double wings,'' ``needles''; \citealt{esposito2016, lee2016, lin2019}) that require the dust rings to be at least moderately eccentric.
The eccentricity is typically ascribed to the gravitational influence of a nearby substellar companion. Although these morphologies are presently only seen at high inclinations, disk eccentricity should be agnostic toward inclination. Thus, a census of low-inclination debris disks would produce an eccentricity distribution that can be applied to high-inclination disks and used to infer the incidence of extreme morphologies like double-winged moths.

Low-inclination disk systems also make prime targets for substellar companion imaging searches, even more so than typical debris disk systems, because companions inside large projected holes are less likely to be conflated with disk emission or caught in conjunction with the star due to their orbital phase. Additionally, once companions are detected, dynamical analyses incorporating the sharpness of a disk edge and its distance from the companion provide constraints on the object's mass independent of evolutionary models \citep{quillen2006, chiang2009, rodigas2014_shepherd}. Overall, polarimetric scattered-light imaging of debris disks should prove especially powerful for investigating planet and brown dwarf interaction signatures in disk morphologies and uncovering the architectures of these planetary systems.

\section{Conclusions}

GPIES observed 104 stars in its polarimetric imaging mode to resolve circumstellar debris disks in $H$-band scattered light. The campaign resulted in 26 debris disk detections, as well as three protoplanetary/transitional disk detections. Overall, we detected 24 of the debris disks in polarized intensity and 18 in total intensity (with two of the disks detected exclusively in total intensity). In general, our data probed projected separations of $0\farcs15$--$1\farcs4$ (out to $1\farcs8$ in the corners of the square field of view), which translate to solar system-like scales of 1--200 au depending on the system's distance.

We presented the first scattered-light images of debris disks around HD~117214 and HD~156623 and quantified their basic properties from MCMC modeling. HD~117214 resembles the narrow dust ring detected around HR~4796~A, though with a smaller inner radius of $\sim$20 au, and HD~156623 is unusual because its dust-scattered light from its radially broad ring is detected all the way inward to the edge of the FPM at 12 au. Over the entire campaign, GPIES resolved a total of seven debris disks for the first time in scattered light. Thirteen (i.e., half) of all GPIES debris disk detections belong to the 7--18 Myr old Scorpius-Centaurus OB association.

We examined select properties of our sample for trends and found several of note. Our detections heavily favored high IR-excess magnitudes, as all of them have \Lir$>10^{-4}$. Morphologically, all of our debris disks can be described as rings with dust-poor inner holes except for HD~156623, which appears radially broader than the other disks and has no defined inner hole. Combining measurements from GPIES and other instruments, we found that the disks' radii of peak surface density as measured in scattered light averaged 1.4 times larger than their peak radii from resolved millimeter imaging, which we take as evidence that micron-size dust is predominantly located exterior to the millimeter-size grains and larger planetesimals that comprise most of the ring's mass. Disks' scattered-light peak radii are also 4.3 times larger than their SED-derived blackbody radii and show marginal evidence of increasing slowly with host star luminosity.

Our results also showed that increasing disk inclination improves scattered-light detectability. This effect is most dominant in total intensity but still present in polarized intensity, indicating that the scattering phase functions of debris disks are fundamentally anisotropic. Polarimetric imaging proved especially effective for imaging disks with $i\lesssim75\degr$, avoiding the filtering of low spatial frequencies that is inherent to ADI-based PSF-subtraction algorithms for total intensity, but also perhaps benefiting from polarization fractions that peak near $60\degr$--$90\degr$ scattering angles. Our polarized intensity data provide the best (and often only) constraints on the scattered-light morphologies of these low-inclination disks, and give complementary views of the higher inclination disks with additional power to break parameter degeneracies in models and other future analyses.

The breadth and uniformity of the GPIES disk sample make it highly complementary to other disk and exoplanet-related data. By comparing two new scattered-light disks around F-type stars that have also been resolved with ALMA, we found evidence for significantly different grain populations based on differing ratios of polarized scattered-light flux to millimeter flux. When considering gas content, we find no clear differences between the average scattered-light properties of the gas-rich and gas-poor debris disks. Along similar lines, we find no significant difference between the rates of GPIES debris disk detections for systems with directly imaged giant planets and those without (albeit without correcting for completeness or sensitivity). Several GPIES disks without known planets have morphologies that suggest disk--planet interaction, such as dust-poor inner holes, stellocentric offsets implying eccentric rings, and warping implying mutual inclination between the inner and outer disks. These make particularly interesting targets for future planet searches.

The GPIES results show that high-contrast polarimetric imaging is, and will continue to be, a powerful tool for debris disk science in terms of both discovery and characterization. We expect there are yet more disks around nearby stars that can be resolved in polarized scattered light if observed by instruments like GPI, SPHERE, and SCExAO/CHARIS \citep{jovanovic2015a, groff2016, groff2017}. The abundance of disks detected in Sco--Cen, in particular, tells us that identifying new young stellar associations, or new members to known associations, within ${\sim}200$ pc would be highly beneficial for guiding those observations. Similarly, the demonstrated synergies between scattered-light and resolved millimeter imaging emphasize the value of pairing GPI-like data with high-angular-resolution ALMA observations for as many disks as possible. Combining polarized and total intensity data with models, ideally at multiple wavelengths, should also prove fruitful for characterizing the disk material, especially once those models incorporate scattering by realistic grain structures like aggregates. Such synthesis may help to place rigorous constraints on dust compositions for the first time and open another window into planetary system construction.

\acknowledgements{
We wish to thank the anonymous referee for their constructive comments that improved this manuscript. This work is based on observations obtained at the Gemini Observatory, which is operated by the Association of Universities for Research in Astronomy, Inc. (AURA), under a cooperative agreement with the National Science Foundation (NSF) on behalf of the Gemini partnership: the NSF (United States), the National Research Council (Canada), CONICYT (Chile), Ministerio de Ciencia, Tecnolog\'ia e Innovaci\'on Productiva (Argentina), and Minist\'erio da Ci\^encia, Tecnologia e Inova\c c\~ao (Brazil). This work made use of data from the European Space Agency mission {\it Gaia} (\url{https://www.cosmos.esa.int/gaia}), processed by the {\it Gaia} Data Processing and Analysis Consortium (DPAC, \url{https://www.cosmos.esa.int/web/gaia/dpac/consortium}). Funding for the DPAC has been provided by national institutions, in particular the institutions participating in the {\it Gaia} Multilateral Agreement. This research made use of the SIMBAD and VizieR databases, operated at CDS, Strasbourg, France, and of ESA's Herschel Science Archive (\url{http://archives.esac.esa.int/hsa/whsa}).

Supported by NSF grants AST-1411868 (E.L.N., K.B.F., B.M., and J.P.), AST-141378 (G.D.), and AST-1518332 (T.M.E., R.J.D.R., J.R.G., P.K., G.D.). Supported by NASA grants NNX14AJ80G (E.L.N., B.M., F.M., and M.P.), NNX15AC89G and NNX15AD95G/NExSS (T.M.E., B.M., R.J.D.R., G.D., J.J.W, J.R.G., P.K.), NN15AB52l (D.S.), and NNX16AD44G (K.M.M.). M.R. is supported by the NSF Graduate Research Fellowship Program under grant number DGE-1752134. J.R. and R.~Doyon acknowledge support from the Fonds de Recherche du Qu\`ebec. J.~Mazoyer's work was performed in part under contract with the California Institute of Technology/Jet Propulsion Laboratory funded by NASA through the Sagan Fellowship Program executed by the NASA Exoplanet Science Institute. M.M.B. and J.~Mazoyer were supported by NASA through Hubble Fellowship grants \#51378.01-A and HST-HF2-51414.001, respectively, and I.C. through Hubble Fellowship grant HST-HF2-51405.001-A, awarded by the Space Telescope Science Institute, which is operated by AURA, for NASA, under contract NAS5-26555. K.W.D. is supported by an NRAO Student Observing Support Award SOSPA3-007. J.J.W. is supported by the Heising-Simons Foundation 51~Pegasi~b postdoctoral fellowship. This work benefited from NASA's Nexus for Exoplanet System Science (NExSS) research coordination network sponsored by NASA's Science Mission Directorate. Portions of this work were also performed under the auspices of the U.S. Department of Energy by Lawrence Livermore National Laboratory under Contract DE-AC52-07NA27344.}

\software{Gemini Planet Imager Data Pipeline (\citealt{perrin2014_drp, perrin2016_drp}, \url{http://ascl.net/1411.018}), pyKLIP (\citealt{wang2015_pyklip}, \url{http://ascl.net/1506.001}), NumPy (\citealt{numpy}, \url{https://numpy.org}), SciPy (\citealt{scipy2020}, \url{http://www.scipy.org/}), Astropy \citep{astropy2018}, matplotlib \citep{matplotlib2007}, iPython \citep{ipython2007}, emcee (\citealt{foreman-mackey2013}, \url{http://ascl.net/1303.002}), corner (\citealt{corner}, \url{http://ascl.net/1702.002})}.

\facilities{Gemini:South}

%% APPENDICES
\appendix

\section{GPIES Data in Stokes \Ur} \label{appendix-a}

\begin{figure*}[h!]
\centering
\includegraphics[width=\textwidth]{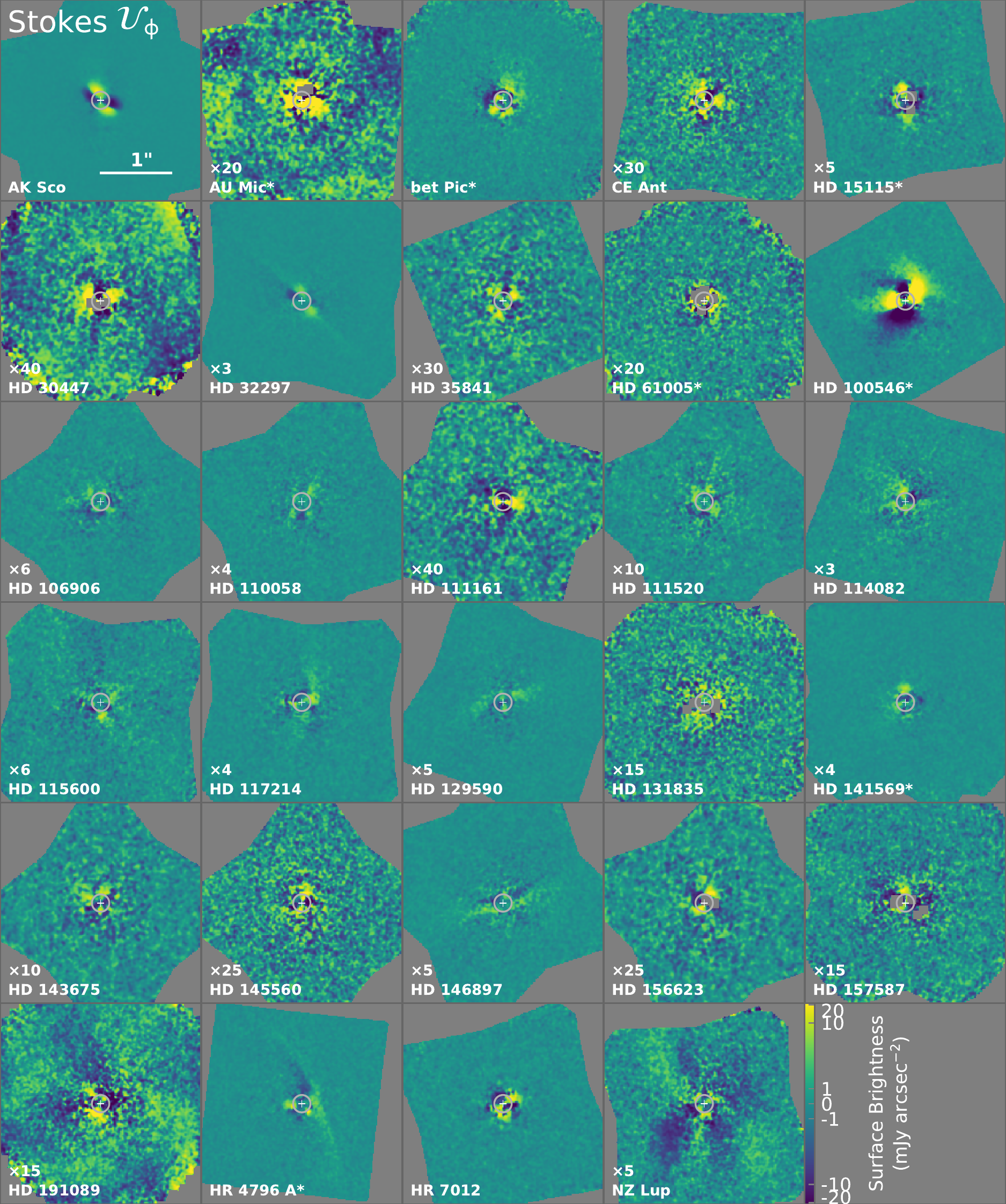}
\caption{All GPIES disk detections as seen in $H$-band Stokes \Ur. North is up, east is left, and all panels are on the same angular size scale. Disks observed during GPI commissioning have asterisks after their names. The brightness scales match those used for \Qr in Figure \ref{fig:det_gallery_Qr} but extend to negative values. Thus, all panels display the disk surface brightness using the same color map that is logarithmic between -20 and 20 mJy arcsec$^{-2}$ except for being linear between -1 and 1 mJy arcsec$^{-2}$; however, all but the brightest disks have been scaled linearly before plotting by a factor noted above the target name. The white circles mark the GPI $H$-band FPM edge, and the crosses mark the star location.}
\label{fig:det_gallery_Ur_all}
\end{figure*}

The Stokes \Ur intensity images for all GPIES disk detections, including those not detected in \Qr, are shown in Figure \ref{fig:det_gallery_Ur_all}. In the case of single scattering of stellar photons by dust grains (often assumed for optically thin disks), we expect polarization vectors to be oriented azimuthally and thus present little to no disk signal in \Ur. This is true for the majority of our debris disk detections, which only show varying degrees of residual instrumental polarization, detector persistence, and random noise in \Ur. The disks that are brightest in \Qr, however, appear to also have faint \Ur signals. The most notable examples are HD~32297, HD~110058, HD~114082, HD~117214, HD~129590, HD~146897, and HR~4796~A. The two brightest non-debris disks, AK~Sco and HD~100546, show strong \Ur signals.

One possible explanation is that there may be a systematic error in the way we extract the linear Stokes vectors. Specifically, inaccuracies using the Mueller matrix to recover the astrophysical polarization signal from our modulated measurements could cause the \Qr signal to ``leak'' into the \Ur channel (via crosstalk between $\mathcal{Q}$ and $\mathcal{U}$). The correlation of the \Ur signal amplitude with disk surface brightness is consistent with this scenario if the amount of leakage is proportional to the \Qr amplitude. We are investigating this possibility but have not reached a conclusive result by the time of publication.

Another explanation is that the \Ur is astrophysical in nature, resulting from a significant fraction of photons experiencing multiple scatterings as they pass through these disks. These could be scatterings by multiple grains \citep{bastien1988, canovas2015} or multiple scatterings within a single grain. In either case, higher dust densities would lead to greater \Ur amplitudes, which is also consistent with the observed trend. If these signals are indeed astrophysical, it may be possible to disentangle the two multiple-scattering scenarios and determine if these disks have relatively high optical depths at near-IR wavelengths or whether their grain properties lead to multiple internal scatterings.

We also note that observations from other instruments show similar \Ur signals. In just a few examples: \citet{garufi2016} reported primarily ${-}$\Ur intensity for HD~100546 in SPHERE/ZIMPOL visible-light polarimetry, \citet{asensio-torres2016} reported primarily ${+}$\Ur intensity for HD~32297 in Subaru/HICIAO $H$-band data, and \citet{avenhaus2014} reported both positive and negative \Ur for the HD~142527 protoplanetary disk in VLT/NaCo $H$- and $Ks$-band data. In the latter case, the \Ur signal was partially removed by correcting for the crosstalk between Stokes vectors.

% Detection gallery Ur

%\clearpage

\section{Photometric Calibration Factors for Polarimetric Data} \label{appendix-b}

We recorded in Table \ref{tab:calfactors} the factors used to convert polarimetric Stokes datacubes from analog-to-digital units (ADU) per second (already converted from ADU coadd$^{-1}$) to physical units of Jansky, as referenced in Section \ref{sect:pol_reduction}. The mean calibration factor is $\num{6.56e-7}$ Jy (ADU/s)$^{-1}$ with a standard deviation of $0.92 \times 10^{-7}$ Jy (ADU/s)$^{-1}$.

To estimate the error on the calibration factor, we first calculate the error on the aperture-integrated flux of each satellite spot in the data set as its Poisson error combined in quadrature with the estimated error on the ``sky'' background flux that is subtracted from the satellite spot flux \citep{hung2016}. We then sum in quadrature all of the satellite spot flux errors, take their average, and divide that by the average satellite spot flux to get the corresponding fractional error. Finally, we compute the quadrature sum of this fractional error on the satellite spot flux and the fractional error in the stellar flux. This final sum gives us the error on the calibration factor shown in Table \ref{tab:calfactors}.

% Photometric calibration table.
%\startlongtable
\begin{deluxetable}{lccD}
%\tablecolumns{6}
%\tablewidth{0pt}
\decimals
\tablecaption{Photometric Calibration Factors}
\tablehead{
% Header Line 1
\colhead{Name} & \colhead{Date} & \colhead{Factor} & \twocolhead{Error} \vspace{-0.3cm} \\
% Header Line 2
 & & \colhead{$10^{-7}$ (Jy (ADU/s)$^{-1}$)} & \twocolhead{$10^{-7}$ (Jy (ADU/s)$^{-1}$) }
}

\startdata
AK Sco & 180811 & 6.54 & 0.315 \\
AU Mic & 140515 & 6.07 & 0.834 \\
bet Pic & 131212 & 5.01 & 0.141 \\
CE Ant & 180405 & 6.84 & 0.228 \\
HD 15115 & 141216 & 5.90 & 0.216 \\
HD 30447 & 160922 & 6.22 & 0.617 \\
HD 32297 & 141218 & 6.42 & 0.283 \\
HD 35841 & 160318 & 6.98 & 0.452 \\
HD 61005 & 140324 & 5.11 & 0.222 \\
HD 100546 & 131212 & 6.61 & 0.475 \\
HD 106906 & 150701 & 6.56 & 0.238 \\
HD 110058 & 160319 & 8.29 & 0.789 \\
HD 111161 & 180310 & 5.99 & 0.388 \\
HD 111520 & 160318 & 7.72 & 0.611 \\
HD 114082 & 170807 & 8.65 & 1.67 \\
HD 115600 & 150703 & 6.96 & 0.755 \\
HD 117214 & 180311 & 6.36 & 0.396 \\
HD 129590 & 170809 & 5.24 & 0.401 \\
HD 131835 & 150501 & 8.12 & 1.03 \\
HD 141569 & 140322 & 5.82 & 0.166 \\
HD 143675 & 180408 & 7.86 & 0.550 \\
HD 145560 & 180812 & 5.95 & 0.402 \\
HD 146897 & 160321 & 6.99 & 0.685 \\
HD 156623 & 190427 & 5.69 & 0.240 \\
HD 157587 & 150829 & 6.90 & 0.600 \\
HD 191089 & 150902 & 6.14 & 0.363 \\
HR 4796 A & 131212 & 7.14 & 0.374 \\
HR 7012 & 180921 & 5.46 & 0.151 \\
NZ Lup & 150408 & 6.78 & 0.689 \\

\enddata
\tablecomments{Targets are listed alphabetically by name. Other column headings: date of observation as YYMMDD, the multiplicative calibration factor to convert data units from ADU s$^{-1}$ to Jy, and estimated error on the calibration factor.\label{tab:calfactors}}
\end{deluxetable}

\clearpage

\bibliographystyle{aasjournal}
\bibliography{disk_exop_refs}

\end{document}